\PassOptionsToPackage{table}{xcolor}
\documentclass[acmtog, nonacm=true]{acmart}
\acmSubmissionID{251}
\usepackage{subcaption}
\usepackage{algorithm}
\usepackage{algpseudocode}
\usepackage{graphicx}
\usepackage{wrapfig}
\usepackage{siunitx}
\usepackage{cancel}
\usepackage{booktabs}
\usepackage[table]{xcolor}
\usepackage{makecell}
\AtBeginDocument{%
  \providecommand\BibTeX{{%
    \normalfont B\kern-0.5em{\scshape i\kern-0.25em b}\kern-0.8em\TeX}}}
    
\makeatletter
\renewcommand*\env@matrix[1][\arraystretch]{%
  \edef\arraystretch{#1}%
  \hskip -\arraycolsep
  \let\@ifnextchar\new@ifnextchar
  \array{*\c@MaxMatrixCols c}}
\makeatother

\newcommand{\rig}{\mathcal{R}}
\newcommand{\trk}{\mathcal{T}}
\newcommand{\trkh}{\mathcal{T}_{\mathcal{H}}}
\newcommand{\trkd}{\mathcal{T}_{D}}
\newcommand{\loss}{\mathcal{L}}

\newcommand{\ess}{\mathcal{S}}
\newcommand{\veee}{\mathcal{V}}
\newcommand{\dd}[2]{\frac{\partial #1}{\partial #2}}

\makeatletter
\def\subsubsection{\@startsection{subsubsection}{3}{\z@}%
  {.5\linespacing \@plus .7\linespacing}{.2\linespacing}%
  {\ACM@NRadjust{\@subsubsecfont}}}
\makeatother

\begin{document}

\title{Improving Facial Rig Semantics for Tracking and Retargeting}

\author{Dalton Omens}
\email{domens@stanford.edu}
\affiliation{%
  \institution{Stanford University}
  \city{Stanford}
  \state{California}
  \country{USA}
}
\affiliation{%
  \institution{Epic Games}
  \city{Cary}
  \state{North Carolina}
  \country{USA}
}

\author{Allise Thurman}
\email{allise@stanford.edu}
\affiliation{%
  \institution{Stanford University}
  \city{Stanford}
  \state{California}
  \country{USA}
}

\author{Jihun Yu}
\email{jihun.yu@epicgames.com}
\affiliation{%
  \institution{Epic Games}
  \city{Cary}
  \state{North Carolina}
  \country{USA}
}

\author{Ronald Fedkiw}
\email{rfedkiw@stanford.edu}
\affiliation{%
  \institution{Stanford University}
  \city{Stanford}
  \state{California}
  \country{USA}
}
\affiliation{%
  \institution{Epic Games}
  \city{Cary}
  \state{North Carolina}
  \country{USA}
}


\begin{abstract}
In this paper, we consider retargeting a tracked facial performance to either another person or to a virtual character in a game or virtual reality (VR) environment. We remove the difficulties associated with identifying and retargeting the semantics of one rig framework to another by utilizing the same framework (3DMM, FLAME, MetaHuman, etc.) for both subjects. Although this does not constrain the choice of framework when retargeting from one person to another, it does force the tracker to use the game/VR character rig when retargeting to a game/VR character. We utilize volumetric morphing in order to fit facial rigs to both performers and targets; in addition, a carefully chosen set of Simon-Says expressions is used to calibrate each rig to the motion signatures of the relevant performer or target. Although a uniform set of Simon-Says expressions can likely be used for all person to person retargeting, we argue that person to game/VR character retargeting benefits from Simon-Says expressions that capture the distinct motion signature of the game/VR character rig. The Simon-Says calibrated rigs tend to produce the desired expressions when exercising animation controls (as expected). Unfortunately, these well-calibrated rigs still lead to undesirable controls when tracking a performance (a well-behaved function can have an arbitrarily ill-conditioned inverse), even though they typically produce acceptable geometry reconstructions. Thus, we propose a fine-tuning approach that modifies the rig used by the tracker in order to promote the output of more semantically meaningful animation controls, facilitating high efficacy retargeting. In order to better address real-world scenarios, the fine-tuning relies on implicit differentiation so that the tracker can be treated as a (potentially non-differentiable) black box.
\end{abstract}

\maketitle


\section{Introduction}
\label{sec:introduction}

\sloppy Faces are ubiquitous in digital media such as games and film. Performance-driven facial animation (as opposed to manually keyframed animation) is often used to improve quality, to reduce manual labor, and to facilitate real-time applications. The impact of performance-driven capture is significantly bolstered by the ability to transfer, or \textit{retarget}, captured performances to other human or character models. Although facial motion retargeting has traditionally been limited to big-budget feature films and AAA games due to the need for artist-driven cleanup and intervention, more robust (and democratized) methods would have an enormous impact on a large variety of real-time applications especially given the surging interest in the so-called metaverse.

The computer vision community has made leaps and bounds in the reconstruction of faces using a variety of techniques, and the higher fidelity approaches tend to utilize parameterized models (such as 3DMM \cite{blanzvetter1999}, FLAME \cite{li2017flame}, etc.) in order to regularize the reconstruction. Although it is not entirely clear whether or not a computer graphics style semantically-meaningful animation rig (such as is available in Maya \cite{maya}, in Blender's Rigify \cite{blender}, or in Unreal Engine's MetaHuman plugins \cite{metahuman}, etc.) could achieve better geometry reconstructions than a PCA-based rig (such as 3DMM, FLAME, etc.), it is clear that most of the progress has occurred with these latter non-semantic rigs that are prevalent in the computer vision community. If the goal were merely to reconstruct a real-world performance in a virtual world, perhaps with different lighting and textures as well as novel views, then non-semantic PCA-style geometry reconstruction is likely sufficient; however, retargeting of that performance even if only to a younger/older version of the performer can be problematic without a semantically-meaningful rig that can disentangle reconstruction geometry from semantic intention.

Whether it be PCA-based or semantic, a rig typically separates the identity parameters $N$ that define the geometry in a neutral expressionless pose from the set of expression parameters $\theta$. Given an image $I$, a geometry reconstruction $\veee$ is obtained by solving an inverse problem to find geometry parameters $\ess(I; N, \theta, \psi)$ where $\psi$ represents parameters internal to the solver. Viewing this from the standpoint of an animator, they use animation controls to determine geometry parameters $\ess$ based on an (actual or mentally envisioned) image $I$. They choose the animation controls based on knowledge of how the rig works (even if this is acquired via trial and error), meaning that $\ess$ still depends on $N$ and $\theta$; in this scenario, $\psi$ represents parameters internal to the brain of the animator. Regardless of whether $\ess$ is determined via a solver or via an animator, $\veee(\mathcal{S}(I; N, \theta, \psi); N, \theta)$ represents the reconstructed geometry. 

Assuming that the same rig framework is used for both subjects, in order to remove the difficulties associated with identifying and retargeting the semantics of one rig framework to another, retargeting can be expressed as $\veee(\mathcal{S}(I; N, \theta, \psi); \hat{N}, \hat{\theta})$ where $\hat{N}$ and $\hat{\theta}$ are the identity and expression parameters of the target. This expression highlights the main issue with retargeting, i.e. that $\ess$ is unaware of $\hat{N}$ and $\hat{\theta}$. To elucidate this, let $\veee_o = \veee(\mathcal{S}(I; N, \theta, \psi); N, \theta)$ be geometry reconstructed from an input image $I$ of a performer, and let $\hat{\veee}_o$ be the preferred geometry on the target; then, $\hat{\veee}_o = \veee(\mathcal{S}(\hat{I}; \hat{N}, \hat{\theta}, \psi); \hat{N}, \hat{\theta})$ where $\hat{I}$ is an image of $\hat{\veee}_o$. Unfortunately, $\veee(\mathcal{S}(I; N, \theta, \psi); \hat{N}, \hat{\theta})$ will typically differ significantly from $\hat{\veee}_o$ due to the disparity between $\ess(I; N, \theta, \psi)$ and $\ess(\hat{I}; \hat{N}, \hat{\theta}, \psi)$. 

An expert animator working with a high quality rig would determine $\ess$ by choosing animation controls while thinking about $N$ and $\theta$ abstractly in a manner that includes a dependence on $\hat{N}$, $\hat{\theta}$, and $\hat{I}$. Some high-end trackers are designed to mimic this approach in various ways; for example, certain degrees of freedom may be preferred or penalized in the objective function. Some trackers aim for an optimal geometry reconstruction in the first pass while subsequently reworking the degrees of freedom with preferences and penalties in a second pass. In fact, it is sometimes useful to allow for a varying identity in the first pass (as this can be used to explain errors in camera extrinsics, etc.), while rigidifying the identity in the second pass; in contrast, rigidifying the identity in the first pass could cause erroneous values for expression parameters as they aim to explain various errors. 

In spite of the various strategies used to improve the output of high-end trackers, expert animators still spend a great deal of time cleaning the output (i.e. modifying $\ess$ via animator controls) before it can be used for retargeting. Thus, in this paper, we aim to improve high-end trackers even further by eliminating the disparity between $\ess(I; N, \theta, \psi)$ and $\ess(\hat{I}; \hat{N}, \hat{\theta}, \psi)$. The maturity of reconstruction techniques means that $N$ and $\hat{N}$ are well-determined. In addition, a high-end tracker likely has a well-engineered solver with optimal parameters $\psi$. Thus, we focus on the expression parameters stressing that motion signatures can vary greatly from person to person. That is, our main idea is to modify the original $\theta$ to a new $\theta_{\ess}$ in order to obtain $\ess(I; N, \theta_{\ess}, \psi) \approx \ess(\hat{I}; \hat{N}, \hat{\theta}, \psi)$. In order to cover the full range of expression parameters, a large number of image pairs $I$ and $\hat{I}$ would be required. Unfortunately, a different $\theta_{\ess}$ needs to be determined for every performer/target pair. Once again, we rely on animator mimicry in order to address this problem. The main idea is to use agreement in animator controls as a proxy for agreement in geometry degrees of freedom. The prior implies the latter but not necessarily vice versa, so this is typically a stricter condition.

In summary, we address the scenario where real-world performance is used to drive a virtual model of another person or game/VR character. In the first case, the same rig framework is volumetrically morphed to 3D reconstructions of both the performer and the target; in the second case, the carefully crafted game/VR character rig is volumetrically morphed to a 3D reconstruction of the performer. A carefully chosen set of Simon-Says expressions (see \cite{zhu2024democratizingcreationanimatablefacial}) is used to calibrate each rig to the motion signatures of the relevant performer or target; for performer to game/VR character retargeting, the Simon-Says expressions should capture the motion signatures of the game/VR character rig. Unfortunately, these well-calibrated rigs still lead to undesirable controls when tracking a performance. Thus, we propose a fine-tuning approach that modifies the expression parameters used by the tracker in order to promote the output of more semantically meaningful animation controls, facilitating high efficacy retargeting. The fine-tuning relies on implicit differentiation so that a high-end potentially non-differentiable black box tracker can be utilized.

We summarize our novel contributions as follows:
\begin{itemize}
    \item We provide a mathematical analysis of rigs and trackers via linearization in order to rigorously demonstrate that a tracker can be thought of as a geometry reconstruction followed by the inverse of a rig operation.
    \item In the context of Simon-Says, we choose a set of expressions that covers the most important rig parameters while still remaining minimal with expressions that are straightforward for an average person to make. In addition, we ensure that the numerical optimization respects the various constraints expert riggers utilize in the rig design process.
    \item Our implicit function theorem style approach allows for modification of the tracker parameters via an implicit approximation of the derivative of the tracker output with respect to those parameters without requiring access to or knowledge of the internal workings of the tracker (the tracker can be treated as a black box) and without requiring differentiability. 
    \item We present a reformulation of Broyden's method illustrating its connection to finite difference estimates and subsequently show how to substitute in more robust finite differences that minimize truncation error while avoiding roundoff errors. These observations allow us to improve the search direction.
    \item We illustrate that our approach is both feasible on and has high efficacy on real-world problems by utilizing best practices at each stage of the process: for example, we propose techniques aimed at obtaining accurate reconstructed geometry, we use volumetric morphing to obtain tight fitting high-end rigs that produce high fidelity motion, we worked with animators and riggers to get domain-expert feedback, etc.
\end{itemize}


\section{Related Work}
\label{sec:relatedwork}

There is a plethora of work aimed at representing faces in digital worlds. Although the seminal efforts relied quite heavily on the ability of artists to sculpt, paint, and animate both virtual characters and digital doubles, the impact of data-driven approaches has increased significantly over the years. A number of works mostly aim primarily to \textit{disentangle the camera view} from a virtual model of the subject of interest, rasterizing photorealistic images from novel views. The most notable, e.g. \cite{gafni2021dynamicneuralradiance,grassal2021neural,zheng2022imavatar,duan2023bakedavatar,zheng2023pointavatar,xu2023gaussianheadavatar,xiang2024flashavatar,cao2022authenticvolumetric,lombardi2018deepappearance,trevithick2023realtime,kirschstein2023nersemble,buehler2023preface,teotia2024hq3davatar,wu2023ganhead,zielonka2022InstantVH,sarkar2023litnerf,yang2024vrrm,buehler2024cafca}, utilize either NeRFs \cite{mildenhall2021nerf} or Gaussian Splats \cite{kerbl2023gaussiansplat}. See also \cite{ma2021pixelcodec}.

Going one step further, a plethora of work has been aimed at geometry reconstruction, which additionally (i.e. in addition to disentangling the camera viewpoint) \textit{disentangles three-dimensional geometry from texture and lighting}. Many such works utilize a PCA representation, typically either a 3DMM \cite{blanzvetter1999,egger20203dmorphable} or FLAME \cite{li2017flame}, for regularization. See e.g. \cite{li2009robustsingleview,laine2017productionlevel,zhang2023accurate3dface,taubner2024flowface,hewitt2024lookmanomarkers,zielonka2022mica,retsinas2024smirk,wang20243dfacereconstruction,danecek2022emoca,wood2022dense}. Other works use anatomical constraints \cite{beeler2014rigidstabilization,wu2016anatomicallyconstrained} or sculpted blend shapes \cite{garrido2013reconstructing} for regularization. Similar in spirit to our approach, \cite{baert2024spark} points out that geometric reconstruction results can be improved by personalizing the FLAME expression coefficients.

Independent of geometry, various works aim to additionally \textit{disentangle texture from lighting}, see e.g. \cite{bharadwaj2023flare,tweari2017mofa,deng2019accurate3d,dib2021towardshigh,dib2021practical,dib2023s2f2,lei2023hrn,zhou2024ultravatar,han2024cora,dib2024mosar,saito2024rgca,athar2024bridging}. The high-end approaches utilize a light stage \cite{debevec2000acquiring,ghosh2011multiviewfacecapture,saito2017photorealistic}.

When retargeting, it is important to \textit{disentangle semantic intention from geometry}. Although there is a plethora of computer vision work interested in semantic intention for the sake of scene interpretation, these works are rarely interested in the 3D reconstruction of facial geometry. Those who are focused on both typically work in the field of computer graphics and make use of so-called facial animation rigs, see e.g. \cite{cani2012afacialriggingsurvey}. Notable work on facial animation rigs includes \cite{bao2019highquality,debevec2000acquiring,zhang2024spacetime,li2010examplebasedfacial,bao2015fullyautomatic,yang2023towardspractical,bailey2020fastanddeepfacial,yang2023animplicitphysical,kavan2024compressed,ming2024highquality,choi2022animatomy}. Publicly available options for facial animation software include MetaHumans in the Unreal Engine \cite{metahuman} as well as various tools suites in Maya \cite{maya} and Blender \cite{blender}.

For a review of prior work on the \textit{retargeting} of facial performances to other persons or characters, see \cite{zhu2024motionretargetingsurvey}. Although some works have found a degree of success by transferring \textit{per-vertex} geometry displacements from one mesh to another, see e.g. \cite{qin2023NFR,noh2001expressioncloning,sumner2004deformationtransfer,seol2012spacetimeexpression,cha2025neuralfaceskinning,choi2025deeplearningretargeting}, most focus on the underlying rig degrees of freedom. Although some works have aimed to find reasonable mappings between \textit{disparate rig frameworks}, see e.g. \cite{song2011characteristicfacial,deng2006crossmapping,wang2004highresolution}, this is still a difficult problem. Various authors have aimed to address this difficult problem via neural network techniques, see e.g. \cite{kim2021deeplearningbased,aneja2018learningtogenerate,zhang2022facialexpression}, while others have relied on constrained numerical optimization, e.g. \cite{chandran2022localanatomically} uses anatomical constraints.

Some approaches have aimed to skip the retargeting of geometry and/or rig controls entirely by instead using \textit{image-based retargeting} followed by reconstruction or rig inversion from the retargeted image. \cite{moser2021semisupervised} uses a network to translate an image of the performer to a semantically similar image of the target and regress animation controls from the new image. \cite{qiu2024freeavatar} encodes an image of the performer into a latent space and subsequently uses a target-specific decoder to obtain animation controls. With the recent surge in popularity of LLMs, there is also some interest in audio-based retargeting followed by reconstruction or rig inversion from the retargeted audio. In this vein, \cite{pan2025VASA} recently proposed a method for regressing MetaHuman animator controls for an audio signal.

High-end approaches tend to use the \textit{same rig frameworks} on both the performer and target, aiming for a one-to-one remapping of rig control parameters. In order to obtain high efficacy results, the semantic intention of the performance needs to be disentangled from the geometry. \cite{tran2024voodooxp} and \cite{bai2024universalfacial} focus on achieving realistic person-to-person retargeting within the constrained input domain of VR headset cameras. \cite{rackovic2024refinedinverserigging} proposes a new optimization framework aiming to obtain semantically-meaningful animation controls that retarget well. Similar in spirit to our approach, \cite{ribera2017facialretargeting} uses a Range of Motion (ROM) video to modify expression blendshapes aiming to obtain semantically-meaningful animation controls. \cite{zhu2024facialmotionretargeting} proposes a new FACS-based rig designed to improve the retargeting efficacy of regressed animation controls.


\section{Preliminaries}
\label{sec:preliminaries}

Similar to $\veee$ (from Section \ref{sec:introduction}), animation rigs $\rig (c; N, \theta_{\rig})$ output geometry. The main difference is that the geometry degrees of freedom in a rig are specified directly (instead of indirectly, i.e. Section \ref{sec:introduction} assumes that $\ess$ is derived from $c$) by the animation controls $c$. Importantly, the animation controls are typically designed to have interpretable semantic meaning, whereas PCA coefficients simply represent geometric modes. Given a 3D geometric reconstruction of the neutral pose $N$ and a set of corresponding animation controls versus geometry pairs $(c_k, v_k)$,
\begin{equation} \label{eq:sculptopt}
    \min_{\theta_{\rig}} \sum_k ||\rig (c_k; N, \theta_{\rig}) - v_k||_2^2
\end{equation}
can be used to determine the internal parameters $\theta_{\rig}$ of the rig. Each of the $v_k$ can either be sculpted by an artist or reconstructed from an image $I_k$ chosen to correspond with $c_k$. Various methods (e.g. volumetric morphing \cite{zhu2024democratizingcreationanimatablefacial}) can be used to retarget a rig to a new character, allowing an artist to co-animate both characters with the same artist controls. Unfortunately, the expressions on the new character will closely resemble those of the old character. This can be remedied by laboriously sculpting geometry $v_k$ for a number of key poses (each specified by a $c_k$) on the new character; then, Equation \ref{eq:sculptopt} can be solved to determine new internal parameters for the new character's rig. When retargeting to another person, sculpting can be avoided by using a Simon-Says approach (see Section \ref{sec:simonsays}) to create semantically correlated images $I_k$ that can be reconstructed into geometry $v_k$ via inverse rendering (see Section \ref{sec:geometryreconstruction}) and other techniques.

Similar to $\ess$ (from Section \ref{sec:introduction}), a tracker $\trk(I; \psi)$ solves an inverse problem to find animation controls $c$ that produce geometry when evaluated in the rig $\rig(c; N, \theta_{\rig})$. Generally speaking, a tracker does not necessarily need to depend on a rig; thus, $N$ and $\theta_{\rig}$ are temporarily (for the sake of this section) absorbed into $\psi$ if the tracker depends on them. Given a set of corresponding animation controls versus image pairs $(c_k, I_k)$,
\begin{equation} \label{eq:trackerobj}
    \min_\psi \sum_k ||\trk (I_k; \psi) - c_k||_2^2
\end{equation}
can be used to determine the internal tracker parameters. In order to improve the tracker's ability to robustly generalize to unseen images, regularization is typically required. This is sometimes accomplished by considering the geometry $\rig (\trk (I; \psi); N, \theta_{\rig}) + \tilde{v}_{\trk }(I; \psi)$ obtained by evaluating the tracker output in the corresponding rig. Here, $\tilde{v}_{\trk }$ is a correction to the geometry, often included when a rig is thought to have limited expressivity. This motivates a modification of Equation \ref{eq:trackerobj} to
\begin{equation} \label{eq:trackerobj2}
\begin{split}
  \min_\psi \sum_k ~ & \gamma_1 ||\trk (I_k; \psi) - c_k||_2^2 ~ + \\[-1.5ex] & \gamma_2 ||\rig (\trk (I_k; \psi); N, \theta_{\rig}) + \tilde{v}_{\trk }(I_k; \psi) -  v_k||_2^2 ~ + \\  & \gamma_{\tilde{v}_{\trk}} || \tilde{v}_{\trk}(I_k; \psi) ||_2^2
\end{split}
\end{equation}
in order to match both the artist controls and the geometry. The last term in Equation \ref{eq:trackerobj2} regularizes $\tilde{v}_{\trk}$ to be small, so that the geometry $v_k$ is mostly matched by the rig. With regard to terminology, a would-be tracker that outputs good geometry but poor artist controls (focusing on the second term in Equation \ref{eq:trackerobj2} at the expense of the first) is typically thought of as a reconstruction method, not an expression tracker.

Note that the first term in Equation \ref{eq:trackerobj2} expresses the ``agreement in animator controls'' mentioned in Section \ref{sec:introduction} as a proxy for agreement in geometry degrees of freedom. That is, changing the minimization in Equation \ref{eq:trackerobj2} to be over ``$\theta_{\ess}$'' instead of $\psi$ would essentially achieve the desired result. Importantly, this only has to be done once per performer/rig combination independent of the number of target characters. 


\section{A Motivational Linearization}
\label{sec:motivationallinearization}


For the sake of simplicity, assume that the rig output $v$ consists of vertex displacements from the neutral pose so that $\rig $ does not necessarily depend on $N$; in addition, assume that the rig is linear in $c$ implying 
\begin{equation} \label{eq:vac0}
v = A(\theta_{\rig}) c
\end{equation}
where $A(\theta_{\rig})$ is a size $m \times n$ matrix. Stacking all the $(c_k, v_k)$ pairs into separate matrices $C$ and $V$ yields  
\begin{equation} \label{eq:vac}
    V = A(\theta_{\rig}) C
\end{equation}
\begin{equation} \label{eq:ctatvt}
    C^T A^T(\theta_{\rig}) = V^T
\end{equation}
\begin{equation} \label{eq:anormaleqs}
    A(\theta_{\rig}) = V C^T (C C^T)^{-T} = V C^T (C C^T)^{-1}
\end{equation}
illustrating that $A$ is unique as long as the $c_k$ are chosen to make $C^T$ full rank. If the expression set does not make $C^T$ full rank, then Equation \ref{eq:ctatvt} can be modified to be full rank by adding an additional set of equations $\epsilon I_{nxn} A^T(\theta_{\rig}) = \epsilon A^T(\theta_0) $ where $\theta_0$ are default values for the rig (perhaps from a template rig). As is typical for Levenberg-Marquardt, $\epsilon$ is chosen to limit the interference of the regularizing equations with $C^T A^T = V^T$. 

Assume that the tracker executes an error-free geometry reconstruction to obtain geometry $v$ from an image $I$; in addition, assume that the tracker is linear in the reconstructed geometry $v$ so that 
\begin{equation} \label{eq:cahatv0}
    c = B(\psi) v
\end{equation}
where $B(\psi)$ is a size $n \times m$ matrix. $B$  satisfies
\begin{equation} \label{eq:cahatv}
    C = B(\psi) V
\end{equation}
\begin{equation}
    C C^T  (C C^T)^{-1} = B(\psi) V C^T (C C^T)^{-1}
\end{equation}
\begin{equation} \label{eq:inxneqba}
    I_{nxn} = B(\psi) A(\theta_{\rig})
\end{equation}
implying that $B$ is a size $n \times m$ left inverse of $A$. \textit{This notion of a tracker consisting of a geometry reconstruction (from $I$ to $v$) followed by the ``inverse'' of a rig operation (Equation \ref{eq:cahatv0}) is conceptually useful.} If $C^T$ is not full rank, the additional rows of $C^T$ and $V^T$ that were added to Equation \ref{eq:ctatvt} need to be included as additional columns in Equation \ref{eq:cahatv} in order to derive Equation \ref{eq:inxneqba}. These can be seen as additional $(c_k, v_k)$ pairs.

Now, instead suppose that the tracker erroneously reconstructs geometry $\hat{v}_k$ from the images $I_k$; then, Equation \ref{eq:cahatv} instead becomes
\begin{equation} \label{eq:cbvhat}
    C = \hat{B}(\psi) \hat{V}
\end{equation}
where $\hat{B}$ rectifies errors in geometry reconstruction. Similar to Equations \ref{eq:vac} to \ref{eq:anormaleqs}, Equation \ref{eq:cbvhat} leads to 
\begin{equation} \label{eq:bnormaleqs}
     \hat{B}(\psi) = C \hat{V}^T (\hat{V} \hat{V}^T)^{-1}
\end{equation}
when $\hat{V}^T$ is full rank; otherwise, equations of the form $\epsilon I_{mxm} \hat{B}^T(\psi) = \epsilon \hat{B}^T(\psi_0)$ can be included. Note that the regularization of $\hat{V}^T$ adds columns to $C$ and $\hat{V}$ that are entirely different from the columns that were added to $C$ and $V$ in order to regularize $C^T$ in Equation \ref{eq:ctatvt}. Thus, there are two reasons that $\hat{B}$ in Equation \ref{eq:bnormaleqs} is no longer a left inverse of $A$ in Equation \ref{eq:anormaleqs}. Not only have reconstruction errors caused $V$ to change to $\hat{V}$, but the rig-solve based regularization used to obtain Equation \ref{eq:inxneqba} is different from the tracker-solve based regularization used to obtain Equation \ref{eq:bnormaleqs}. Since $\hat{B}$ is no longer a left inverse of $A$, the rig used for the ``inverse'' operation in the tracker \textit{should} be inconsistent with the animation rig.

Rewriting Equation \ref{eq:cbvhat} from the standpoint of a solver that ``inverts'' the rig, the tracker solves a linear system of the form 
\begin{equation} \label{eq:bcv0}
    \hat{A}(\theta_{\trk}) C = \hat{V}
\end{equation}
to determine $C$. This leads to
\begin{equation} \label{eq:ahatisnota}
    \hat{A}(\theta_{\trk}) = \hat{V} C^T (C C^T)^{-1} \neq V C^T (C C^T)^{-1} = A(\theta_{\rig})
\end{equation}
where the regularization of $C$ and $\hat{V}$ is the same as the regularization of $C$ and $V$ in Equation \ref{eq:ctatvt}; thus, the only difference between the left hand side and right hand side of Equation \ref{eq:ahatisnota} is $\hat{V}$ versus $V$. Here, $\hat{A}$ denotes the perturbed rig used by the tracker; thus, $\hat{A}$ is written to depend on parameters $\theta_{\trk} \neq \theta_{\rig}$. Starting with Equation \ref{eq:bcv0} instead of Equation \ref{eq:vac} leads to an Equation \ref{eq:inxneqba} assertion that $B$ should be the left inverse of $\hat{A}$. However, actually finding such a $B$ (equivalent to running a tracker) requires regularization similar to what was used to obtain Equation \ref{eq:bnormaleqs}. Thus, $\hat{A}$ should be perturbed even further (beyond Equation \ref{eq:ahatisnota}) in order to account for regularization in the solver.


\section{Determining a Rig for the Tracker's ``Solve''}
\label{sec:determiningarig}

Going forward, we will write $\trk (I; \theta_{\trk}, \psi)$ to stress that the tracker depends on the rig parameters in some perhaps unknown or complex fashion. We drop the explicit mention of the dependence on the neutral $N$ since it has no bearing on the discussion. In fact, $\psi$ could be omitted as well, since only $\theta_{\trk}$ will be modified; however, $\psi$ is left in as a placeholder for non-modified tracker parameters (technically, $N$ could be absorbed into $\psi$).

Although the linearization in section \ref{sec:motivationallinearization} provides motivation, the vast majority of state-of-the-art trackers will be nonlinear, complex, and not readily differentiable. We only assume that the tracker has been trained to work via Equation \ref{eq:trackerobj2}, or a similar in spirit approach. Thus, we propose using a modified version of Equation \ref{eq:trackerobj2}, i.e. 
\begin{equation} \label{eq:trackerobj4}
\begin{split}
   \min_{\theta_{\trk}} \sum_k \Big(  & \gamma_1 ||\trk (I_k; \theta_{\trk}, \psi) - c_k||_2^2 ~ + \\[-1.5ex] & \gamma_2 ||\rig (\trk (I_k; \theta_{\trk}, \psi); \theta_{\rig}) - v_k||_2^2 \Big)  ~ + \\ & \gamma_{\epsilon} || \theta_{\trk} - \theta_{\rig} ||_2^2
\end{split}
\end{equation}
in order to fine-tune the tracker by modifying the $\theta_{\trk}$ parameters of the underlying animation rig. Note that $\tilde{v}_{\trk}$ has been eliminated, so that modifying $\theta_{\trk}$ does not encourage the rig to drift away from its ability to match the ground-truth geometry; in addition, this removes the need to estimate derivatives of $\tilde{v}_{\trk}$ with respect to $\theta_{\trk}$. Finally, the $\gamma_{\epsilon}$ term is added.

In our experience, most trackers have been trained to provide good geometric reconstruction. This means that the $\gamma_2$ term in Equation \ref{eq:trackerobj4} will be small when $\theta_{\trk} = \theta_{\rig}$. However, in our experience, trackers often struggle to output semantically correct rig controls. This means that the $\gamma_1$ term in Equation \ref{eq:trackerobj4} will typically be large when $\theta_{\trk} = \theta_{\rig}$. Thus, our goal is to minimize the $\gamma_1$ term while staying close to the constraint surface that keeps the $\gamma_2$ term small. The $\gamma_{\epsilon}$ term emphasizes that the rig is being fine-tuned, softly constraining the minimization of the $\gamma_1$ term on or near the $\gamma_2$ constraint surface to occur within a ball about $\theta_{\rig}$; otherwise, $\theta_{\trk}$ can drift too much (which we have observed in our experiments).

The $\gamma_2$ term is important since it emphasizes that the output of the tracker will subsequently be used in the animation rig, which depends on $\theta_{\rig}$. The observant reader might wonder whether the rig used internal to the tracker also requires geometry regularization. This can be accomplished by including 
\begin{equation} \label{eq:trackerobj6}
    \gamma_3 ||\rig (\trk (I_k; \theta_{\trk}, \psi); \theta_{\trk}) - v_k||_2^2
\end{equation}
in Equation \ref{eq:trackerobj4}. Our preliminary tests (see Section \ref{sub:example_2}) obviate the need for this term.


\section{Differentiating the Tracker}
\label{sec:differentiatingtracker}

In some scenarios, one may not have access to the internal workings of the tracker. In other scenarios, the tracker may be open-source, albeit nonlinear, complex, and/or not readily differentiable. In either scenario, $\dd{\trk }{\theta}$ cannot be computed explicitly. Thus, in the spirit of Equation \ref{eq:trackerobj6}, we make the reasonable ansatz that the tracker actually uses its internal rig in order to provide some sort of regularization; that is, \textit{we assume that the controls output by the tracker yield credible geometry when evaluated on the rig internal to the tracker.} This can be written as
\begin{equation} \label{eq:rigoutputpluserror}
    \hat{v}(I; \theta_{\trk}, \psi) = v(I) + \tilde{v}(I; \theta_{\trk}, \psi) = \rig (\trk (I; \theta_{\trk}, \psi); \theta_{\trk})
\end{equation}
where $v$ is the ground-truth geometry and $\tilde{v}$ is an erroneous perturbation away from the ground-truth geometry; then, the ansatz that the tracker uses its internal rig to provide reasonable regularization can be restated as a claim that $\tilde{v}$ should be small. Differentiating Equation \ref{eq:rigoutputpluserror} with respect to $\theta$ leads to
\begin{equation} \label{eq:definedvhat}
\begin{split}
    \left. \dd{\tilde{v}(I; \theta, \psi)}{\theta} \right|_{\theta = \theta_{\trk}} \hspace{-0.5em} = \left. \dd{\rig (c; \theta_{\trk})}{c} \right|_{c=\trk (I; \theta_{\trk}, \psi)} \hspace{-0.5em} \left. \dd{ \trk (I; \theta, \psi)}{\theta} \right|_{\theta = \theta_{\trk}} +  \\ \left. \dd{\rig (\trk (I; \theta_{\trk}, \psi); \theta)}{\theta} \right|_{\theta = \theta_{\trk}}
\end{split}
\end{equation}
and thus 
\begin{equation} \label{eq:solvedtdtheta}
\begin{split}
     \left. \dd{\rig (c; \theta_{\trk})}{c} \right|_{c=\trk (I; \theta_{\trk}, \psi)} \left. \dd{\trk (I; \theta, \psi)}{\theta} \right|_{\theta = \theta_{\trk}} = &  \\ -  \left. \dd{ \rig (\trk (I; \theta_{\trk}, \psi); \theta)}{\theta} \right|_{\theta = \theta_{\trk}} + & \left. \dd{\tilde{v}(I; \theta, \psi)}{\theta} \right|_{\theta = \theta_{\trk}}
\end{split}
\end{equation}
as an implicit equation for $\dd{\trk }{\theta}$. Note that $\dd{v}{\theta} = 0$, since $v$ is the ground-truth geometry. When $\frac{\partial \rig }{\partial c}$ is not full rank, equations of the form $\epsilon I_{nxn} \frac{\partial \trk}{\partial \theta} = 0$ can be included in order to regularize the gradients of the controls in the unconstrained subspace; alternatively, when tractable, the pseudo-inverse can be used as well.

Although one could assume that the tracker produces a $\hat{v}$ very close to $v$ making $\tilde{v}$ small enough to remove it from Equation \ref{eq:rigoutputpluserror}, subsequently removing its derivative from Equations \ref{eq:definedvhat} and \ref{eq:solvedtdtheta}, it is also possible to estimate $\dd{\tilde{v}}{\theta}$. In order to estimate $\frac{\partial \tilde{v}}{\partial \theta}$, we utilize the finite difference update at the heart of Broyden's method \cite{broyden1965}, which is also the motivation for many other optimization schemes including SR1 \cite{nodecalwright2006numerical}, DFP \cite{dfp1,dfp2}, BFGS \cite{fletcher1987practical}, and L-BFGS \cite{liu1989lbfgs}.

For the sake of exposition, let $u$ be some instance of $\tilde{v}$ for a given $I$ and $\psi$, so that $u$ only varies with $\theta_{\trk}$ during optimization.  Given distinct values $\theta_{\alpha}$ and $\theta_{\beta}$ with $\Delta \theta = \theta_{\beta} - \theta_{\alpha}$, an estimate to $\frac{\partial u}{\partial \theta}$ at $\theta_{\alpha}$ can be updated via 
\begin{equation} \label{eq:rankoneupdate}
\begin{split}
    \left( \left. \frac{\partial u}{\partial \theta} \right|_{\theta = \theta_{\alpha}} \right)^{new} = \left( \left. \frac{\partial u}{\partial \theta} \right|_{\theta = \theta_{\alpha}} \right)^{old} + \qquad\qquad\qquad\qquad\quad \\ \frac{1}{(\Delta \theta)^T \Delta \theta} \left( u(\theta_{\beta}) - u(\theta_{\alpha}) - \left( \left. \frac{\partial u}{\partial \theta} \right|_{\theta = \theta_{\alpha}} \right)^{old} \Delta \theta \right) (\Delta \theta)^T
\end{split}
\end{equation}
so that
\begin{equation} \label{eq:secant}
    \left( \left. \frac{\partial u}{\partial \theta} \right|_{\theta = \theta_{\alpha}} \right)^{new} \Delta \theta = u(\theta_{\beta}) - u(\theta_{\alpha})
\end{equation}
is satisfied. Equation \ref{eq:secant} shows that the estimate for the derivative satisfies a secant-type equation in the direction of $\Delta \theta$, as is typical for Broyden-style methods. In fact, letting $s = ||\Delta \theta||_2 = \sqrt{(\Delta \theta)^T \Delta \theta}$ allows Equations \ref{eq:rankoneupdate} and \ref{eq:secant} to be written as
\begin{equation} \label{eq:rankoneupdatewiths}
\begin{split}
    \left( \left. \frac{\partial u}{\partial \theta} \right|_{\theta = \theta_{\alpha}} \right)^{new} = \left( \left. \frac{\partial u}{\partial \theta} \right|_{\theta = \theta_{\alpha}} \right)^{old}  + \qquad\qquad\qquad\qquad\quad \\ \left( \frac{u(\theta_{\alpha} + s \widehat{\Delta \theta}) - u(\theta_{\alpha})}{s} - \left( \left. \frac{\partial u}{\partial \theta} \right|_{\theta = \theta_{\alpha}} \right)^{old} \widehat{\Delta \theta} \right) \widehat{\Delta \theta}^T
\end{split}
\end{equation}
\begin{equation} \label{eq:secantwiths}
    \left( \left. \frac{\partial u}{\partial \theta} \right|_{\theta = \theta_{\alpha}} \right)^{new} \widehat{\Delta \theta} = \frac{u(\theta_{\alpha} + s \widehat{\Delta \theta}) - u(\theta_{\alpha})}{s}
\end{equation}
where $\widehat{\Delta \theta}$ is a unit vector in the direction of $\Delta \theta$. Equations \ref{eq:rankoneupdatewiths} and \ref{eq:secantwiths} illustrate that Equation \ref{eq:rankoneupdate} is updating $\dd{u}{\theta}$ to be consistent with a finite difference estimate of $\dd{u}{\theta}$ in the direction $\widehat{\Delta \theta}$.

Unlike the typical approach which compiles a running estimate of $\frac{\partial u}{\partial \theta}$ as the optimization steps through parameter space, we use Equation \ref{eq:rankoneupdatewiths} (equivalent to Equation \ref{eq:rankoneupdate}) in order to construct an estimate of $\dd{u}{\theta}$ at $\theta_{\alpha}$. This is accomplished by using Equation \ref{eq:rankoneupdatewiths} repeatedly with different $\widehat{\Delta \theta}$ directions. The various $\widehat{\Delta \theta}$ can be chosen randomly, importance sampled, etc., noting that the estimate of $\frac{\partial u}{\partial \theta}$ is merely used to aid in the choice of the search direction; thus, this approach is perhaps best described as a predictor-corrector method similar in spirit to Nesterov or second-order Runge-Kutta. Finally, note that each $\widehat{\Delta \theta}$ direction can use a varying $s$ in order to ascertain a reasonable finite difference approximation that avoids round-off error while minimizing truncation error. 

The initial guess for $\dd{u}{\theta}$ in the first iteration of Equation \ref{eq:rankoneupdatewiths} can be set to be the final value obtained for $\dd{u}{\theta}$ when iterating Equation \ref{eq:rankoneupdatewiths} in the prior optimization step. This warm start bears resemblance to Broyden's method, and it greatly accelerated convergence in all of our tests. In the first optimization iteration where there is no prior estimate for $\dd{u}{\theta}$, an initial guess of zero can be used. This differs from Broyden's method, which uses the identity because the estimate is typically used in conjunction with matrix inversion.

Note that $u$ cannot actually be an instance of $\tilde{v}$, since $\tilde{v}$ represents unknown error; however, letting $u$ be an instance of $\hat{v}$, which is equivalent to the rig output (see Equation \ref{eq:rigoutputpluserror}), still leads to an estimate of $\frac{\partial \tilde{v}}{\partial \theta}$ since the exact solution $v$ does not vary with $\theta$. Formally, 
\begin{equation} \label{eq:solvedtdthetawithvhat}
\begin{split}
     \left. \frac{\partial \rig (c; \theta_{\trk})}{\partial c} \right|_{c=\trk (I; \theta_{\trk}, \psi)} \left. \frac{\partial \trk (I; \theta, \psi)}{\partial \theta} \right|_{\theta = \theta_{\trk}} = & \\ - \left. \frac{\partial \rig (\trk (I; \theta_{\trk}, \psi); \theta)}{\partial \theta} \right|_{\theta = \theta_{\trk}}  + & \left. \frac{\partial \hat{v}(I; \theta, \psi)}{\partial \theta} \right|_{\theta = \theta_{\trk}}
\end{split}
\end{equation}
can be used as a replacement for Equation \ref{eq:solvedtdtheta}.

Note that it may be more appropriate to use
\begin{equation} \label{eq:replacerigrhs}
\begin{split}
    \rig (\trk (I; \theta_{\trk}, \psi); \theta_{\trk}) + \tilde{v}_{\trk}(I; \theta_{\trk}, \psi)
\end{split}
\end{equation}
on the far right side of Equation \ref{eq:rigoutputpluserror} when $\tilde{v}_{\trk}$ is used as a geometry correction for rigs with limited expressivity (see the discussion leading up to Equation \ref{eq:trackerobj2}). This leads to 
\begin{equation} \label{eq:rigoutputpluserrorrewrite}
\begin{split}
    \hat{v}(I; \theta_{\trk}, \psi) - \tilde{v}_{\trk}(I; \theta_{\trk}, \psi) &= v(I) + \tilde{v}(I; \theta_{\trk}, \psi) - \tilde{v}_{\trk}(I; \theta_{\trk}, \psi) \\ &= \rig (\trk (I; \theta_{\trk}, \psi); \theta_{\trk})
\end{split}
\end{equation}
emphasizing that $\tilde{v} - \tilde{v}_{\trk}$ plays the same role here as $\tilde{v}$ played in Equations \ref{eq:rigoutputpluserror} and \ref{eq:solvedtdtheta}. In some scenarios, $\dd{\tilde{v}_{\trk}}{\theta}$ might be known, and $\tilde{v}$ might be negligible when compared to $\tilde{v}_{\trk}$. Otherwise, $\dd{}{\theta} (\tilde{v} - \tilde{v}_{\trk})$ can be estimated by letting $u$ be an instance of $\tilde{v} - \tilde{v}_{\trk}$ in the prior discussion. Of course, similar to the discussion proceeding Equation \ref{eq:solvedtdthetawithvhat}, $u$ would actually need to be an instance of $\hat{v} - \tilde{v}_{\trk}$, which is the rig output from Equation \ref{eq:rigoutputpluserrorrewrite}, in order to accomplish this.


\section{Simple Linear Examples}
\label{sec:simplelinearexamples}

Assuming linearity of the rig according to Equation \ref{eq:vac0} gives
\begin{equation} \label{eq:dRdc1}
    \frac{\partial \rig(c, \theta)}{\partial c} = A(\theta)
\end{equation}
\begin{equation} \label{eq:dRdtheta1}
    \frac{\partial \rig(c, \theta)}{\partial \theta} =  \begin{bmatrix}
        c^T & 0_{1x3} & 0_{1x3} \\
        0_{1x3} & c^T & 0_{1x3} \\
        0_{1x3} & 0_{1x3} & c^T
    \end{bmatrix}
\end{equation}
where $\dd{\rig}{\theta}$ is size $3 \times 9$ and $\theta_{\ell}$ for $\ell = 1 \dots 9$ are the entries of A enumerated in row-major order. Equation \ref{eq:trackerobj4}, including Equation \ref{eq:trackerobj6}, can be written as
\begin{equation} \label{eq:trackerobj5}
\begin{split}
    \min_{\theta_{\trk}} \sum_k \Big( & \gamma_1 ||\trk (v_k; \theta_{\trk}) - c_k||_2^2 + \\[-1.5ex] & \gamma_2 ||A (\theta_{\rig}) \trk (v_k; \theta_{\trk}) - v_k||_2^2 + \\ & \gamma_3 ||\hat{A} (\theta_{\trk}) \trk (v_k; \theta_{\trk}) - v_k||_2^2 \Big) + \\ & \gamma_{\epsilon} || \theta_{\trk} - \theta_{\rig} ||_2^2
\end{split}
\end{equation}
after removing the unused tracker parameters $\psi$ and assuming that the tracker perfectly reconstructs $v$ from $I$, eliminating $I$ entirely. Recall that $\hat{A} \neq A$ from Equation \ref{eq:ahatisnota} and the discussion thereafter. 


Assuming linearity of the tracker according to Equation \ref{eq:cahatv0} led to the notion of a tracker that solves a linear system $\hat{A} c = v$. Letting $\hat{A}^{-1} v$ represent the result obtained by the tracker when solving this linear system leads to 
\begin{equation} \label{eq:trackerobj8}
\begin{split}
    \min_{\theta_{\trk}} \sum_k \Big( & \gamma_1 ||\hat{A}^{-1}(\theta_{\trk}) v_k - c_k||_2^2 + \\[-1.5ex] & \gamma_2 ||A (\theta_{\rig}) \hat{A}^{-1}(\theta_{\trk}) v_k - v_k||_2^2 + \\ & \gamma_3 ||\hat{A} (\theta_{\trk}) \hat{A}^{-1}(\theta_{\trk}) v_k - v_k||_2^2 \Big) + \\ & \gamma_{\epsilon} || \theta_{\trk} - \theta_{\rig} ||_2^2
\end{split}
\end{equation}
in place of Equation \ref{eq:trackerobj5}. Note that errors in the tracker solve, regularization, etc. typically cause $\hat{A}$ and $\hat{A}^{-1}$ to not necessarily cancel. In the space spanned by the $v_k$, the second term aims to make $\hat{A}^{-1}$ the right inverse of $A$ (driving $\theta_{\trk}$ to $\theta_{\rig}$ in a subspace), while the third term aims to make $\hat{A}$ and $\hat{A}^{-1}$ the left/right inverses of each other. When $A(\theta_{\rig}) c_k = v_k$, the second term is merely a scaling of the first term. When $A(\theta_{\rig}) c_k \neq v_k$, which arises when an inconsistent set of $(c_k, v_k)$ pairs makes Equation \ref{eq:ctatvt} overdetermined, the second term provides information beyond that of the first term (penalizing the solution to better match the geometry). 


Solving Equation \ref{eq:trackerobj5} requires the computation of $\dd{\trk}{\theta}$ for each of the $k$ terms in the sum. In this linearized scenario, Equations \ref{eq:rigoutputpluserror} and \ref{eq:solvedtdthetawithvhat} become
\begin{equation} \label{eq:vhateqahattrk}
    \hat{v}(v; \theta_{\trk}) = \hat{A}(\theta_{\trk})\trk (v; \theta_{\trk})
\end{equation}
\begin{equation} \label{eq:simpleexamplesolvedtdtheta}
\begin{split}
    \hat{A}(\theta_{\trk}) \dd{\trk (v; \theta_{\trk})}{\theta_{\trk}} = \qquad\qquad\qquad\qquad\qquad\qquad\qquad\qquad \\ - \begin{bmatrix} \trk (v; \theta_{\trk})^T & 0_{1x3} & 0_{1x3} \\ 0_{1x3} & \trk (v; \theta_{\trk})^T & 0_{1x3} \\ 0_{1x3} & 0_{1x3} & \trk (v; \theta_{\trk})^T \end{bmatrix} + \dd{\hat{v}(v; \theta_{\trk})}{\theta_{\trk}} 
\end{split}
\end{equation}
where $\trk(v; \theta_{\trk})$ is found by solving $\hat{A} c = v$. This solve can be accomplished via the inverse (when it exists), least squares (when $\hat{A}$ is full rank), or the pseudo-inverse (to obtain the minimum norm solution, when $\hat{A}$ is not full rank); optionally, the equations can be augmented with Levenberg-Marquardt regularization and solved via the normal equations. In order to estimate $\dd{\hat{v}}{\theta}$, Equation \ref{eq:rankoneupdatewiths} is iterated for various $\widehat{\Delta\theta}$ directions; afterwards, Equation \ref{eq:simpleexamplesolvedtdtheta} can be solved independently for each column of $\dd{\trk}{\theta}$. 

For completeness, the derivatives of the four terms in Equation \ref{eq:trackerobj5} are 
\begin{equation} \label{eq:dl_gamma1_dtheta}
    \dd{\loss_{\gamma_1}}{\theta_{\trk}} = 2 \gamma_1 \sum_{k} (\trk(v_k; \theta_{\trk}) - c_k)^T \dd{\trk(v_k; \theta_{\trk})}{\theta_{\trk}} 
\end{equation}
\begin{equation}
\begin{split}
    \dd{\loss_{\gamma_2}}{\theta_{\trk}} = 2 \gamma_2 \sum_{k} (A(\theta_{\rig}) \trk(v_k; \theta_{\trk}) - v_k)^T A(\theta_{\rig}) \dd{\trk(v_k; \theta_{\trk})}{\theta_{\trk}}
\end{split}
\end{equation}
\begin{equation} \label{eq:dloss3dthetat}
\begin{split}
    \dd{\loss_{\gamma_3}}{\theta_{\trk}} = 2 \gamma_3 \sum_{k} (\hat{A}(\theta_{\trk}) \trk(v_k; \theta_{\trk}) - v_k)^T \Bigg( \hat{A}(\theta_{\trk}) \dd{\trk(v_k; \theta_{\trk})}{\theta_{\trk}} + \\ \begin{bmatrix}
        \trk(v_k; \theta_{\trk})^T & 0_{1x3} & 0_{1x3} \\
        0_{1x3} & \trk(v_k; \theta_{\trk})^T & 0_{1x3} \\
        0_{1x3} & 0_{1x3} & \trk(v_k; \theta_{\trk})^T
    \end{bmatrix}  \Bigg) \qquad
\end{split}
\end{equation}
\begin{equation} \label{eq:dloss4dthetat}
    \dd{\loss_{\gamma_{\epsilon}}}{\theta_{\trk}} = 2 \gamma_{\epsilon} (\theta_{\trk} - \theta_{\rig})^T
\end{equation}
where the term in parentheses in Equation \ref{eq:dloss3dthetat} would be replaced by
\begin{equation}
\begin{split}
    \dd{\hat{A}}{\theta_{\trk}}  = \left. \dd{\hat{A}(c; \theta_{\trk})}{c}  \right|_{c = \trk(v_k; \theta_{\trk})} \hspace{-3ex} \dd{\trk(v_k; \theta_{\trk})}{\theta_{\trk}}  +  \left.  \dd{\hat{A}( c; \theta_{\trk})}{\theta_{\trk}} \right|_{c = \trk(v_k; \theta_{\trk})} \hspace{-3ex}
\end{split}
\end{equation}
when $\hat{A}(\trk(v_k; \theta_{\trk}); \theta_{\trk})$ is nonlinear.

Motivated by the $\gamma_1$ term in Equation \ref{eq:trackerobj8}, the $(c_k, v_k)$ pairs can be stacked together to obtain $\hat{A}^{-1} V - C$ where $V^T \hat{A}^{-T} = C^T$ can be solved separately for each column of $\hat{A}^{-T}$. Since Equation \ref{eq:trackerobj5} solves for $\hat{A}$, not $\hat{A}^{-1}$, a better alternative is to solve $C^T \hat{A}^T = V^T$ using least squares, minimum norm, or Levenberg-Marquardt. This direct method minimizes 
\begin{equation} \label{eq:loss_leastsquares}
    \loss_{D} = \sum_{i} || C^T \hat{A}^T(\theta_{\trk}) e_i - V^T e_i ||_2^2
\end{equation}
where $i$ loops through the rows of $\hat{A}$. Equation \ref{eq:loss_leastsquares} will be used to evaluate the efficacy of various approaches to solving Equation \ref{eq:trackerobj5}.


\subsection{Avoiding \texorpdfstring{$\infty$}{infinity} times 0 singularities}
\label{sub:example_1}

\begin{figure}[!htb]
\begin{subfigure}{.5\linewidth}
    \centering
    \includegraphics[width=\linewidth]{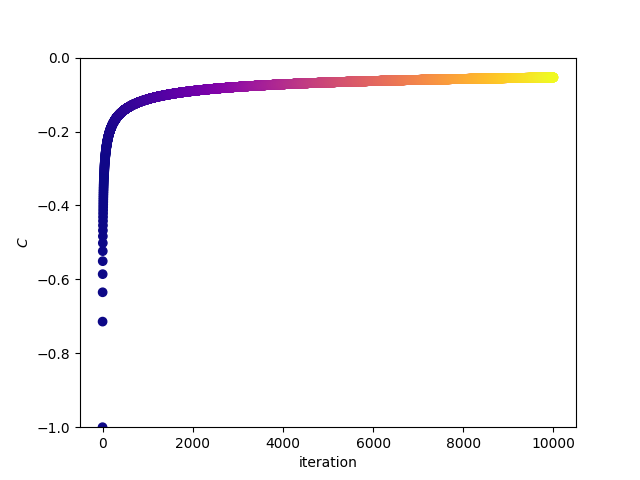}
    \label{fig:case1.1_subfig1}
\end{subfigure}%
\begin{subfigure}{.5\linewidth}
    \centering
    \includegraphics[width=\linewidth]{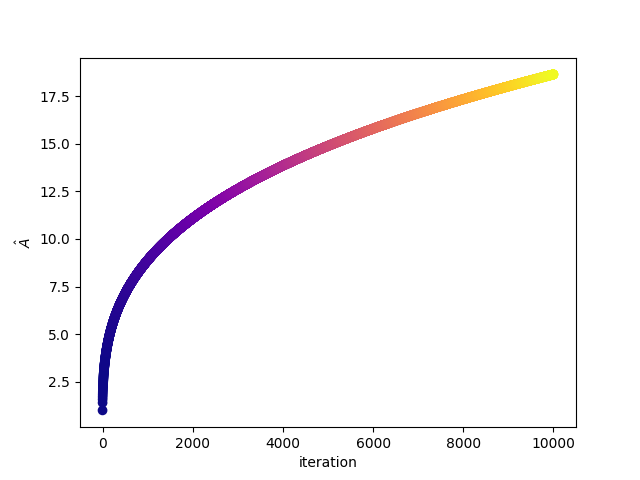}
    \label{fig:case1.1_subfig2}
\end{subfigure}
\vspace*{-10mm}
\caption{As $C \rightarrow \begin{bmatrix}0\end{bmatrix}$, $\hat{A} \rightarrow \begin{bmatrix}\infty\end{bmatrix}$.}
\Description{plots 1}
\label{fig:case1.1}
\end{figure}

Consider a single pair $(c_1, v_1) = (1, -1)$ where $\hat{A} = \begin{bmatrix}-1\end{bmatrix}$. For now, consider only the $\gamma_1$ term. Starting with an initial guess of $\hat{A}_0 = 1$, the tracker solves $\hat{A}_0 C_0 = V$ to obtain $C_0 = \begin{bmatrix}-1\end{bmatrix}$; then, the optimization attempts to drive $C$ from $\begin{bmatrix}-1\end{bmatrix}$ to $\begin{bmatrix}1\end{bmatrix}$ by sending $\hat{A} \rightarrow \begin{bmatrix}\infty\end{bmatrix}$ to obtain $C \rightarrow \begin{bmatrix}0\end{bmatrix}$. See Figure \ref{fig:case1.1}. Unfortunately, the origin is a non-removable singularity, and there is no mechanism to cross over the origin changing the entry in $C$ from negative to positive while flipping the entry in $\hat{A}$ from $\infty$ to $-\infty$.

\begin{figure}[!htb]
\begin{subfigure}{.5\linewidth}
    \centering
    \includegraphics[width=\linewidth]{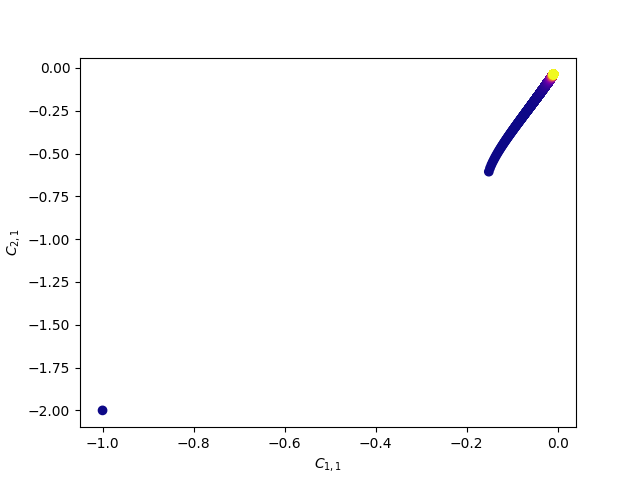}
    \label{fig:case1.2_subfig1}
\end{subfigure}%
\begin{subfigure}{.5\linewidth}
    \centering
    \includegraphics[width=\linewidth]{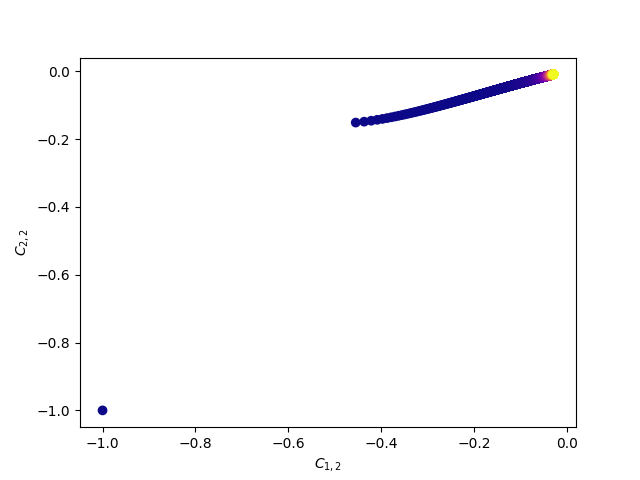}
    \label{fig:case1.2_subfig2}
\end{subfigure}
\vspace*{-10mm}
\caption{Starting with $\hat{A}_0 = I$ leads to $C_0 = V$. The left/right sub-figure shows the first/second column of $C$ with each row entry on a separate axis. As the iteration proceeds, the results are color-coded from blue to yellow. All four entries are driven to zero; likewise, all four entries of $\hat{A}$ blow up.}
\Description{plots 2}
\label{fig:case1.2}
\end{figure}

Next, consider 
\begin{equation} \label{eq:example1ahatcv}
\hat{A} = \begin{bmatrix}-1 & 0 \\ 0 & -1\end{bmatrix}, C = \begin{bmatrix} 1 & 1 \\ 2 & 1\end{bmatrix}, V = \begin{bmatrix}-1  & -1\\ -2 & -1 \end{bmatrix}
\end{equation}
where $\hat{A}C = V$. When starting with an initial guess of $\hat{A}_0 = I$, the optimization suffers from the same issue as was discussed in the prior example. See Figure \ref{fig:case1.2}. 

\begin{figure}[!htb]
\begin{subfigure}{.5\linewidth}
    \centering
    \includegraphics[width=\linewidth]{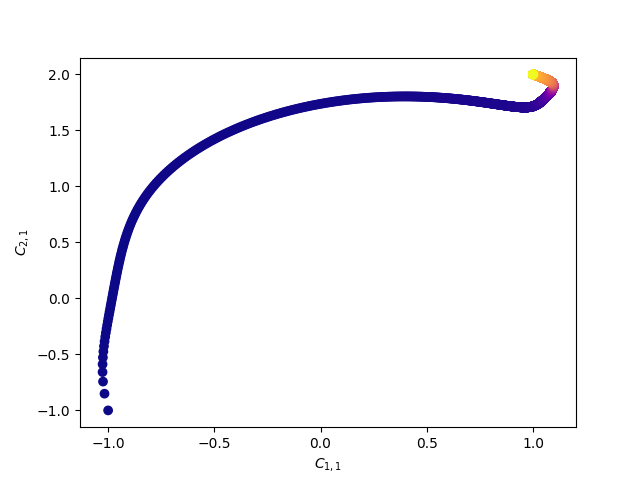}
    \label{fig:case1.3_subfig1}
\end{subfigure}%
\begin{subfigure}{.5\linewidth}
    \centering
    \includegraphics[width=\linewidth]{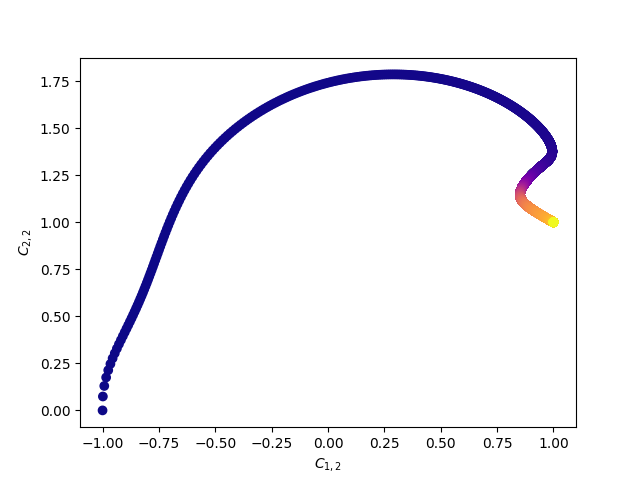}
    \label{fig:case1.3_subfig2}
\end{subfigure}
\vspace*{-10mm}
\caption{Changing the initial guess allows both $C$ and $\hat{A}$ to converge correctly.}
\Description{plots 3}
\label{fig:case1.3}
\end{figure}

Changing the initial guess to
\begin{equation}
    \hat{A} = \begin{bmatrix}1 & 1 \\ 0 & 1 \end{bmatrix}
\end{equation}
achieves the expected results, since the origin is avoided. See Figure \ref{fig:case1.3}. Note that perturbing the initial guess is a viable strategy because the optimization is done offline (and only once) in order to determine a suitable tracking rig. Moreover, using $\theta_{\rig}$ as an initial guess for $\theta_{\trk}$ is likely a good strategy, especially when the tracking rig does not need to be perturbed too much from the animation rig.

\subsection{Solver Efficacy}
\label{sub:example_2}

Consider the set of pairs
\begin{equation}
C = \begin{bmatrix}
        1 & 2 & 3 & 1 \\ 2 & -1 & 1 & 1 \\ 3 & -1 & -2 & 1 \\
    \end{bmatrix}, V = \begin{bmatrix}
        -1 & -2 & -3 & -1 \\ 4 & -2 & 2 & 2 \\ -2 & \frac{2}{3} & \frac{4}{3} & -\frac{2}{3} 
    \end{bmatrix}
\end{equation}
consistent with a rig
\begin{equation}
A = \begin{bmatrix}
        -1 & 0 & 0 \\ 0 & 2 & 0 \\ 0 & 0 & -\frac{2}{3}
    \end{bmatrix}
\end{equation}
in order to demonstrate the efficacy of Equation \ref{eq:trackerobj5} and our approach to solving it. Starting with $\hat{A} = I$, Table \ref{table:solverefficacy1} demonstrates that $\hat{A} \rightarrow A$ using either the $\gamma_1$ term or the $\gamma_2$ term; moreover, either term drives all three of the $\gamma_1$, $\gamma_2$, and $\gamma_3$ terms to zero. The results obtained via the direct least squares method are shown for the sake of comparison.

\begin{table}[!htb]
\footnotesize
\begin{tabular}{r|ccc}
\toprule
& $\loss_D$ & $\loss_{\gamma_1} +  \loss_{\gamma_2} + \loss_{\gamma_3}$ & $||\hat{A} - A||_F^2$ \\
\midrule
Direct & $4.58\textsc{e}{-12}$ & $4.85\textsc{e}{-12}$ & $4.26\textsc{e}{-13}$  \\
$\gamma_1$ only & $2.05\textsc{e}{-8}$ & $2.58\textsc{e}{-8}$ & $1.31\textsc{e}{-8}$  \\
$\gamma_2$ only & $1.54\textsc{e}{-9}$ & $4.15\textsc{e}{-9}$ & $9.82\textsc{e}{-10}$ \\
\bottomrule
\end{tabular}
\caption{}
\vspace*{-3ex}
\label{table:solverefficacy1}
\end{table}

The results in Table \ref{table:solverefficacy1} were obtained using all four columns of $C$ and $V$. The results obtained using only three columns of $C$ and $V$ are similar. Table \ref{table:solverefficacy2} shows the results of using only the first two columns of $C$ and $V$ with initial guess $\hat{A} = I$. Here, $A_2$ is the minimum norm solution, unobtainable by Equation \ref{eq:trackerobj5} without regularizing $\theta_{\trk} \rightarrow 0$ and undesirable since $\theta_{\trk} \rightarrow \theta_{\rig}$ is preferred. The $\gamma_{\epsilon}$ term regularizes the unconstrained degree of freedom resulting in $\hat{A} \rightarrow A$.

\begin{table}[!htb]
\footnotesize
\begin{tabular}{r|ccccc}
\toprule
& $\loss_D$ & $\loss_{\gamma_1} +  \loss_{\gamma_2} + \loss_{\gamma_3}$ & $||\hat{A} - A_2||_F^2$ &  $||\hat{A} - A||_F^2$ \\
\midrule
Direct & $1.71\textsc{e}{-12}$ & $6.89\textsc{e}{-13}$ & 0 & $2.77\textsc{e}{+0}$ \\
$\gamma_1$ only &  $4.228\textsc{e}{-9}$ & $4.90\textsc{e}{-12}$ & $7.72\textsc{e}{+1}$ & $6.75\textsc{e}{+1}$ \\
$\gamma_2$ only &  $2.56\textsc{e}{-8}$ & $1.66\textsc{e}{-7}$ & $1.62\textsc{e}{+2}$ & $1.46\textsc{e}{+2}$  \\
$\gamma_1 ~\& ~\gamma_{\epsilon}$ & $1.43\textsc{e}{-12}$ & $7.33\textsc{e}{-12}$ & $2.77\textsc{e}{0}$ & $5.74\textsc{e}{-9}$ \\
$\gamma_2 ~\& ~\gamma_{\epsilon}$ & $9.94\textsc{e}{-14}$ & $2.13\textsc{e}{-12}$ & $2.77\textsc{e}{0}$ & $2.05\textsc{e}{-10}$ \\
\bottomrule
\end{tabular}
\caption{}
\vspace*{-3ex}
\label{table:solverefficacy2}
\end{table}

Next, consider randomly perturbing the entries of $V$ by 1\% to obtain
\begin{equation}
\hat{V} = \begin{bmatrix}
        -0.990 & -2.00 & -2.97 & -1.00 \\ 3.97 & -1.98 & 1.98 & 2.01 \\ -1.98 & 0.667 & 1.33 & -0.667
    \end{bmatrix}
\end{equation}
so that pairing $C$ with $\hat{V}$ would yield a perturbation of $A$. The results obtained starting with $\hat{A} = I$ are shown in Table \ref{table:solverefficacy3}, omitting $\loss_{\gamma_3}$ since it was on the order of round-off error. Since the pairs over-determine $\hat{A}$, the direct method cannot achieve a small residual and the $\gamma_1$ term is bounded from below; however, they both result in the correct least squares solution (specified by $\hat{A}_4$ in the table). Using only the first three columns of $C$ and $\hat{V}$ allows both the direct method and the $\gamma_1$ term to achieve significantly smaller errors (as expected) since $\hat{A}$ is no longer over-determined. The $\gamma_2$ drives $\hat{A} \rightarrow A$ as expected.

\begin{table}[!htb]
\footnotesize
\begin{tabular}{r|ccccc}
\toprule
& $\loss_D$ & $\loss_{\gamma_1}$ & $\loss_{\gamma_2}$ & $||\hat{A} - \hat{A}_4||_F^2$ & $||\hat{A} - A||_F^2$ \\
\midrule
Direct & $6.60\textsc{e}{-4}$ & $3.47\textsc{e}{-4}$ & $2.55\textsc{e}{-3}$ & 0 & $5.57\textsc{e}{-4}$ \\
$\gamma_1$ only & $6.60\textsc{e}{-4}$ & $3.47\textsc{e}{-4}$ & $2.54\textsc{e}{-3}$ & $9.25\textsc{e}{-10}$ & $5.57\textsc{e}{-4}$ \\
$\gamma_2$ only & $3.21\textsc{e}{-3}$ & $2.38\textsc{e}{-3}$ & $5.65\textsc{e}{-9}$ & $5.58\textsc{e}{-4}$ & $3.63\textsc{e}{-9}$  \\
\bottomrule
\end{tabular}
\caption{}
\vspace*{-3ex}
\label{table:solverefficacy3}
\end{table}

Table \ref{table:solverefficacy4} shows the results obtained using only the first two columns of $C$ and $\hat{V}$ with an initial guess $\hat{A} = I$, again omitting $\loss_{\gamma_3}$ since it was on the order of round-off error. The minimum norm solution is unobtainable via Equation \ref{eq:trackerobj5}, as discussed above. Instead, $\hat{A}_{2^*}$ is computed by specifying a third independent column of $C$, multiplying it by $A$, and adding it to $\hat{V}$ before solving $C^T \hat{A}_{2^*}^T = \hat{V}^T$. This makes $\hat{A}_{2^*}^T$ agree with the minimum norm solution in the two constrained degrees of freedom and agree with $A$ in the remaining degree of freedom. The $\gamma_1$ and $\gamma_2$ terms converge as expected, and neither converges to $A$ or $\hat{A}_{2^*}$, due to the unconstrained degree of freedom; however, an initial guess of $\hat{A} = A$ does result in the $\gamma_1$ term and the  $\gamma_2$ terms driving $\hat{A} \rightarrow \hat{A}_{2^*}$ and $\hat{A} \rightarrow A$ respectively (as expected). The $\gamma_\epsilon$ term regularizes the unconstrained degree of freedom driving $\hat{A} \rightarrow A$; however, it competes with the $\gamma_1$ term that aims to drive $\hat{A} \rightarrow \hat{A}_{2^*}$ in a two-dimensional subspace. 

\begin{table}[!htb]
\footnotesize
\begin{tabular}{r|ccccc}
\toprule
& $\loss_D$ & $\loss_{\gamma_1}$ & $\loss_{\gamma_2}$ &  $||\hat{A} - \hat{A}_{2^*}||_F^2$  &   $||\hat{A} - A||_F^2$  \\
\midrule
Direct & $1.71\textsc{e}{-12}$ & $1.20\textsc{e}{-12}$ & $1.80\textsc{e}{-3}$ &  $2.77\textsc{e}{0}$  & $2.77\textsc{e}{0}$ \\
$\gamma_1$ only &  $9.89\textsc{e}{-9}$ & $8.56\textsc{e}{-9}$ & $1.80\textsc{e}{-3}$ & $6.66\textsc{e}{+1}$  & $6.66\textsc{e}{+1}$ \\
$\gamma_2$ only & $1.13\textsc{e}{-2}$ & $1.33\textsc{e}{-3}$ & $8.24\textsc{e}{-9}$ &  $1.13\textsc{e}{+2}$ & $1.13\textsc{e}{+2}$  \\
$\gamma_1 ~ \& ~\gamma_{\epsilon}$ & $7.19\textsc{e}{-5}$ & $1.97\textsc{e}{-5}$ & $1.24\textsc{e}{-3}$ & $7.23\textsc{e}{-6}$  & $9.26\textsc{e}{-5}$ \\
$\gamma_2 ~ \& ~\gamma_{\epsilon}$ & $1.80\textsc{e}{-3}$ & $1.33\textsc{e}{-3}$ & $7.62\textsc{e}{-11}$ & $1.39\textsc{e}{-4}$ & $8.44\textsc{e}{-9}$ \\
\bottomrule
\end{tabular}
\caption{}
\vspace*{-3ex}
\label{table:solverefficacy4}
\end{table}


\subsection{Improving the search direction via \texorpdfstring{$\dd{\hat{v}}{\theta_{\trk}}$}{d v / d theta} }
\label{sub:examples3and4}


Consider a tracker $\trk(v; \theta_{\trk}) = \hat{A}^{\dagger} v$ where $\hat{A}^{\dagger}$ depends on the method being used to solve $\hat{A} c = v$. Next, perturb the output of the tracker to
\begin{equation} \label{eq:trackerwithperturbv1}
    \trk_1(v; \theta_{\trk}) = \hat{A}^{\dagger} v + \tilde{c}
\end{equation}
where $\tilde{c}$ captures the fact that trackers often (intentionally, for robustness) do not correctly/precisely invert the rig.  Equation \ref{eq:rigoutputpluserror} becomes
\begin{equation} \label{eq:vhatwithperturbv1}
    \hat{v}(v; \theta_{\trk}) = v + \tilde{v}(v; \theta_{\trk}) = \hat{A}(\theta_{\trk}) \left(\hat{A}^{\dagger}(\theta_{\trk}) v + \tilde{c}(v)\right)
\end{equation}
where $\tilde{v} \neq 0$. When $\hat{A}$ happens to be invertible,
\begin{equation} \label{eq:vtildeeqahatctilde}
    \tilde{v}(v; \theta_{\trk}) = \hat{A}(\theta_{\trk}) \tilde{c}(v)
\end{equation}
\begin{equation} \label{eq:dvhatkdthetatv1}
    \dd{\hat{v}(v; \theta_{\trk})}{\theta_{\trk}} = \dd{\tilde{v}(v; \theta_{\trk})}{\theta_{\trk}} = \begin{bmatrix}
        \tilde{c}(v)^T & 0_{1x3} & 0_{1x3} \\
        0_{1x3} & \tilde{c}(v)^T & 0_{1x3} \\
        0_{1x3} & 0_{1x3} & \tilde{c}(v)^T
    \end{bmatrix}
\end{equation}
leading to 
\begin{equation} \label{eq:lvhatprimev1}
    \loss_{\hat{v}'} = \sum_{k} \left|\left| \dd{\hat{v}_k(v_k; \theta_{\trk})}{\theta_{\trk}} -\begin{bmatrix}
        \tilde{c}_k(v_k)^T & 0_{1x3} & 0_{1x3} \\
        0_{1x3} & \tilde{c}_k(v_k)^T & 0_{1x3} \\
        0_{1x3} & 0_{1x3} & \tilde{c}_k(v_k)^T
    \end{bmatrix} \right|\right|_F^2
\end{equation}
as a way of evaluating the efficacy of approximations to $\dd{\hat{v}}{\theta_{\trk}}$.

Plugging Equation \ref{eq:trackerwithperturbv1} into the $\gamma_1$ term in Equation \ref{eq:trackerobj5} and stacking columns leads to $\hat{A}^{\dagger} V + \tilde{C} - C = 0$ or $V  - \hat{A} C + \hat{A} \tilde{C} = 0$ when $\hat{A}$ happens to be invertible; thus, minimizing the $\gamma_1$ term leads to a small value for $V  - \hat{A} C + \hat{A} \tilde{C}$ instead of a small value for $V  - \hat{A} C$. Table \ref{table:example3_1} shows the results obtained with and without using $\dd{\hat{v}}{\theta_{\trk}}$ to improve the search direction. In this simple example, auto-diff can be used to compute $\dd{\hat{v}}{\theta_{\trk}}$. The average $\loss_{\hat{v}'}$ error obtained when using auto-diff for $\dd{\hat{v}}{\theta_{\trk}}$ was $6.44\textsc{e}{-10}$; for the sake of comparison, the average value of $\sum_{k} \left|\left| \dd{\hat{v}_k}{\theta_{\trk}} \right|\right|_F^2$ was $7.05\textsc{e}{-3}$. Only the first three columns of $C$ and $V$ were used, $\tilde{C} = A^{-1} (\hat{V} - V)$ was chosen so that the perturbations in $C$ resemble those used for $V$, and the initial guess was $\hat{A} = I$. The $\gamma_2$ term gave similar results, as expected since $AC = V$ here.

\begin{table}[!htb]
\begin{tabular}{r|ccccc}
\toprule
& $||V - \hat{A} C + \hat{A} \tilde{C}||_F^2$  & iters \\
\midrule  
$\gamma_1$ only &  $3.86\textsc{e}{-8}$ & $13868$ \\
$\gamma_1$ \& $\dd{\hat{v}}{\theta_{\trk}}$ & $3.82\textsc{e}{-8}$ & $12960$ \\
\bottomrule
\end{tabular}
\caption{}
\vspace*{-3ex}
\label{table:example3_1}
\end{table}


Next, consider perturbing the output of the tracker to
\begin{equation} \label{eq:trackerperturb2}
    \trk_2(v; \theta_{\trk}) = \hat{A}^{\dagger} v + \hat{A}(\theta_{\trk}) \tilde{c}(v)
\end{equation}
instead of Equation \ref{eq:trackerwithperturbv1}. Equation \ref{eq:rigoutputpluserror} becomes 
\begin{equation} \label{eq:vhatwithperturbv2}
    \hat{v}(v; \theta_{\trk}) = v + \tilde{v}(v; \theta_{\trk}) = \hat{A}(\theta_{\trk}) (\hat{A}^{\dagger}(\theta_{\trk}) v + \hat{A}(\theta_{\trk})\tilde{c}(v))
\end{equation}
and
\begin{equation}
    \tilde{v}(v; \theta_{\trk}) = \hat{A}(\theta_{\trk}) \hat{A}(\theta_{\trk}) \tilde{c}(v)
\end{equation}
\begin{equation} \label{eq:dvhatkdthetatv2}
\begin{split}
    \dd{\hat{v}(v; \theta_{\trk})}{\theta_{\trk}} = \hat{A}(\theta_{\trk}) \begin{bmatrix}
        \tilde{c}(v)^T & 0_{1x3} & 0_{1x3} \\
        0_{1x3} & \tilde{c}(v)^T & 0_{1x3} \\
        0_{1x3} & 0_{1x3} & \tilde{c}(v)^T
    \end{bmatrix} + \qquad\qquad \\ \begin{bmatrix}
        \tilde{c}(v)^T \hat{A}^T(\theta_{\trk}) & 0_{1x3} & 0_{1x3} \\
        0_{1x3} & \tilde{c}(v)^T \hat{A}^T(\theta_{\trk}) & 0_{1x3} \\
        0_{1x3} & 0_{1x3} & \tilde{c}(v)^T \hat{A}^T(\theta_{\trk})
    \end{bmatrix}
\end{split}
\end{equation}
replaces Equations \ref{eq:vtildeeqahatctilde} and \ref{eq:dvhatkdthetatv1} when $\hat{A}$ happens to be invertible. Equation \ref{eq:dvhatkdthetatv2} can be used to define a new $\loss_{\hat{v}'}$ that replaces Equation \ref{eq:lvhatprimev1}. Plugging Equation \ref{eq:trackerperturb2} into the $\gamma_1$ term in Equation \ref{eq:trackerobj5} and stacking columns leads to the notion that minimizing the $\gamma_1$ term should lead to a small value of $V - \hat{A} C + \hat{A}^2 \tilde{C}$, assuming that $\hat{A}$ happens to be invertible. Table 6 shows the results obtained with and without using $\dd{\hat{v}}{\theta_{\trk}}$ to improve the search direction. Once again, this example is simple enough to use auto-diff. The average $\loss_{\hat{v}'}$ error obtained when using auto-diff for $\dd{\hat{v}}{\theta_{\trk}}$ was $2.54\textsc{e}{-10}$; for the sake of comparison, the average value of $\sum_{k} \left|\left| \dd{\hat{v}_k}{\theta_{\trk}} \right|\right|_F^2$ was $1.91\textsc{e}{-1}$. 

\begin{table}[!htb]
\begin{tabular}{r|ccccc}
\toprule
&  $||V - \hat{A} C + \hat{A}^2 \tilde{C}||_F^2$  & iters \\
\midrule  
$\gamma_1$ only & $3.82\textsc{e}{-8}$ & $29083$ \\
$\gamma_1$ \& $\dd{\hat{v}}{\theta_{\trk}}$ & $3.79\textsc{e}{-8}$ & $5329$ \\
\bottomrule
\end{tabular}
\caption{}
\vspace*{-3ex}
\label{table:example4_1}
\end{table}

\subsection{Estimating  \texorpdfstring{$\dd{\hat{v}}{\theta_{\trk}}$}{d v / d theta} }
\label{sub:estimatingdvhatdthetat}

A general black-box tracker is not necessarily amenable to auto-diff; in such a scenario, $\dd{\hat{v}}{\theta_{\trk}}$ would need to be estimated via Equation \ref{eq:rankoneupdatewiths}. In order to demonstrate the efficacy of Equation \ref{eq:rankoneupdatewiths}, consider minimizing the $\gamma_1$ term using either $\trk_1$ from Equation \ref{eq:trackerwithperturbv1} or $\trk_2$ from Equation \ref{eq:vhatwithperturbv1} along with the analytic values for $\dd{\hat{v}}{\theta_{\trk}}$ given in Equations \ref{eq:dvhatkdthetatv1} and \ref{eq:dvhatkdthetatv2}. After about $13000$ optimization steps, the analytic solution for $\dd{\hat{v}_k}{\theta_{\trk}}$ is used about $39000$ times since $C$ and $V$ have three columns. 

\begin{figure}[!htb]
\begin{subfigure}{.5\linewidth}
    \centering
    \includegraphics[width=\linewidth]{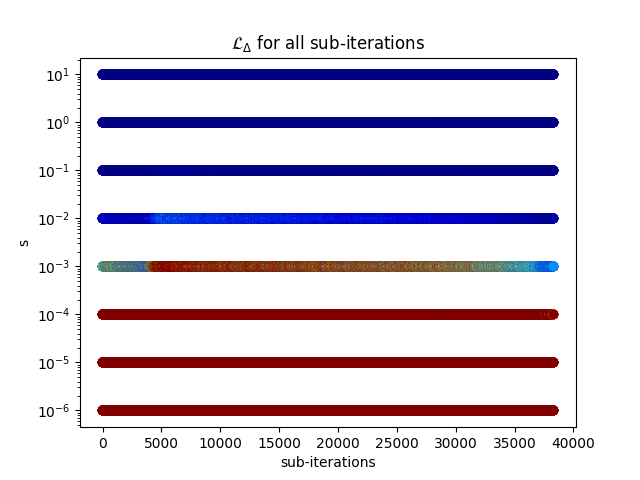}
    \label{fig:example3ldelta_subfig1}
\end{subfigure}%
\begin{subfigure}{.5\linewidth}
    \centering
    \includegraphics[width=\linewidth]{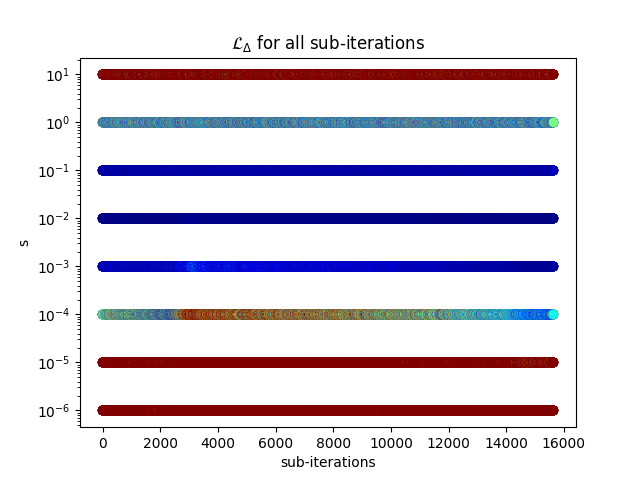}
    \label{fig:example4ldelta_subfig1}
\end{subfigure}
\vspace*{-10mm}
\caption{Color-coded from lower error (blue) to higher error (red). Left: $\trk_1$. Right: $\trk_2$. }
\Description{plots 4}
\label{fig:examples3and4ldelta}
\end{figure}

In order to ascertain the sensitivity of the parameter $s$ in the finite difference approach, an approximation to $\dd{\hat{v}_k}{\theta_{\trk}}$ was computed via one iteration of Equation \ref{eq:rankoneupdatewiths} using a randomly chosen $\widehat{\Delta\theta}$ direction and the following values of $s$: $1\textsc{e}{-7}$, $1\textsc{e}{-6}$, $1\textsc{e}{-5}$, $1\textsc{e}{-4}$, $1\textsc{e}{-3}$, $1\textsc{e}{-2}$, $1\textsc{e}{-1}$, $1\textsc{e}{0}$, $1\textsc{e}{1}$, $1\textsc{e}{2}$. The finite difference approximations for any two values of $s$ can be compared via a Frobenius norm. Let $\loss_{\Delta}(s; \theta_{\trk}, \widehat{\Delta \theta})$ be the sum of the Frobenius norms obtained by comparing $s$ to both its next higher and next lower values; then, $\loss_{\Delta}$ indicates the relative sensitivity of the finite difference approximations to perturbations in $s$. Figure \ref{fig:examples3and4ldelta} shows the results obtained on all $39000$ sub-iterations.  The red regions at the bottom of the figures represent round-off errors, and the red region at the top of the right figure represents truncation error (as expected, see e.g. \cite{Gear_1971}). There is no red region at the top of the left figure, since $\dd{\hat{v}_k}{\theta_{\trk}}$ is constant (see Equation \ref{eq:dvhatkdthetatv1}). The large blue region illustrates the ease at which $s$ can be chosen.

Next, consider the same tests with the analytic solution replaced by various approximations to the $\dd{\hat{v}_k}{\theta_{\trk}}$. For the finite difference approximations, $s$ was chosen to minimize $\loss_\Delta$. Tables \ref{table:example3_constantperturb} and \ref{table:example3_variableperturb} show the results obtained for $\trk_1$ and $\trk_2$ respectively using either a number of randomly-chosen $\widehat{\Delta\theta}$ directions, auto-diff, or a single $\widehat{\Delta\theta}$ chosen in the direction of steepest descent. The third column of the tables averages the error (see Equations \ref{eq:lvhatprimev1} and \ref{eq:dvhatkdthetatv2}) over all iterations; for the sake of comparison, the second column averages the solution. As expected, increasing the number of randomly chosen $\widehat{\Delta\theta}$ improves the estimate of $\dd{\hat{v}_k}{\theta_{\trk}}$; however, improved estimates of $\dd{\hat{v}_k}{\theta_{\trk}}$ did not tend to improve convergence. On the other hand, choosing $\widehat{\Delta\theta}$ in the steepest descent direction did improve convergence. This makes sense from the viewpoint of a predictor-corrector approach, even though $\dd{\hat{v}_k}{\theta_{\trk}}$ is poorly estimated as compared to the other approaches. Unfortunately, iterating on steepest descent choices for $\widehat{\Delta\theta}$ did not further improve convergence.

\begin{table}[!htb] 
  \begin{tabular}{r|cccccc} 
  \toprule
  $\trk_1 \quad$ & iters & $\mu \left(\sum_{k} \left|\left| \dd{\hat{v}_k}{\theta_{\trk}} \right|\right|_F^2 \right)$ &  $\mu(\loss_{\hat{v}'})$\\
  \midrule
  1 random      & $12781$ & $7.04\textsc{e}{-3}$ & $4.57\textsc{e}{-6}$ \\
  10 random     & $12921$ & $7.05\textsc{e}{-3}$ & $2.86\textsc{e}{-7}$ \\
  100 random    & $12964$ & $7.05\textsc{e}{-3}$ & $4.63\textsc{e}{-8}$ \\
  auto-diff     & $12960$ & $7.05\textsc{e}{-3}$ & $1.74\textsc{e}{-11}$\\
  steepest descent   & $9089$ & $2.09\textsc{e}{-3}$ & $6.41\textsc{e}{-3}$ \\
  \bottomrule
  \end{tabular}
  \caption{}
  \label{table:example3_constantperturb}
\end{table}

\begin{table}[!htb] 
  \begin{tabular}{r|ccccc} 
  \toprule
  $\trk_2 \quad$ &  iters & $\mu \left(\sum_{k} \left|\left| \dd{\hat{v}_k}{\theta_{\trk}} \right|\right|_F^2 \right)$ & $\mu(\loss_{\hat{v}'})$ \\
  \midrule
  1 random      & $4763$ & $1.71\textsc{e}{-1}$ & $3.50\textsc{e}{-4}$ \\
  10 random     & $5307$ & $1.93\textsc{e}{-1}$ & $6.32\textsc{e}{-6}$ \\
  100 random    & $5329$ & $1.94\textsc{e}{-1}$ & $1.24\textsc{e}{-6}$ \\
  auto-diff     & $5329$ & $1.94\textsc{e}{-1}$ & $2.57\textsc{e}{-10}$ \\
  steepest descent    & $4825$ & $7.28\textsc{e}{-2}$ & $1.93\textsc{e}{-1}$ \\
  \bottomrule
  \end{tabular}
  \caption{}
  \label{table:example3_variableperturb}
\end{table}

\section{Simon Says Animation Rig Creation}
\label{sec:simonsays}

The industry-standard rig calibration pipeline begins by placing an actor into a light-stage where they are asked to create a variety of expressions, each of which can be defined by a set of animator controls $c$. Afterwards, the data is used to create 3D geometry $v$ corresponding to each captured expression. This is a time-consuming process that requires both artist expertise and subjectivity. In fact, some of the expressions are inferred and/or sculpted as opposed to being reconstructed, e.g. \texttt{smile} would be split into \texttt{smile\_left} and \texttt{smile\_right}. To simplify rig calibration, we utilize the Simon-Says approach proposed in \cite{zhu2024democratizingcreationanimatablefacial}.

In order to guide the user, a version of ``Simon" is required. This can be achieved by reconstructing the user's neutral geometry $N$ and subsequently fitting it with an animation rig $\rig$. The ability to reconstruct neutral geometry is becoming more common, see Section \ref{sec:relatedwork}, and commercial software is becoming available. Given a rig fitted to any preexisting geometry, volumetric morphing can be used to transfer that rig to any newly constructed geometry once a topological correspondence is determined. See e.g. \cite{zhu2024democratizingcreationanimatablefacial} for more details. In particular, we utilize the MetaHuman animation rig framework (see \cite{metahuman}); however, since their framework includes a blendshape offset in the neutral pose that can lead to various inaccuracies when making expressions (see \cite{zhu2024democratizingcreationanimatablefacial}), we volumetrically morph the rig to tightly fit the actual geometry. Note that auxiliary attributes such as textures, hair, eyebrows, etc. are selected from a pre-defined database since they are not the focus of this paper.

\begin{figure}[!htb]
\includegraphics[width=\linewidth]{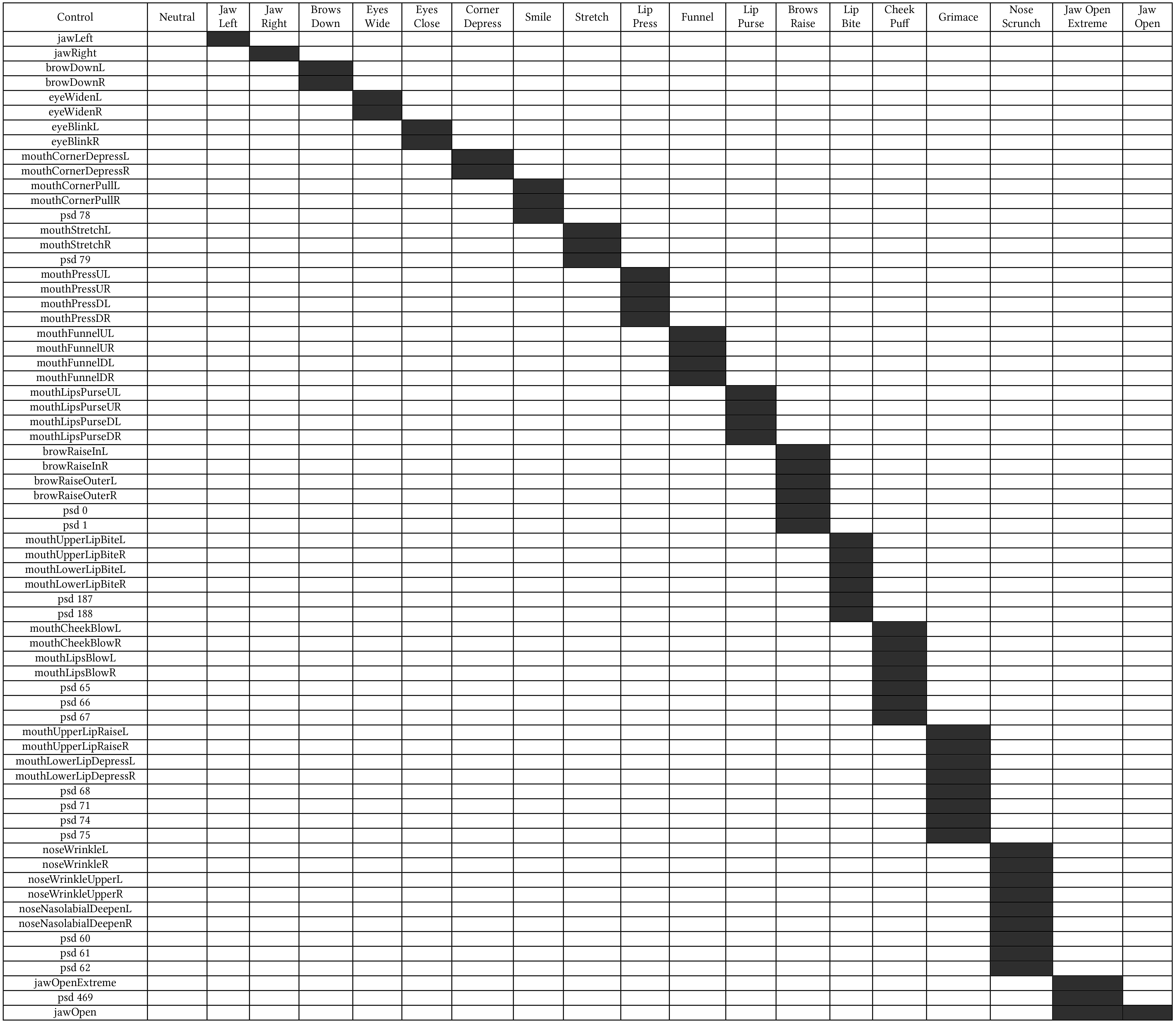}
\vspace*{-5ex}
\caption{For each expression, the associated column shows the active MetaHuman PSD rig controls.}
\Description{DESCRIPTON}
\label{fig:simonsays_chart}
\end{figure}

A carefully chosen set of Simon-Says expressions is used to calibrate each rig to the motion signatures of the relevant user. The expression set should cover the most important rig parameters while still remaining minimal with expressions that are straightforward for an average person to make. In addition, expressions that would typically require artist intervention need to be omitted. In total, nineteen expressions
\begin{itemize}
   \item neutral, brows down, brows up, eyes wide, eyes close, nose wrinkle, cheek puff, teeth grimace, corner pull, mouth stretch, corner depress, lip press, pursed lips, mouth funnel, lip bite, jaw open, jaw open extreme, jaw left, jaw right
 \end{itemize}
are used for person-to-person retargeting. The nineteen expressions along with their associated animation controls are shown in Figure \ref{fig:simonsays_chart}.  For person-to-game/VR character retargeting, the expressions should be chosen in a manner that captures the distinct motion signatures of the game/VR character.

\begin{figure}[!htb]
\includegraphics[width=\linewidth]{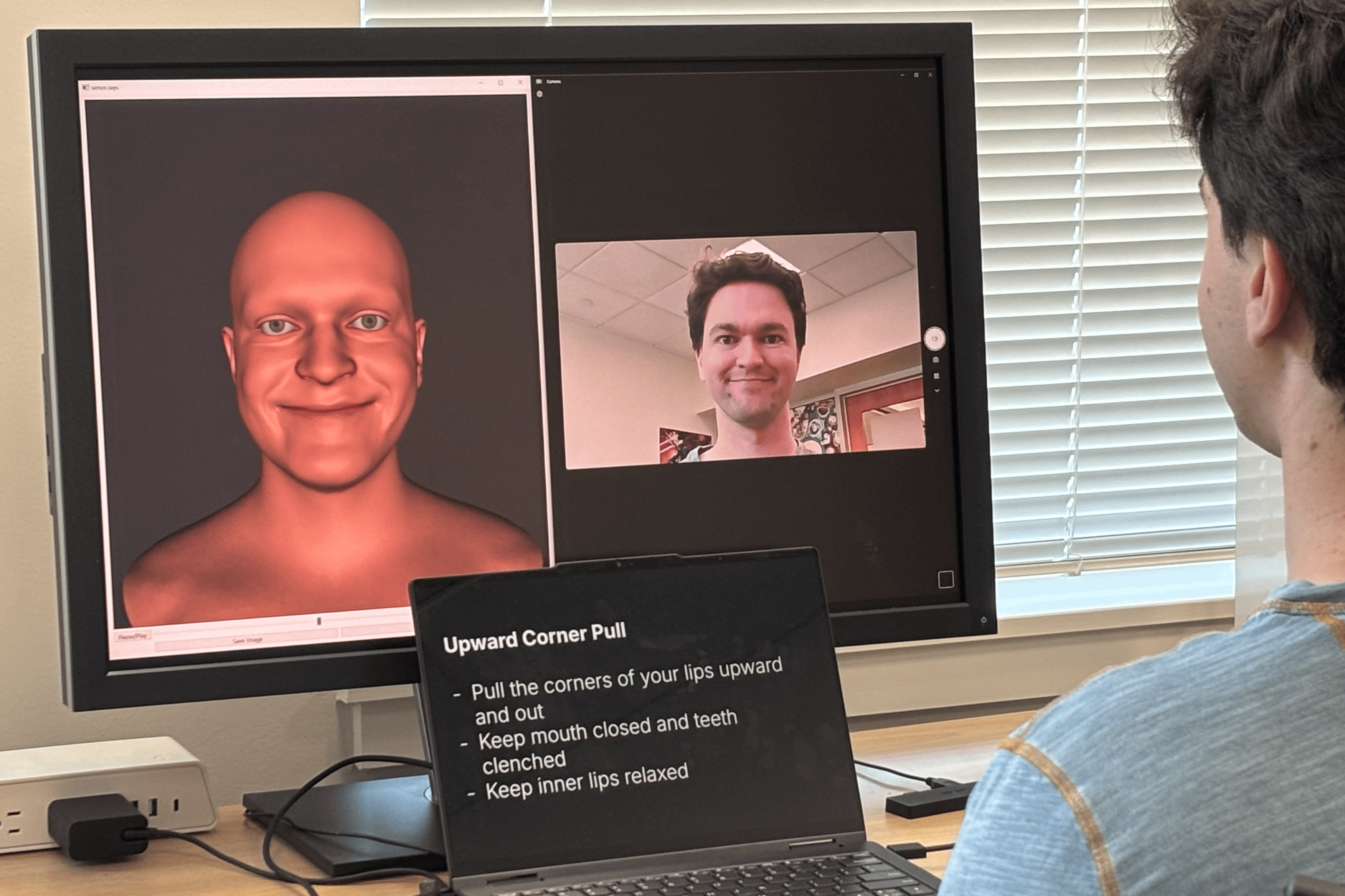}
\vspace*{-5mm}
\caption{Our proposed Simon-Says capture setup.}
\Description{DESCRIPTION}
\label{fig:simon_says}
\end{figure}

For each expression, an image $\phi(\rig (c; N, \theta_{\rig}))$ is shown to the user. Here, $\phi$ represents a renderer with camera parameters set to mimic the user's webcam. The user then attempts to replicate each expression. When they feel as though they are correctly reproducing the expression, an image $I$ of their face is recorded. See Figure \ref{fig:simon_says}. Visual cues  enable non-experts to quickly understand each expression, especially when the cues are similar in appearance to the user (as they are here). Importantly, since the animation rig is unlikely to already contain the correct semantic motion signatures of the user, it is incorrect for the user to simply aim to match each image. For example, \texttt{jaw\_open} and \texttt{jaw\_open\_extreme} should match the user's actual biomechanics, which may differ significantly from those in the rig. Thus, concise verbal descriptions are also provided for each pose. 

Once the Simon-Says session is complete, the captured images can be used to reconstruct geometry (see section \ref{sec:geometryreconstruction}). Given the $(c, v)$ pairs, Equation \ref{eq:sculptopt} can be solved to determine rig parameters $\theta_{\rig}$ that better capture the user's motion signatures. Expert riggers typically utilize a number of constraints during the rig design process. For example, \texttt{jaw\_open} plus \texttt{lips\_together} should properly seal the lips together. It is straightforward to incorporate such constraints by including additional $(c, v)$ pairs, representing these constraints, into the Equation \ref{eq:sculptopt} solve.  Similarly, we can also include surrogate $(c, v)$ pairs for any expressions that were omitted because they would typically require artist intervention. Although these surrogate pairs would not be user-specific, they do help with regularization.

\subsection{Incorrect Expressions}
\label{sec:incorrect_expressions}

When the user struggles to make the correct expression (such as accidentally flexing eyebrows while smiling), the reconstructed geometry $v$ will not correctly correspond to the expression specified by $c$. It is possible to roughly identify the additional controls that need to be activated in order to explain these spurious geometric deformations. Using the current best guess for the animation rig parameters $\theta_{\rig}$, a tracker can be used to compute
\begin{equation} \label{eq:updatecko}
    c^+ = c + (I_{n \times n} - \mathcal{H}(c)) \hspace{.03cm}  \trk (I, \theta_{\rig})
\end{equation}
where $\mathcal{H}(c)$ is a diagonal matrix of Heaviside functions $H(|c_{i}|)$. Equation \ref{eq:updatecko} leaves the nonzero entries of $c$ unmodified while augmenting the other entries to agree with $\trk(I, \theta_{\rig})$. The newly added $(c^+, v)$ pairs help to prevent the inappropriate modifications to $\theta_{\rig}$ that would have resulted from the $(c, v)$ pairs that they replace when solving Equation \ref{eq:sculptopt}.  Alternatively, the geometry can be constrained to mask out regions of the face that should not deform for the given expression, replacing $(c, v)$ pairs with $(c, v^-)$ pairs.

The augmentation of $c$ to  $c^+$ can be problematic when the added controls interfere with the primary expression under consideration; thus, it is desirable to leave interfering controls unmodified. An alternative strategy consists of constraining $\theta_{\rig}$ so that the degrees of freedom irrelevant to $c$ do not change. This is plausible when solving Equation \ref{eq:sculptopt}, since $\rig$ is typically a straightforward deterministic function; however, it can be daunting when solving Equation \ref{eq:trackerobj4}, since $\trk$ contains both an ill-posed geometric reconstruction and an inverse problem (rig inversion).

\subsection{Rig Co-Animation}
\label{sec:rig_co_animation}

It is worth briefly addressing the importance of well-calibrated rigs when it comes to retargeting. It makes no sense to strive for an accurate tracking solve on a performer if those controls are not going to be sensical on the target. This consideration emphasizes the importance of the Simon-Says rig calibration. For example, if the performer opens their jaw as far as they can and \texttt{jaw\_open} is $0.8$ instead of $1$, then the performer will be unable to fully open the jaw of the target. If the performer misunderstood the instructions and did not open their jaw fully, then (in the spirit of Equation \ref{eq:updatecko}) the $\texttt{jaw\_open} = 1$ in $c$ should be lowered to a more appropriate value.

\subsection{Validation with Synthetic Data}
\label{sub:synthetic_test}

In this section, we describe the results of an experiment that used a MetaHuman with hand-crafted animation rig parameters $\theta_{GT}$ that could be considered a ground truth (or close to it). Starting with a different MetaHuman, we morphed its rig onto the MetaHuman being used for this test. Labeling the morphed rig parameters as $\theta_M$, we used a Simon-Says expression set to generate geometry with $\theta_{GT}$ and subsequently optimized $\theta_M \rightarrow \theta_S$. For the Simon-Says expressions, $\theta_{GT}$ and $\theta_S$ behaved similarly, as expected. Generalization of our Simon-Says approach can be tested by comparing the geometry produced by $\theta_{GT}$ to that produced by $\theta_S$ for other expressions unseen by the optimization. In particular, the benefits of the approach can be quantified as the decrease in errors obtained by using $\theta_S$ instead of $\theta_M$. See Table \ref{table:simon_synthetic}.

\begin{table}[!htb]
\footnotesize
\begin{tabular}{r|cc}
\toprule
Expression & $\theta_M$ & $\theta_S$  \\
\midrule
Happy Speaking     & $0.487$  & $0.285$ \\
Partial Stretch    & $0.693$  & $0.370$ \\
Funnel Asymmetric  & $0.308$  & $0.143$ \\
Funneled Speech    & $0.393$  & $0.183$ \\
``Ur''             & $0.507$  & $0.337$ \\
``Em''             & $0.211$  & $0.176$ \\
``Vee''            & $0.247$  & $0.208$ \\
Frown              & $0.329$  & $0.119$ \\
\bottomrule
\end{tabular}
\caption{Average error in mm for synthetic expressions unseen by the Simon-Says optimization.}
\vspace*{-3ex}
\label{table:simon_synthetic}
\end{table}

\section{Expression Capture}
\label{sec:geometryreconstruction}

\begin{figure*}[!htb]
\includegraphics[width=\linewidth]{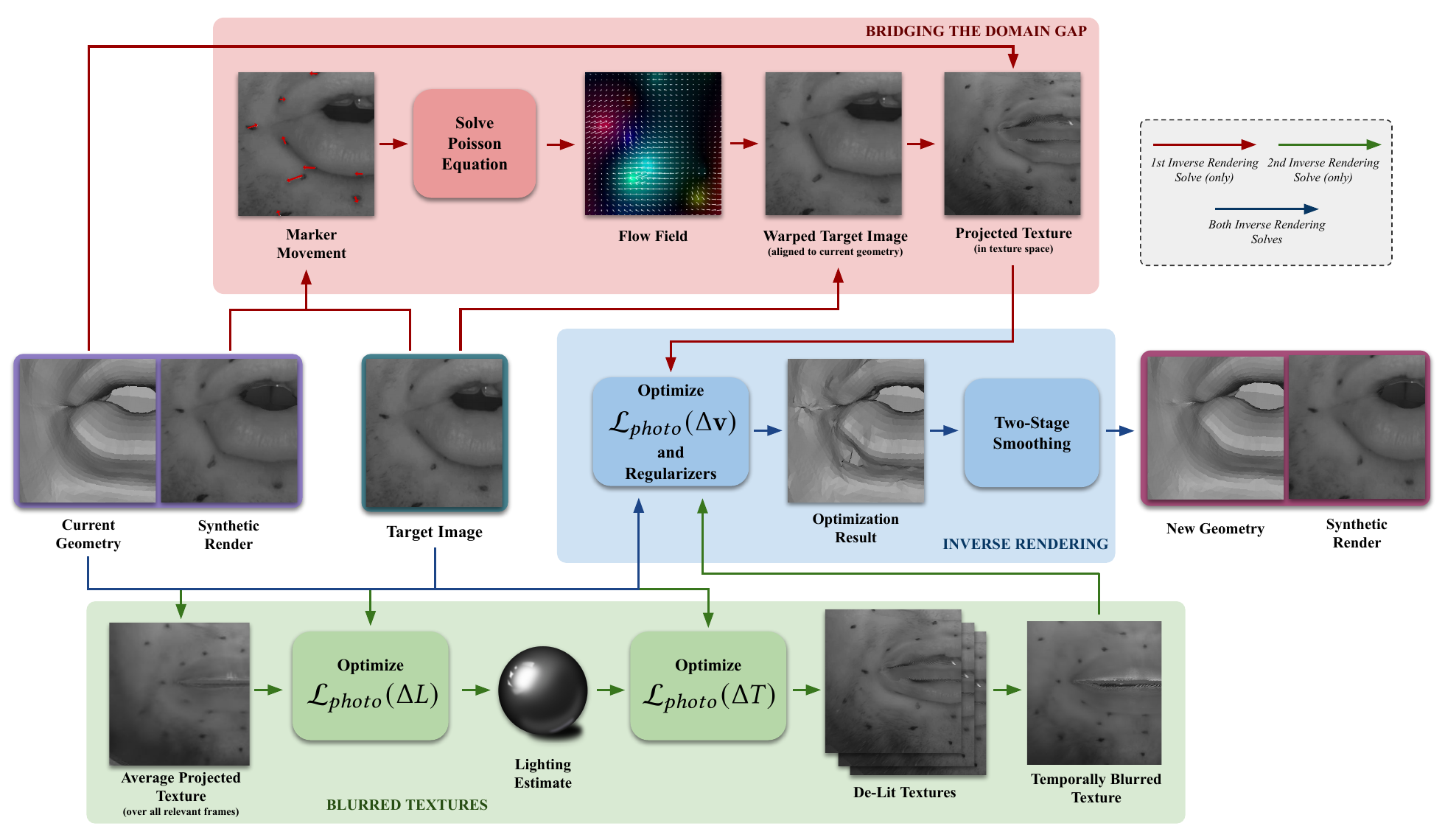}
\label{fig:geo_overview1}
\vspace*{-5mm}
\caption{A high-level overview of our expression capture pipeline.}
\Description{geo 1}
\label{fig:geo_overview}
\end{figure*}

\begin{figure}[!htb]
\includegraphics[width=\linewidth]{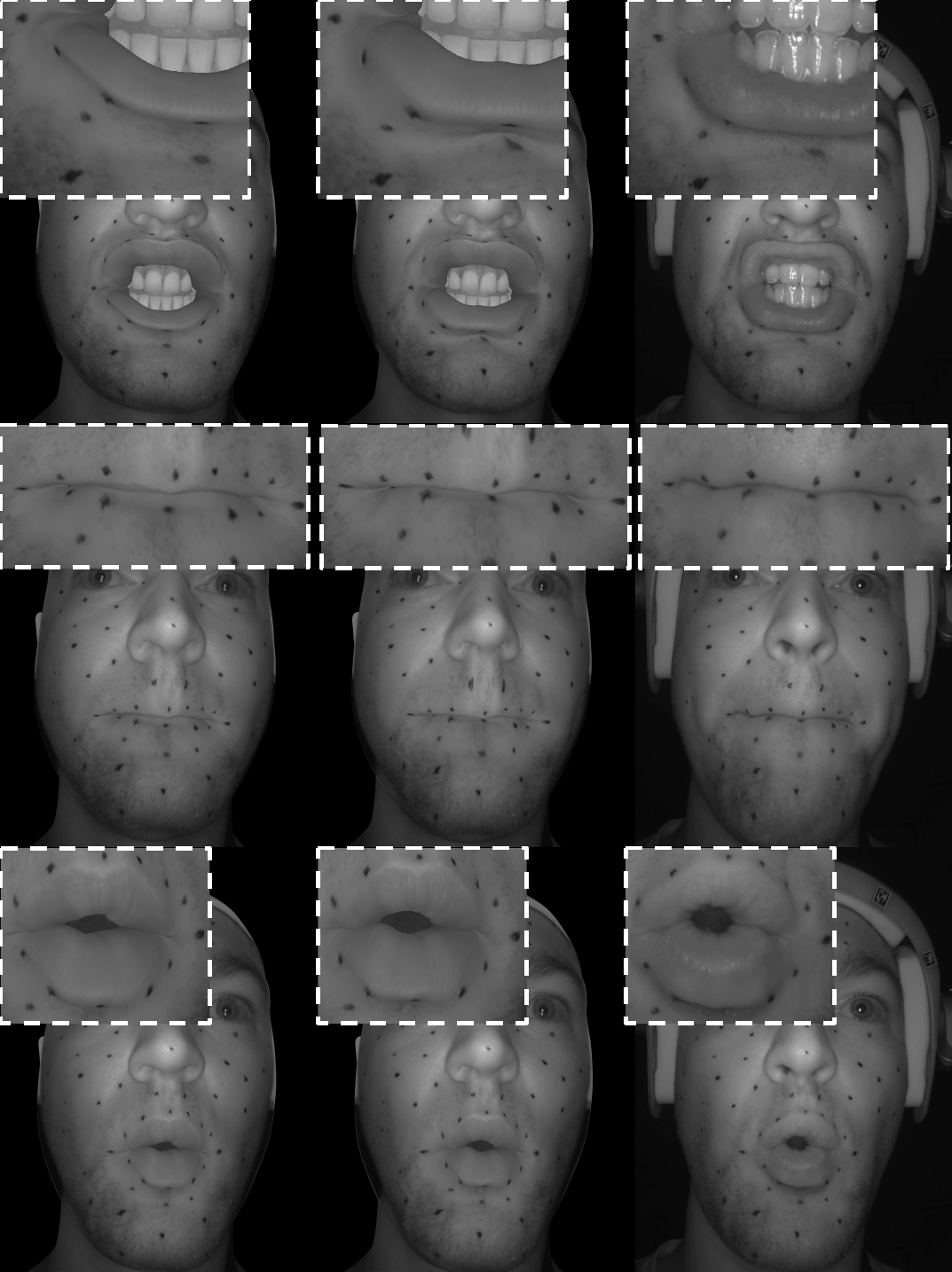}
\vspace*{-5ex}
\caption{Geometry reconstruction results on some difficult expressions. 1st column: Initial reconstructions (from MetaHuman Animator). 2nd column: Our results. 3rd column: Target images. 1st row: The lip bulge has been corrected to better match a ``funnel'' expression. 2nd row: The inward lip roll has been increased to better match a ``lip bite'' expression. 3rd row: The inner lip lines have been corrected to better match an ``oh'' expression.}
\Description{DESCRIPTON}
\label{fig:geo_examples}
\end{figure}

Assuming the existence of a well-constructed neutral geometry, this section focuses on reconstructing geometry that corresponds to the images captured in a Simon-Says session. Although any of the approaches discussed in Section \ref{sec:relatedwork} could be used, higher quality geometry reconstruction likely leads to an animation rig that better captures the user's motion signature. Thus, we briefly address geometry reconstruction in this section. Moreover, we discuss geometry reconstruction in the context of improving upon (or bootstrapping from) state-of-the-art techniques, as opposed to starting from scratch. See Figure \ref{fig:geo_examples} for some representative results, and Figure \ref{fig:geo_overview} for an overall outline of the approach. Finally, although we intentionally treat trackers via a black box ansatz, it is worth noting that improved geometry reconstruction techniques could be used to compute $\tilde{v}_{\trk}$ (see Section \ref{sec:preliminaries}).

Even though the light-stage is often discussed as being the top end data acquisition method for constructing neutral geometry, textures, and animation rigs, the overly confining equipment can adversely impact the authenticity of the user's expressions, hindering the ability to construct an accurate basis for their motion signatures. In fact, we prefer monocular, multi-camera, or head-mounted camera (HMC) setups because they are all less restrictive on the user's movements and expressions. A monocular approach is the least preferable option, since the data term (without robust priors) only includes feature alignment in the image plane instead of in $\mathbb{R}^3$. 

\subsection{Bridging the Domain Gap}
\label{sec:bridgingthedomaingap}

It can be difficult to ascertain correspondences between synthetic geometry and real-world images, and the lack of a proper correspondence leads to inaccuracies in the geometry reconstruction. Even after calibrating the virtual camera to match the real-world camera's intrinsics and estimate the scene extrinsics (typically with the aid of a landmark detector), inaccuracies in the texture and lighting lead to a domain gap that hinders proper correspondence. 

We address the domain gap via the screen-space warping technique from \cite{zhu2024democratizingcreationanimatablefacial}. Given a synthetically rendered image created to match a real-world image as closely as possible, a neural segmentation of both images is used to construct an optical flow field that warps the real-world image to better match the synthetic image; then, the photon mapping method from \cite{lin2022leveragingdeepfakesclosedomain} is used to project samples from the warped real-world image into the synthetic geometry's texture where the samples can be robustly gathered to texels. This baking of the entangled real-world texture and lighting into the synthetic geometry's texture better enables an inverse rendering pipeline to bridge the domain gap. Typically, this process is repeated every time (e.g. every iteration of optimization) the geometry is updated.  
 
Since the Simon-Says calibration of the animation rig (and our modification of the tracking rig) only needs to be done once, it is worth considering a slightly more invasive, yet still democratizable, approach. Easy-to-remove dots can be placed on the face using liquid eyeliner or other makeup. After manually selecting their screen-space locations in the neutral pose, their locations can subsequently be tracked throughout a video (see e.g. \cite{karaev2023cotracker, doersch2023tapir}). During extreme poses, dot locations may be lost and require re-labeling; alternatively, problematic frames can be sandwiched by tracking forward and backward through time starting from frames of higher confidence. Dot correspondences between the real and synthetic images can be used in the place of neural segmentation in order to construct an improved optical flow field for warping. This requires that dots must also be present in the synthetic geometry's texture. Although one could label them manually, we instead use a well-aligned neutral expression in order to project the dots from the real-world image into the synthetic geometry's texture (note that this assumes an accurate neutral geometry). 

The dot-based optical flow field is constructed by solving a Poisson equation (using a standard 5-point stencil) on the synthetic image with Neumann boundary conditions on the image edges and Dirichlet boundary conditions at the locations of the synthetic dots. It is important to note that point-based Dirichlet boundary conditions rapidly decay (as compared to line-based conditions) in $\mathbb{R}^2$, as was discussed in \cite{bao2015fullyautomatic}. We developed a number of techniques aimed at alleviating this issue to some degree. Firstly, and perhaps most importantly, the dots should be placed in the key areas of interest where sliding would lead to large errors. For example, dots should be placed around the edge of the lip line and near the nasolabial folds. In addition, dots should be placed with regular density throughout the face. Secondly, in order to increase the support and decrease the falloff, the entire dot (not just the dot center) should be used for Dirichlet boundary conditions. Since only a single pixel is tracked for each dot, the velocity from that pixel is redistributed to all the surrounding pixels associated with that dot in order to provide values for the Dirichlet boundary conditions.

\subsection{Inverse Rendering}
\label{sec:inverserendering}

Given a sequence of frames, the MetaHuman Animator \cite{metahuman} is used to initialize a rough approximation to the reconstructed geometry on every frame. Each iteration, every real-world image in the sequence is warped and projected onto its corresponding synthetic geometry (see Section \ref{sec:bridgingthedomaingap}). Since differentiable renderers work better with smaller perturbations, only a fraction (increasing linearly from 10\% to 100\% as the iterations proceed) of the optical flow is used for the warp. Next, an off-the-shelf differentiable renderer \cite{ravi2020pytorch3d} is used to improve the current estimate to the reconstructed geometry in each frame. The data term is a per-pixel photometric loss $\mathcal{L}_{photo}(\Delta \mathbf{v}) =  \sum_{i} \big[ \Psi(\mathbf{v} + \Delta \mathbf{v}, T, L)_i - \mathcal{I}^R_{i}  \big]^2$ that compares the synthetic image obtained from a differentiable renderer $\Psi$ to the real-world image $\mathcal{I}^R$. Here, $\mathbf{v}$ are the vertex positions of the geometry, and $\Delta \mathbf{v}$ is a perturbation to the vertex positions. Since the texture $T$ computed according to Section \ref{sec:bridgingthedomaingap} already contains baked-in lighting, only a simple model of ambient lighting is required for $L$; importantly, the ambient lighting should leave the texture colors unchanged. In addition to the data term, a number of regularization terms are used: a penalty on changes in edge lengths (as compared to the initial rough approximation to the geometry), a penalty on the Laplacian (using a one-ring stencil) of vertex displacements, a penalty on deviations of face normals (as compared to the initial rough approximation to the geometry), and a penalty on the strain energy of interior tetrahedra when a volumetric model is used. The strain energy term is quite useful for preserving lip volume and thus for creating lip bulge.

The regularization terms tend to overly prevent the data term from reaching a result consistent with matching the marker positions between the real and synthetic images. Thus, we use weaker weights on the regularization terms sacrificing mesh quality in order to obtain a better result for marker matching. Afterwards, a post-process is used to improve the quality of the mesh without adversely affecting the matching of the markers. This is achieved via a two-stage Laplacian smoothing algorithm. In the first stage, the vertex deltas corresponding to any vertices within a topological threshold distance to a facial marker are fixed in place while all other vertex deltas are smoothed. In the second stage, the complementary set of vertex deltas are held fixed, and the marker-associated deltas are smoothed. This results in a mesh with high marker reconstruction accuracy without high-frequency artifacts.

\subsection{Blurred Textures}
\label{sec:blurredtextures}

Although the process outlined in Section \ref{sec:inverserendering} creates geometry that matches the marker positions well, some regions (such as the inner lip) end up with overly noisy geometry. This can be alleviated by a second inverse rendering solve, which is only used to optimize specific problematic regions of the face such as the inner lip area.

For each frame, the real-world image is (optionally) warped before projecting it onto the corresponding initial rough approximation to the geometry. Then, a single blurred texture is created for a sequence of frames by averaging the projected texel values over all relevant frames. This blurred texture is used along with the rough approximation to the geometry in order to determine a lighting approximation $L$ for the sequence by minimizing $\mathcal{L}_{photo}(\Delta L) = \sum_{i} \big[ \Psi(\mathbf{v}, T, L + \Delta L)_i - \mathcal{I}^R_{i} \big]^2$ across all frames. A spherical harmonics approximation \cite{ramamoorthi2001sh} to the lighting $L$ was used. In texels where the projected values do not vary much over the sequence, the averaged texture is close to the unaveraged values causing the lighting model to aim for an ambient-only explanation. In texels where the averaged texture varies greatly from the unaveraged values, the diffuse components of the lighting model strive to explain those variations in a manner that is consistent with the animated normals of the geometry. Similarly, specular components (if used) would strive to explain variations in a manner consistent with both animated normals and the camera viewing angle.

After estimating the lighting $L$, a per-frame de-lit texture is calculated by minimizing $\mathcal{L}_{photo}(\Delta T) =  \sum_{i} \big[ \Psi(\mathbf{v}, T + \Delta T, L)_i - \mathcal{I}^R_{i}  \big]^2$. Subsequently, each per-frame de-lit texture is blurred by temporally averaging (with a kernel) the texel values in a time window. Generally speaking, temporally blurred textures can improve the results obtained via inverse rendering by creating additional feature overlap (similar in spirit to the incremental warping) and by smoothing out spurious sharp features (such as specular highlights) facilitating differentiability in image space. For similar reasons, it may also be helpful to blur parts of the target image.

The second inverse rendering solve uses $\loss_{photo}(\Delta \mathbf{v})$ with the lighting obtained from $\loss_{photo}(\Delta L)$ and the texture obtained from $\loss_{photo}(\Delta T)$. It is important to stress that temporal blurring typically hinders the all-important matching of marker positions. This is why we restrict the second inverse rendering solve to problematic regions of the face such as the lip region where high curvature can lead to ill-conditioned specular highlights and the inner lip region where the occlusion boundary can cause discontinuities in the projected texture.  

\subsection{Inner Mouth and Teeth}

When the mouth opens, the ability to match a synthetic rendering of the inner mouth and teeth with the corresponding areas of the real-world image helps to avoid confusing those areas with occlusion boundaries of the lips. Thus, we use template teeth, gums, and inner mouth geometry and textures. For each frame, scalar intensity modifiers are calculated and used to uniformly scale the template textures. One scalar is calculated for the teeth/gums region and a separate scalar is calculated for the remainder of the inner mouth. Although segmentation of the ground-truth image works well to identify the inner occlusion boundaries of the lips, allowing for a straightforward comparison of the inner mouth with the corresponding pixels in the image, segmenting out the teeth and/or the teeth and gums together proved to be more difficult. Thus, the scalar multiplier for the teeth/gums textures is calculated by assuming that the teeth/gums region is aligned with the associated region in the image; obviously, this leads to a source of error.

\section{Optimizing the Tracking Rig}
\label{sec:optimizing_the_tracking_rig}

Assuming that the performer has had their rig calibrated to some $\theta_{\rig}$ using Simon-Says in Section \ref{sec:simonsays}, we further optimize that rig to a $\theta_{\trk}$ more appropriate for tracking. The Simon-Says images or any other images can be used for this, as long as aspirational $(I, c)$ pairs exist. Of course, $c$ should be modified to $c^+$ via Equation \ref{eq:updatecko} and other modifications, as discussed in Section \ref{sec:incorrect_expressions}. Using the $(I, c^+)$ pairs, Equation \ref{eq:trackerobj4} can be written as 
\begin{equation} \label{eq:trackeropt11}
\begin{split}
   \min_{\theta_{\trk}} \sum_k \Big(  & \gamma_1 ||\trk (I_k; \theta_{\trk}) - c^+_k||_2^2 ~ + \\[-1.5ex] & \gamma_2 ||\rig (\trk (I_k; \theta_{\trk}); \theta_{\rig}) - v^+_k||_2^2 \Big)  ~ + \\ & \gamma_{\epsilon} || \theta_{\trk} - \theta_{\rig} ||_2^2
\end{split}
\end{equation}
where $v^+ = \rig (c^+; \theta_{\rig})$. Equations \ref{eq:rigoutputpluserror} and \ref{eq:solvedtdthetawithvhat} can be written as
\begin{subequations} \label{eq:vhatandderivs_subequations}
    \begin{equation}
        \hat{v}(I; \theta_{\trk}) = \rig (\trk (I; \theta_{\trk}); \theta_{\trk})
    \end{equation}
    \begin{equation} 
    \begin{split}
        \label{eq:dtdthetawithoutcontractions}
         \left. \frac{\partial \rig (c; \theta_{\trk})}{\partial c} \right|_{c=\trk (I; \theta_{\trk})} \left. \frac{\partial \trk (I; \theta)}{\partial \theta} \right|_{\theta = \theta_{\trk}} = \qquad\qquad\qquad\qquad \\ - \left. \frac{\partial \rig (\trk (I; \theta_{\trk}); \theta)}{\partial \theta} \right|_{\theta = \theta_{\trk}} + \left. \frac{\partial \hat{v}(I; \theta)}{\partial \theta} \right|_{\theta = \theta_{\trk}}
    \end{split}
    \end{equation}
\end{subequations}
in order to differentiate Equation \ref{eq:trackeropt11}. The pseudo-inverse of the normal equations can be used to solve for $\dd{\trk}{\theta}$ in Equation \ref{eq:dtdthetawithoutcontractions}.

\subsection{MetaHuman Framework}
\label{sec:metahuman_tracker}

The MetaHuman Animator (MHA) tools suite was chosen because its size and complexity is representative of similar industry frameworks. For the animation controls, $c \in \mathbb{R}^{174}$ can be divided into 97 primary controls and 77 so-called ``tweaker'' controls. They are combined in various ways to create $814$ pose-space deformation (PSD) controls. The MetaHuman animation rig controls 24,049 vertices via $870$ joints, each with $9$ degrees of freedom ($3$ translation, $3$ rotation, $3$ scaling). For the sake of implementation, we implement the rig to output the 7,830 degrees of freedom associated with the joints instead of the 72,147 degrees of freedom associated with the vertices. This optimization ignores the fact that the PSD controls can also affect blendshape correctives; instead, it assumes that the final mesh can be obtained merely by skinning the joints. The discussions throughout this paper are valid in either case, but this simplification makes the optimization more tractable. The animation rig determines the joint degrees of freedom by multiplying the joint matrix by the PSD controls. The size $7,\hspace{-0.3ex}830 \times 814$ joint matrix has only 745,284 nonzero entries, meaning that $\theta_\rig \in \mathbb{R}^{745284}$.

In order to differentiate the tracker via Equation \ref{eq:dtdthetawithoutcontractions}, $\dd{\rig}{c} \in \mathbb{R}^{(7830, 174)}$ and $\dd{\rig}{\theta} \in \mathbb{R}^{(7830, 745284)}$ would need to be computed. In order to make $\dd{\rig}{\theta}$ tractable, each term in the sum in Equation \ref{eq:trackeropt11} is minimized over only the nonzero $\theta$ in the columns of the joint matrix corresponding to the PSD controls that are not identically zero according to $c^+$. This greatly reduces the dimension of $\theta$ for any reasonable $(c^+, v^+)$ pair. For example, the nineteen expressions (ignoring the neutral) used for the Simon-Says capture have their dimensionality reduced to a far more tractable 1589, 6456, 1470, 1470, 9470, 15252, 11465, 7890, 7939, 2868, 4050, 3198, 12936, 4051, 7262, 3758, 7506, 2594, 2594, respectively. Of course, reducing each expression to a few thousand parameters means that hundreds of expressions (depending on expression overlap) would be needed to cover the full rig. We circumvent this by limiting our expression set to cover only the most important controls, noting that this leaves $\theta_{\trk}$ fixed to its $\theta_{\rig}$ values for the parameters that do not appear in the expression set (in the spirit of the last term of Equation \ref{eq:trackeropt11}).

\subsection{Tracker Variants}
\label{sec:tracker_variants}

Our optimization strategy will make use of the following variations of Equation \ref{eq:trackeropt11}. For each expression, i.e. each term in the sum in Equation \ref{eq:trackeropt11}, let $\mathcal{H}$ be a diagonal matrix of Heaviside functions $H(|c^+_{i}|)$. Let $\trk_{\mathcal{H}} = \mathcal{H} \hspace{.03cm}  \trk$ represent the filtering of the tracker to zero out entries that are zero in $c^+$. Let $\mathcal{H}_D$ be the decimation of $\mathcal{H}$ into a wide matrix via the elimination of rows that are entirely full of zeros; then, $c_D = \mathcal{H}_D c$ is the subset of $c$ containing all relevant controls, and $\mathcal{H}_D \trk$ is a similarly decimated tracker. Although $\trk_\mathcal{H}$ and $\mathcal{H}_D \trk$ are essentially equivalent, $\mathcal{H}_D \trk$ facilitates code optimizations. Let $\theta_D$ be the subset of the rig parameters that depends on $c_D$, let $\rig_D(c_D; \theta_D)$ be the subset of the rig that can be modified by $c_D$, and let $\trk_D(I; \theta_D)$ represent a reduced tracker that only considers the reduced degrees of freedom from $c_D$. In contrast to $\trk_{\mathcal{H}}$, $\trk_D$ is not allowed to use filtered out degrees of freedom in order to explain the geometry. 

The $\gamma_1$ term in Equation \ref{eq:trackeropt11} can use any of the three trackers, decimating $c^+$ for $\trk_D$. The $\gamma_2$ term in Equation \ref{eq:trackeropt11} can use any of the three trackers modifying $\rig$ to $\rig_D$ for $\trk_D$; in addition, $v^+$ should be decimated if the output of $\rig_D$ is a decimated version of the output of $\rig$. Superscripts on $\gamma$ (i.e. $\gamma$, $\gamma^{\mathcal{H}}$, $\gamma^D$) will be used to indicate which tracker was used in a specific term in Equation \ref{eq:trackeropt11}. Equation \ref{eq:vhatandderivs_subequations} can be used for any of the three trackers, modifying $\rig$ to $\rig_D$ for $\trk_D$; in addition, $v^+$ should be decimated if the output of $\rig_D$ is. Inserting $\trkh$ into Equation \ref{eq:dtdthetawithoutcontractions} and using the normal equations on $\dd{\rig}{c} \mathcal{H}$ leads to a coefficient matrix with nonzero entries corresponding to the normal equations for $\rig_D$, i.e. $\dd{\trkh}{\theta}$ agrees with $\dd{\trkd}{\theta}$ when it is nonzero. This highlights the aforementioned code optimizations for $\dd{\trkh}{\theta}$, which consist of replacing $\rig$ with $\rig_D$ and $\trkh$ with $\mathcal{H}_D \trk$.

\subsection{Optimization Strategy}
\label{sec:optimization_strategy}

A typical rig contains primary controls that retarget well and auxiliary controls that one might want to ignore when retargeting to a new character. Unfortunately, when a tracker ignores the auxiliary controls, the primary controls end up polluted as they work too hard to explain geometry that would have been better explained by auxiliary controls. Thus, a tracker should aim to use the entire set of controls; however, it should lean more heavily on the primary controls as opposed to the auxiliary controls. The trick is to figure out how to use just enough of the auxiliary controls in order to obtain non-polluted values for the primary controls. Motivated by this, we propose modifying the rig used by the tracker (i.e. $\theta_{\trk}$) so that the primary controls properly explain the geometry and so that the auxiliary controls do not attempt to explain the geometry. This can be seen as an attempt to ``orthogonalize'' the rig used by the tracker.

\section{Open-Source Tracker}
\label{sec:opensourcetracker}

\begin{figure}[!htb]
\includegraphics[width=\linewidth]{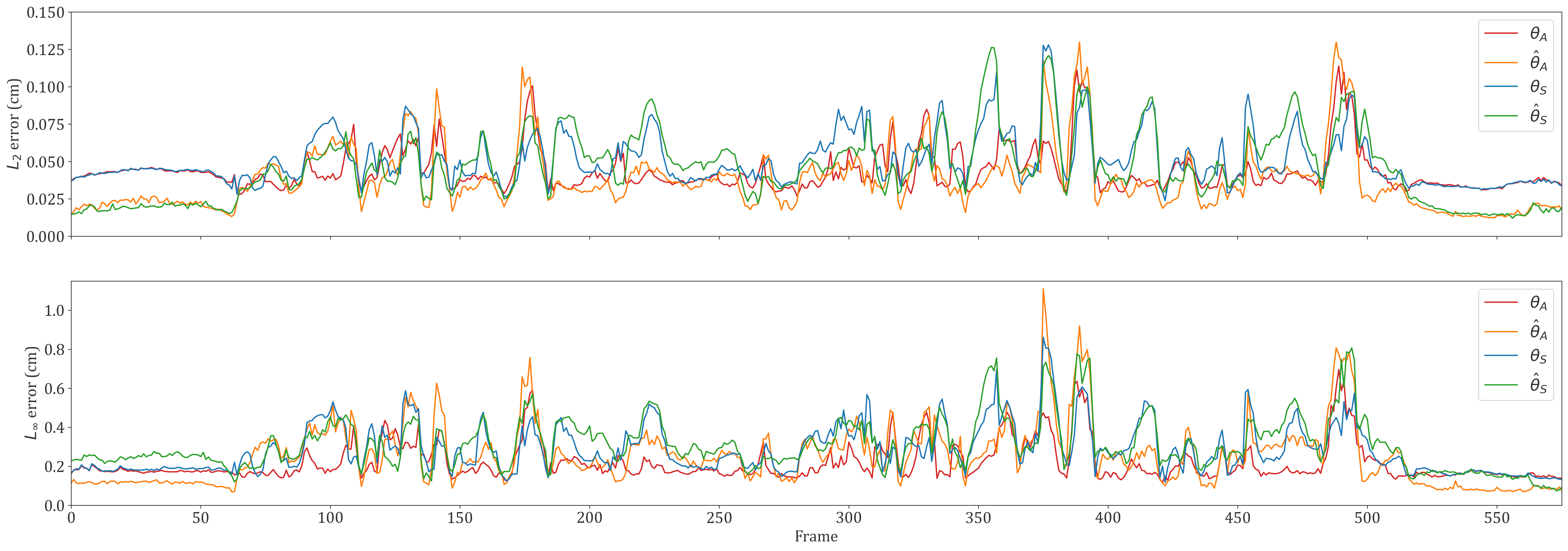}
\vspace*{-5ex}
\caption{A 573 frame pangram was used to test the tracker's ability to invert the rig and match the reconstructed geometry. The tracker was mostly able to adequately minimize geometry errors using any of $\theta_A$, $\hat{\theta}_A$, $\theta_S$, $\hat{\theta}_S$, even though the output controls (and thus the semantic interpretation) can vary significantly. See also Figure \ref{fig:opensource_geometryreconstruction_viz}.}
\Description{DESCRIPTION}
\label{fig:opensource_geometryreconstruction}
\end{figure}

\begin{figure}[!htb]
\includegraphics[width=\linewidth]{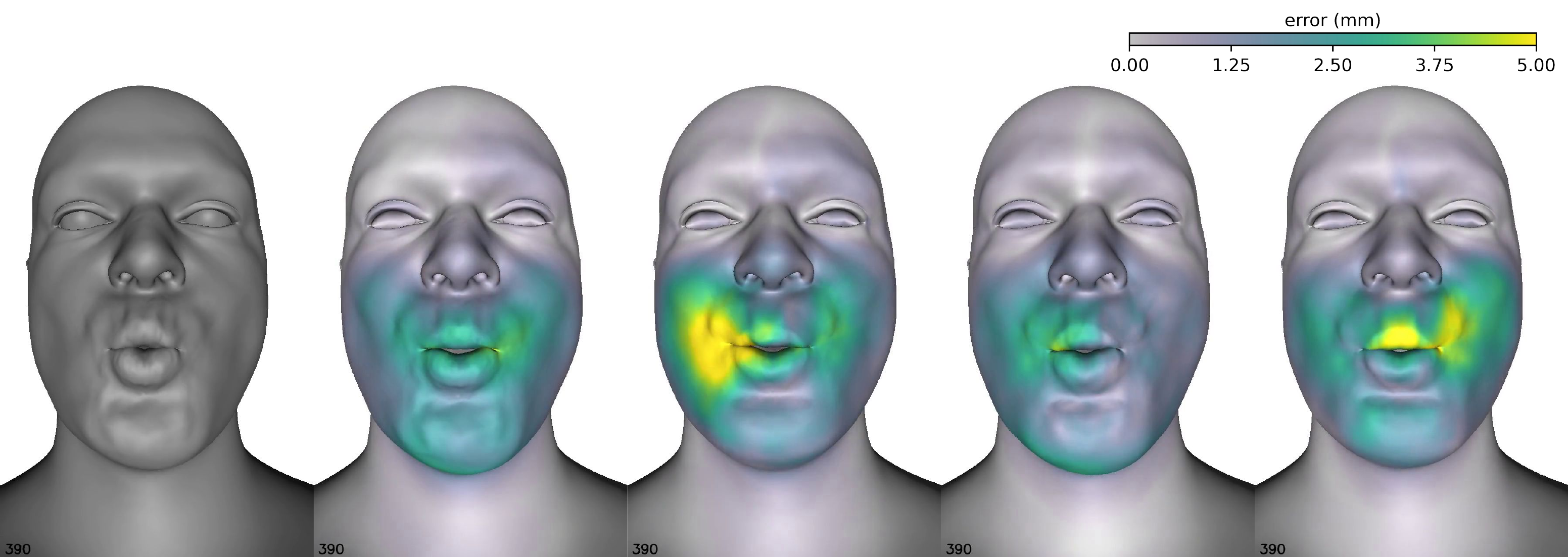}
\caption{Ground-truth reconstructed geometry (left) as compared to the geometry output by the tracker using $\theta_A$, $\hat{\theta}_A$, $\theta_S$, $\hat{\theta}_S$, respectively. This corresponds to frame 390, which was chosen because it has relatively large errors, of Figure \ref{fig:opensource_geometryreconstruction}. Colored regions indicate reconstruction error according to the colorbar.}
\Description{DESCRIPTION}
\label{fig:opensource_geometryreconstruction_viz}
\end{figure}

We begin by considering an open-source tracker (using gradient descent, L-BFGS \cite{fletcher1987practical}, etc.) where one is able to modify $\trk$ to $\trkd$; in Section \ref{sec:blackboxtracker}, we consider a black-box tracker where one cannot make use of $\trkd$.  Let $\theta_A$ be the internal rig parameters obtained from the MetaHuman auto-rigging service, and let $\theta_S$ be the result obtained by modifying the rig parameters via Simon-Says (see Section \ref{sec:simonsays}). Either $\theta_A$ or $\theta_S$ can be used as $\theta_\rig$ in the rig (and as $\theta_{\rig}$ in the $\gamma_2$ term); in addition, either can be used as the initial guess for $\theta_{\trk}$. In this section, we show how to improve the semantics of the tracker by modifying $\theta_A \rightarrow \hat{\theta}_A$ and $\theta_S \rightarrow \hat{\theta}_S$. All the trackers use the same reconstructed geometry from Titan, again assuming that obtaining that geometry reconstruction is more straightforward than rig inversion; thus, the trackers will be functions of the geometry, not the images. As shown in Figures \ref{fig:opensource_geometryreconstruction} and \ref{fig:opensource_geometryreconstruction_viz}, changing $\theta_A \rightarrow \theta_S$ via Simon-Says or modifying $\theta_A \rightarrow \hat{\theta}_A$ and $\theta_S \rightarrow \hat{\theta}_S$ has little effect on the ability of the tracker to invert the rig and match the geometry; however, the tracker output and thus the semantic interpretation can vary significantly. All the Simon-Says expressions were used in the optimization.

\begin{figure}[!htb]
\includegraphics[width=\linewidth]{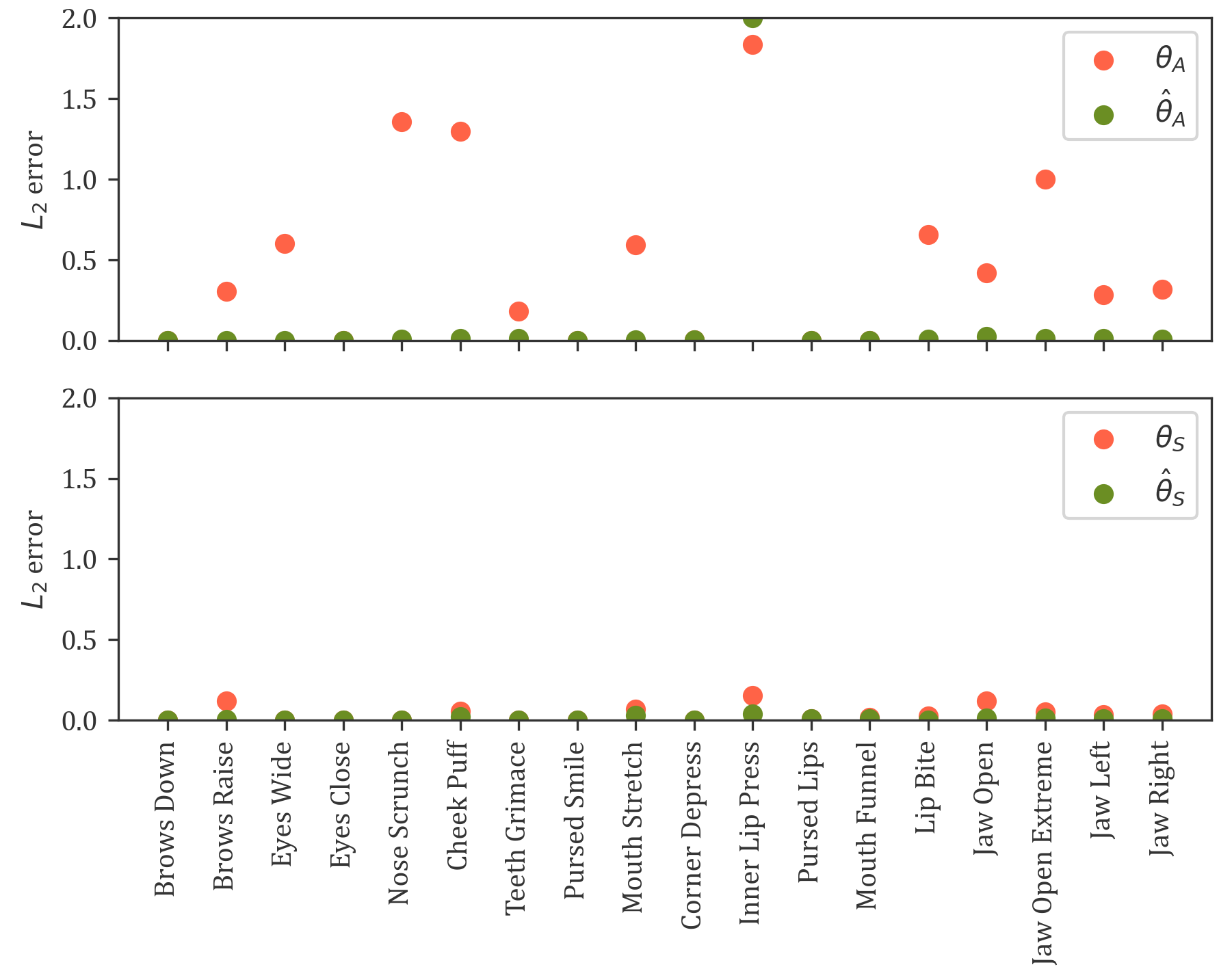}
\vspace*{-5ex}
\caption{$L_2$ errors (according to $\gamma_1^D$) on the various expressions before (red) and after (green) optimizing $\theta_A \rightarrow \hat{\theta}_A$ (top) and $\theta_S \rightarrow \hat{\theta}_S$ (bottom) using only the primary controls on each expression via $\gamma_1^D$ and $\gamma_2^D$.}
\Description{DESCRIPTION}
\label{fig:opensource_stage1}
\end{figure}

Our proposed strategy for improving the tracker, either $\theta_A \rightarrow \hat{\theta}_A$ or $\theta_S \rightarrow \hat{\theta}_S$, begins by utilizing a limited set of rig parameters $\theta_D$ along with $\rig_D$ and $\trkd$ in order to explain the geometry via a limited set of controls $c_D$. Thus, $\gamma_1^D$ and $\gamma_2^D$ are used in Equation \ref{eq:trackeropt11}. See Figure \ref{fig:opensource_stage1}.  Importantly, we do not treat the results of Equation \ref{eq:trackeropt11} as the final solution along the lines of solving an inverse problem. Instead, motivated by machine learning, we use the optimization simply to generate samples in parameter space. In machine learning applications, holdout data is used to pick the most salient sample. We instead use the full tracker $\trk$ in order to supervise the process. That is, modifying $\theta_{\trk}$ to $\hat{\theta}_{\trk}$ by replacing the parameters corresponding to $\theta_D$ with those of a generated sample, the various $\trk(v; \hat{\theta}_{\trk})$ can be analyzed to determine the best $\hat{\theta}_\trk$. Given the best $\hat{\theta}_\trk$, as supervised by $\trk$, this is used to bootstrap improvement of $\trk$ itself.

\begin{figure}[!htb]
\includegraphics[width=\linewidth]{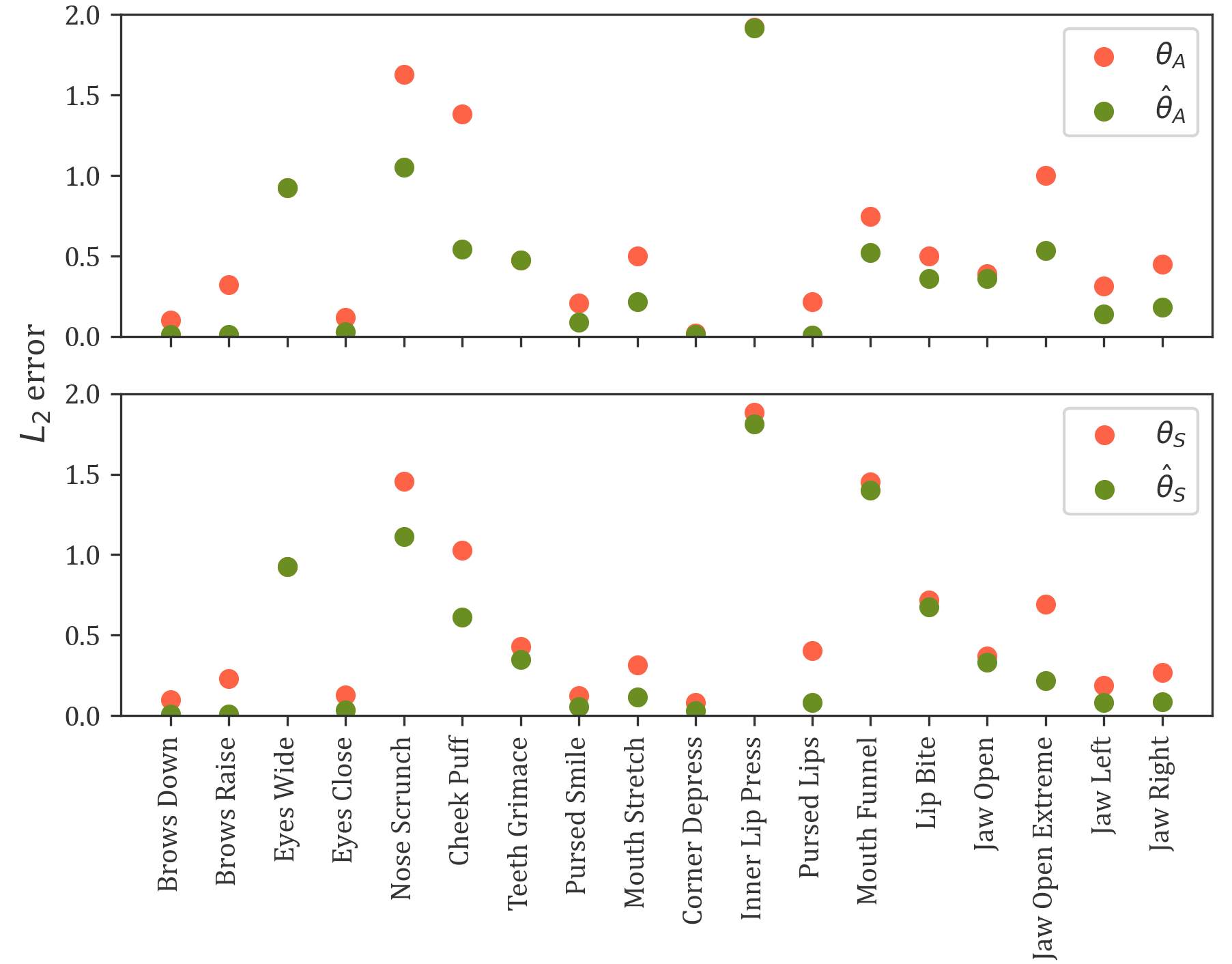}
\vspace*{-5ex}
\caption{$L_2$ errors (according to $\gamma_1^{\mathcal{H}}$) on the various expressions before (red) and after (green) optimizing $\theta_A \rightarrow \hat{\theta}_A$ (top) and $\theta_S \rightarrow \hat{\theta}_S$ (bottom) using the full rig while filtering the primary controls on each expression via $\gamma_1^{\mathcal{H}}$ and $\gamma_2^{\mathcal{H}}$.}
\Description{DESCRIPTION}
\label{fig:opensource_stage2}
\end{figure}

\begin{figure}[!htb]
\includegraphics[width=\linewidth]{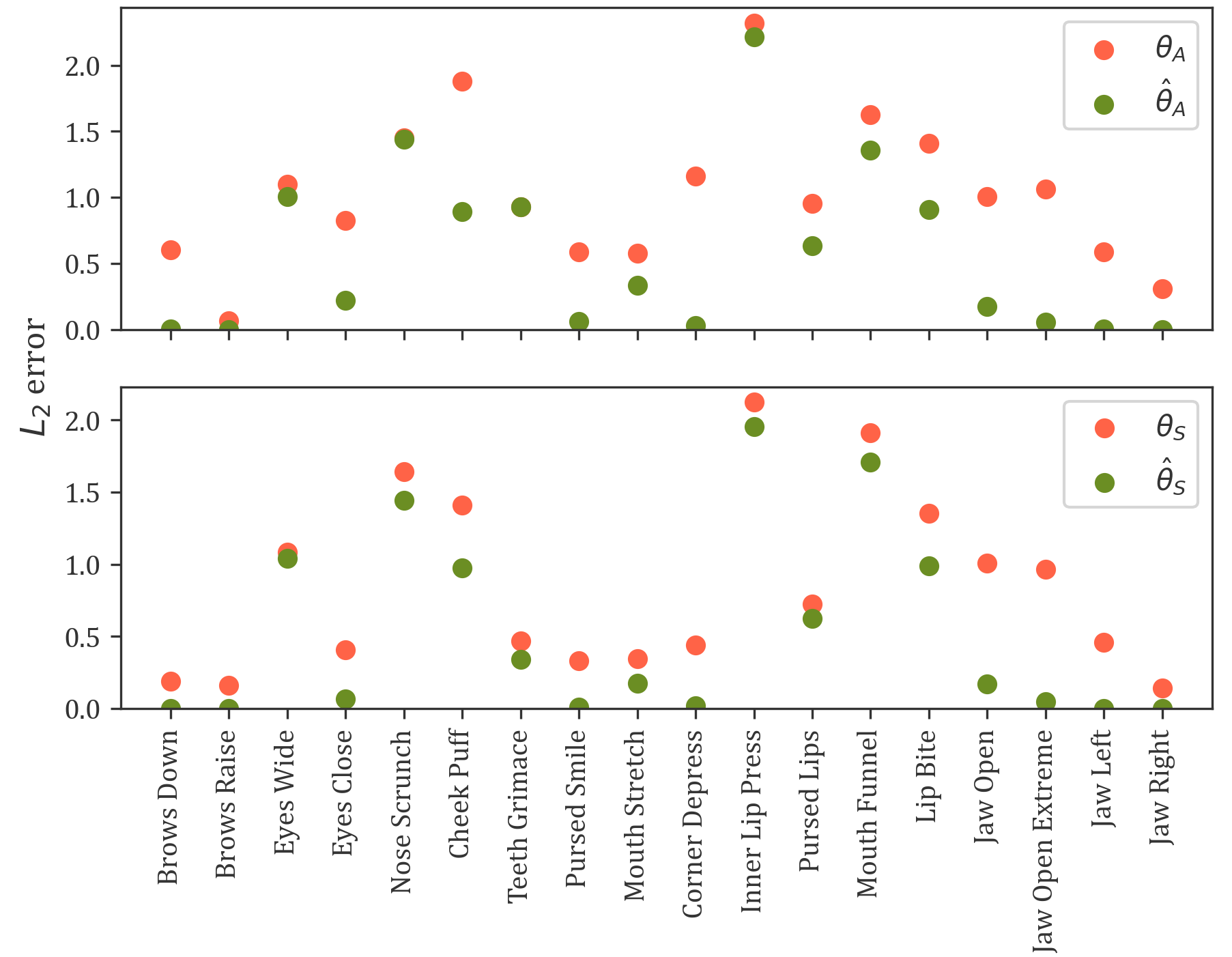}
\vspace*{-5ex}
\caption{$L_2$ errors (according to all $\gamma_1^{\mathcal{H}}$ terms) on the various expressions before (red) and after (green) optimizing $\theta_A \rightarrow \hat{\theta}_A$ (top) and $\theta_S \rightarrow \hat{\theta}_S$ (bottom) using the full rig while filtering the primary controls on each expression via $\gamma_1^{\mathcal{H}}$ and $\gamma_2^{\mathcal{H}}$ and additionally filtering various spurious controls via an additional $\gamma_1^{\mathcal{H}}$ term. Note that the minimization is still only considering the columns corresponding to the primary controls. }
\Description{DESCRIPTION}
\label{fig:opensource_stage3}
\end{figure}

\begin{figure}[!htb]
\includegraphics[width=\linewidth]{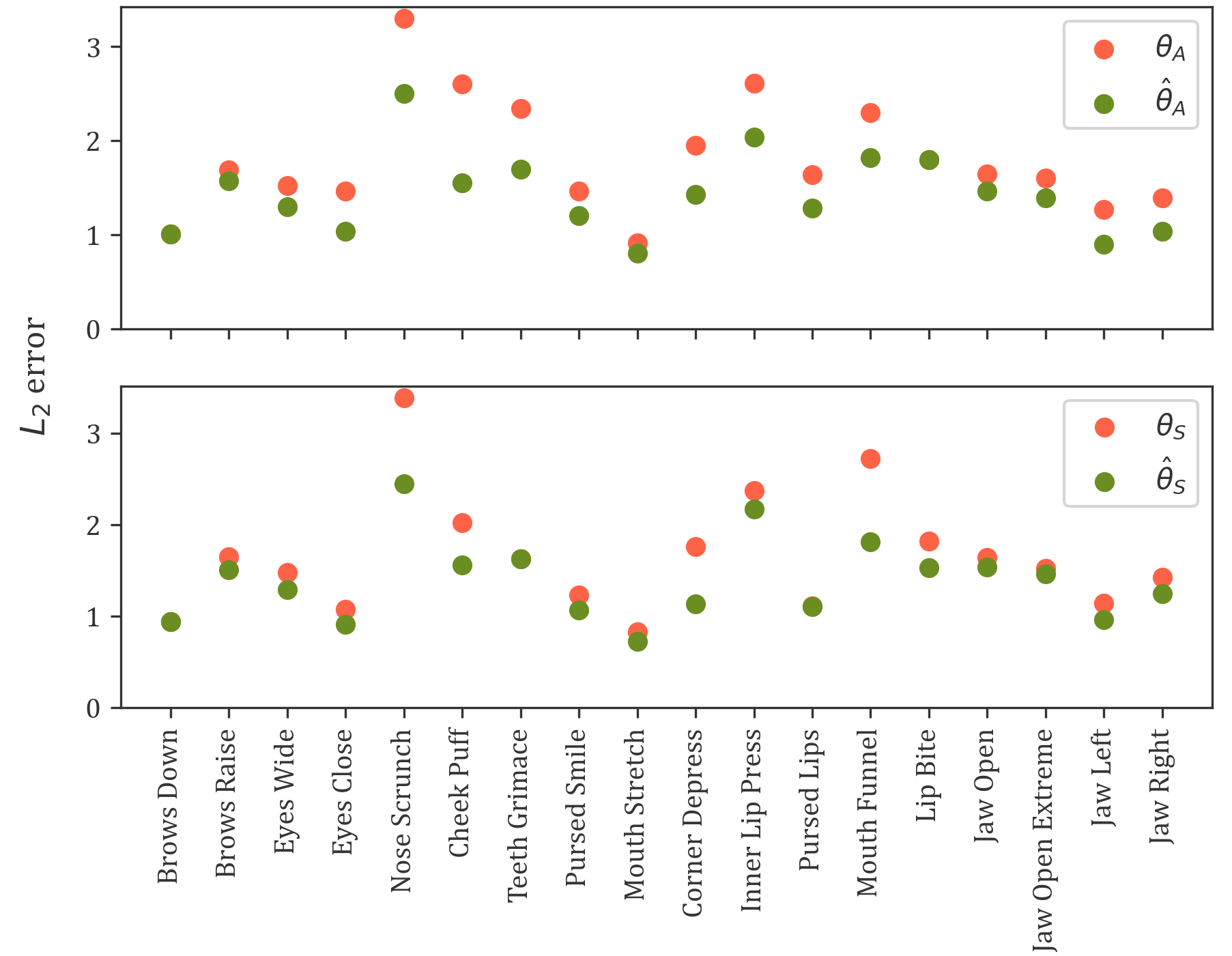}
\vspace*{-5ex}
\caption{$L_2$ errors (according to all $\gamma_1$ terms) on the various expressions before (red) and after (green) optimizing $\theta_A \rightarrow \hat{\theta}_A$ (top) and $\theta_S \rightarrow \hat{\theta}_S$ (bottom) using the full rig while filtering the primary controls on each expression via $\gamma_1^{\mathcal{H}}$ and $\gamma_2^{\mathcal{H}}$ and additionally filtering various spurious controls via an additional $\gamma_1^{\mathcal{H}}$ term. The difference between this figure and Figure \ref{fig:opensource_stage3} is that the minimization is now considering the columns of the spurious controls while freezing the columns of the primary controls.}
\Description{DESCRIPTION}
\label{fig:opensource_stage4}
\end{figure}

In the second stage of the process, we switch from $\trk_D$ to $\trkh$, still optimizing only the same columns as before (i.e. for the primary controls only) but aiming to get the full tracker to also give the desired output. Thus, $\gamma_1^{\mathcal{H}}$ and $\gamma_2^{\mathcal{H}}$, where $\mathcal{H}$ selects only the primary controls, are used in Equation \ref{eq:trackeropt11}. See Figure \ref{fig:opensource_stage2}. In the third stage, again optimizing the same columns, we aim to minimize the contributions of the spurious controls by adding a second $\gamma_1^{\mathcal{H}}$ term that selects problematic spurious controls aiming to set them to zero. Multiple such $\gamma_1^{\mathcal{H}}$ terms can be used to strategically drive spurious controls to zero based on various priorities. Note that the $\gamma_2^{\mathcal{H}}$ term is not changed from the previous stage, i.e. it still only selects the primary controls. See Figure \ref{fig:opensource_stage3}. At this point in the process, the hope would be that the columns corresponding to the primary controls have been adjusted to explain as much of the expression as possible. Since the Simon-Says expression set was chosen to be minimal for the sake of efficiency, the expressions have minimal overlap allowing each term in Equation \ref{eq:trackeropt11} to be run separately and in parallel. The only overlap in Figure \ref{fig:simonsays_chart} is between \texttt{jaw\_open} and \texttt{jaw\_open\_extreme}, and it was not problematic to consider them independently. In the final (fourth) stage of the process, columns corresponding to spurious controls are considered. Since the spurious controls can have a lot of overlap, every relevant expression should be included in the sum in Equation \ref{eq:trackeropt11} when considering them. Again, multiple $\gamma_1^{\mathcal{H}}$ terms can be used in order to select the controls of interest. Note that the $\gamma_2^{\mathcal{H}}$ still remains unchanged, i.e. it still only selects the primary controls. See Figure \ref{fig:opensource_stage4}.

\begin{figure}[!htb]
\includegraphics[width=\linewidth]{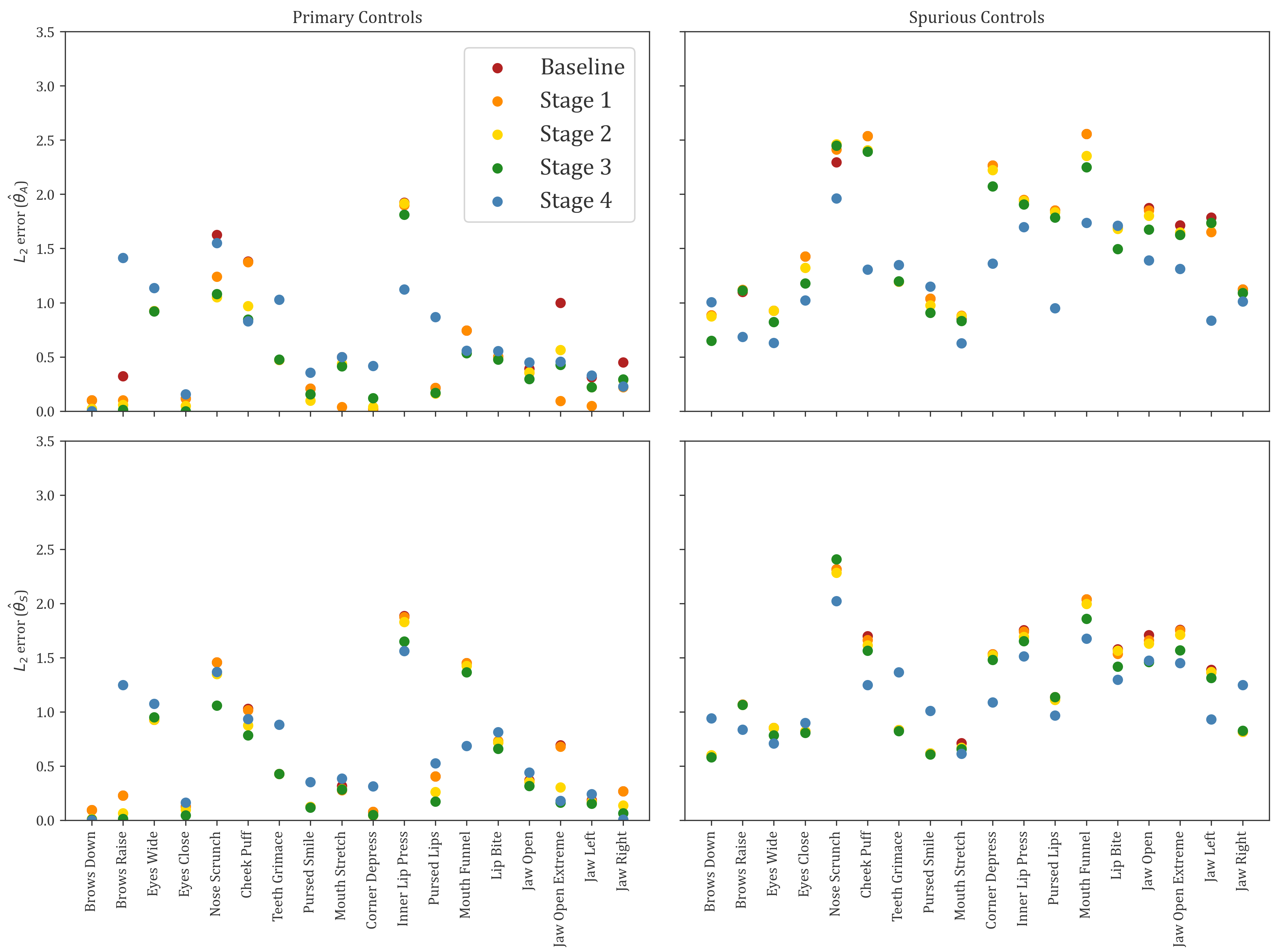}
\vspace*{-5ex}
\caption{Summary of the improvement in the $L_2$ errors (according to $\gamma_1$) throughout all four stages of the process (see Figures \ref{fig:opensource_stage1} to \ref{fig:opensource_stage4}). The improvement in the primary controls is shown to the left, and the improvement in the spurious controls is shown to the right. Optimizing $\theta_A \rightarrow \hat{\theta}_A$ is shown on the top, and optimizing $\theta_S \rightarrow \hat{\theta}_S$ is shown on the bottom.}
\Description{DESCRIPTION}
\label{fig:opensource_allstages}
\end{figure}

Figure \ref{fig:opensource_allstages} shows a summary of how the full tracker $\trk$ improves over all stages. At this point, it is not possible to state the optimal strategy for improving the tracker, for example, it may be beneficial to optimize the columns corresponding to spurious controls first instead of last; however, these results do demonstrate that our approach is both effective and tractable. In particular, an optimal strategy would put more focus on animator preferences, e.g. they tend to prefer the minimization of spurious controls over maximal usage of primary controls; therefore, a proposed so-called ``optimal'' strategy is best left for future work.

\subsection{Validation With Synthetic Data}
\label{sub:retargeting_synthetic}

First, starting with the hand-crafted animation rig parameters $\theta_{GT}$ discussed in Section \ref{sub:synthetic_test}, we again use the Simon-Says expression set to generate ground-truth geometry for the tracker. As usual, the tracker is not able to properly invert the rig to successfully obtain the controls used to create the geometry. The errors are shown in the first column of Table \ref{table:retargeting_synthetic}. Using the optimization to improve $\theta_{GT} \rightarrow \hat{\theta}_{GT}$ can improve the results, as shown in the second column of Table \ref{table:retargeting_synthetic}. Both the morphed rig and the Simon-Says rig can also be improved via optimizing $\theta_M \rightarrow \hat{\theta}_M$ and $\theta_S \rightarrow \hat{\theta}_S$, respectively. Notably, Table \ref{table:retargeting_synthetic} shows that optimizing the morphed rig via Simon-Says $\theta_M \rightarrow \theta_S$ and subsequently improving $\theta_S \rightarrow \hat{\theta}_S$ via tracker optimization results in a $\hat{\theta}_S$ that can give lower errors than $\theta_{GT}$ on unseen expressions.

\begin{table}[!htb]
\footnotesize
\begin{tabular}{r|cccccc}
\toprule
Expression & $\theta_{GT}$ & $\hat{\theta}_{GT}$ & $\theta_M$ & $\hat{\theta}_M$ & $\theta_S$ & $\hat{\theta}_S$   \\
\midrule
Funnel Jaw Open     & $0.0260$ & $0.0230$ & $0.0364$ & $0.0351$ & $0.0268$ & $0.0241$  \\
Asymmetric Funneler & $0.0203$ & $0.0179$ & $0.0304$ & $0.0269$ & $0.0226$ & $0.0189$  \\
Happy Speaking      & $0.0215$ & $0.0191$ & $0.0400$ & $0.0328$ & $0.0256$ & $0.0202$  \\
``Dee''             & $0.0229$ & $0.0206$ & $0.0352$ & $0.0344$ & $0.0252$ & $0.0228$  \\
\bottomrule
\end{tabular}
\caption{Average error in the controls for synthetic expressions that were not included in the Simon-Says expression set when optimizing rig parameters for the tracker.}
\vspace*{-3ex}
\label{table:retargeting_synthetic}
\end{table}

\section{Black Box Tracker}
\label{sec:blackboxtracker}

\begin{figure}[!htb]
\includegraphics[width=\linewidth]{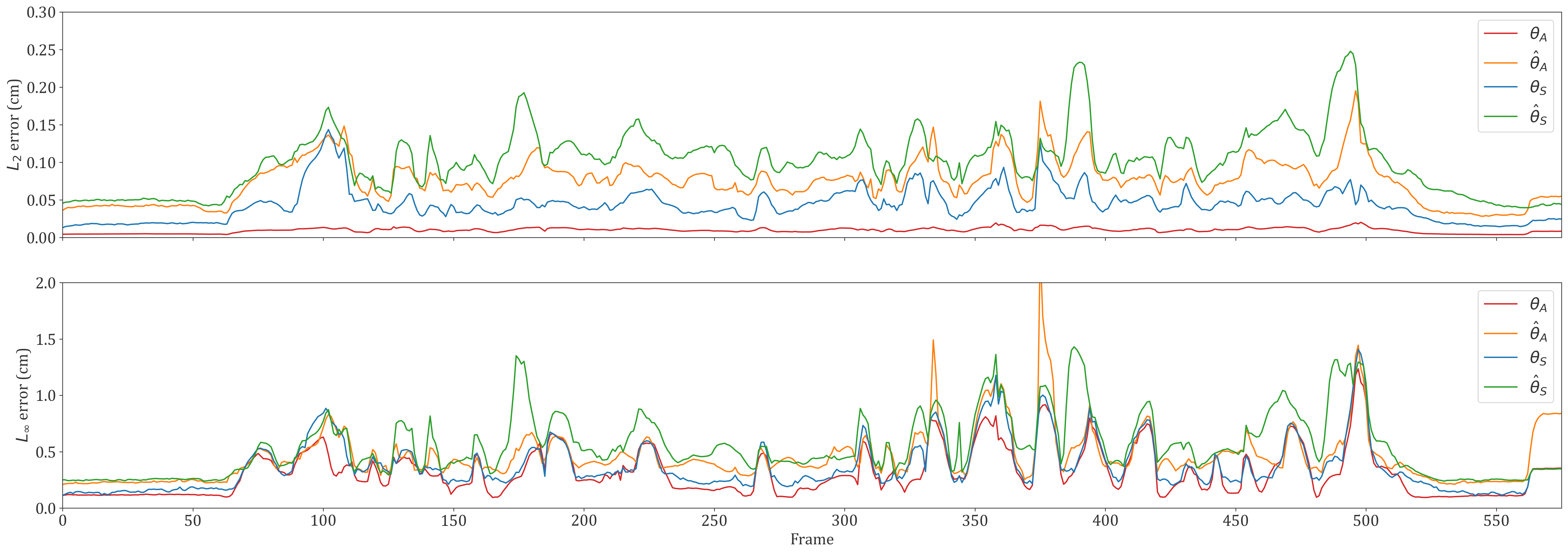}
\vspace*{-5ex}
\caption{A 573 frame pangram was used to test the tracker's ability to invert the rig and match the reconstructed geometry. The tracker was mostly able to adequately minimize geometry errors using any of $\theta_A$, $\hat{\theta}_A$, $\theta_S$, $\hat{\theta}_S$, even though the output controls (and thus the semantic interpretation) can vary significantly. See also Figure \ref{fig:closedsource_geometryreconstruction_viz}.}
\Description{DESCRIPTION}
\label{fig:closedsource_geometryreconstruction}
\end{figure}

\begin{figure}[!htb]
\includegraphics[width=\linewidth]{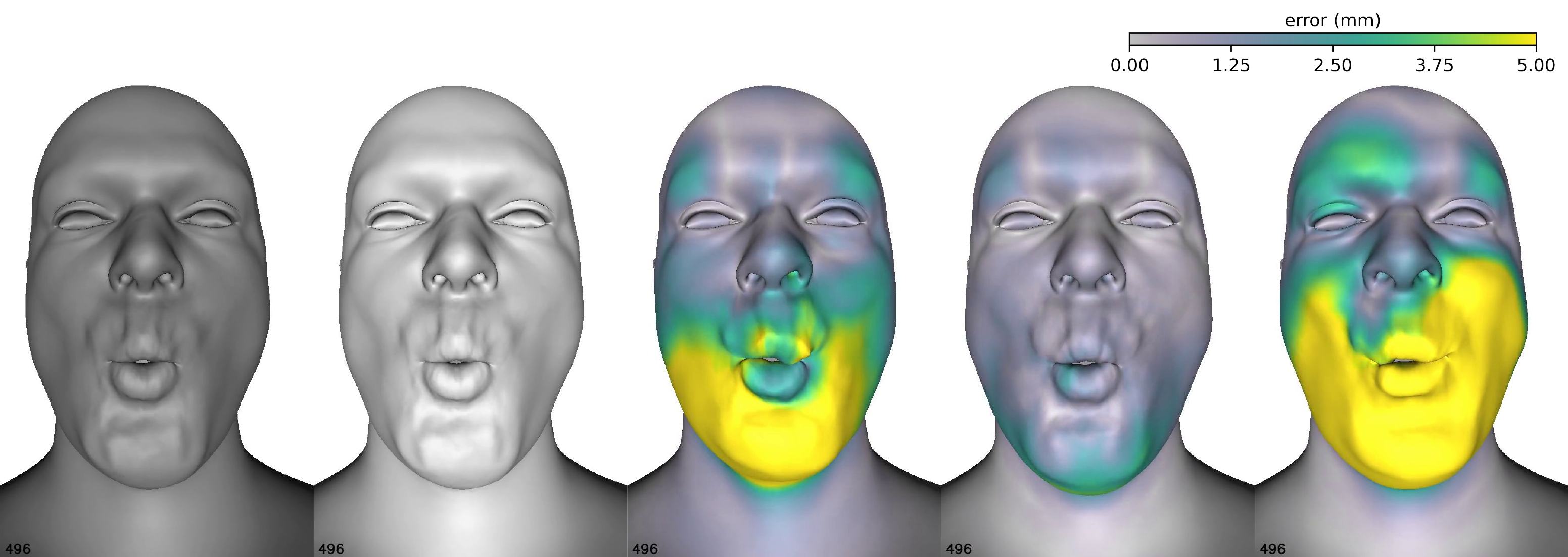}
\caption{Ground-truth reconstructed geometry (left) as compared to the geometry output by the tracker using $\theta_A$, $\hat{\theta}_A$, $\theta_S$, $\hat{\theta}_S$, respectively. This corresponds to frame 100, which was chosen because it has relatively large errors, of Figure \ref{fig:closedsource_geometryreconstruction}. Colored regions indicate reconstruction error according to the colorbar.}
\Description{DESCRIPTION}
\label{fig:closedsource_geometryreconstruction_viz}
\end{figure}

\begin{figure}[!htb]
\includegraphics[width=\linewidth]{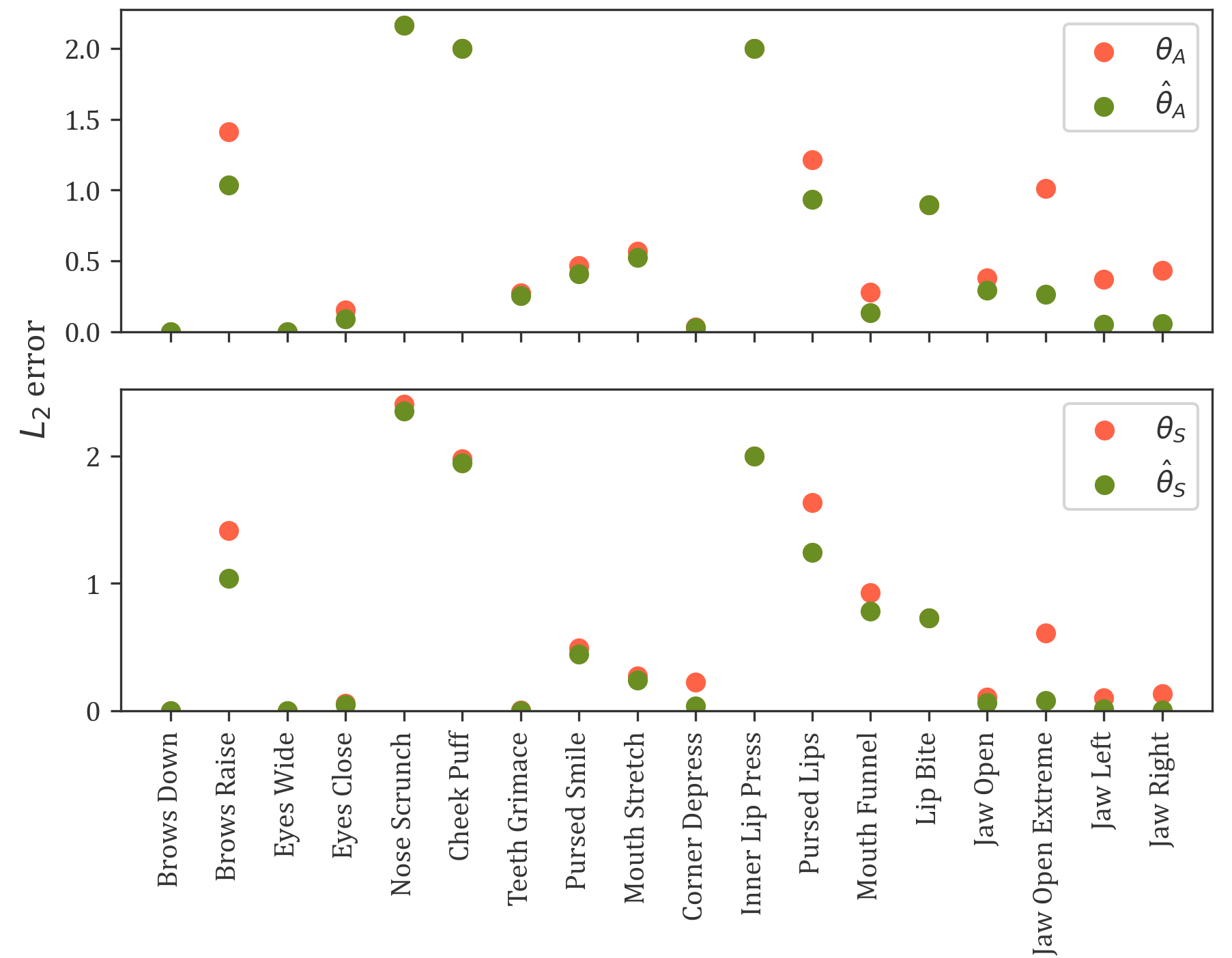}
\vspace*{-5ex}
\caption{$L_2$ errors (according to $\gamma_1^{\mathcal{H}}$) on the various expressions before (red) and after (green) optimizing $\theta_A \rightarrow \hat{\theta}_A$ (top) and $\theta_S \rightarrow \hat{\theta}_S$ (bottom) using the full rig while filtering the primary controls on each expression via $\gamma_1^{\mathcal{H}}$ and $\gamma_2^{\mathcal{H}}$.}
\Description{DESCRIPTION}
\label{fig:closedsource_stage2}
\end{figure}

\begin{figure}[!htb]
\includegraphics[width=\linewidth]{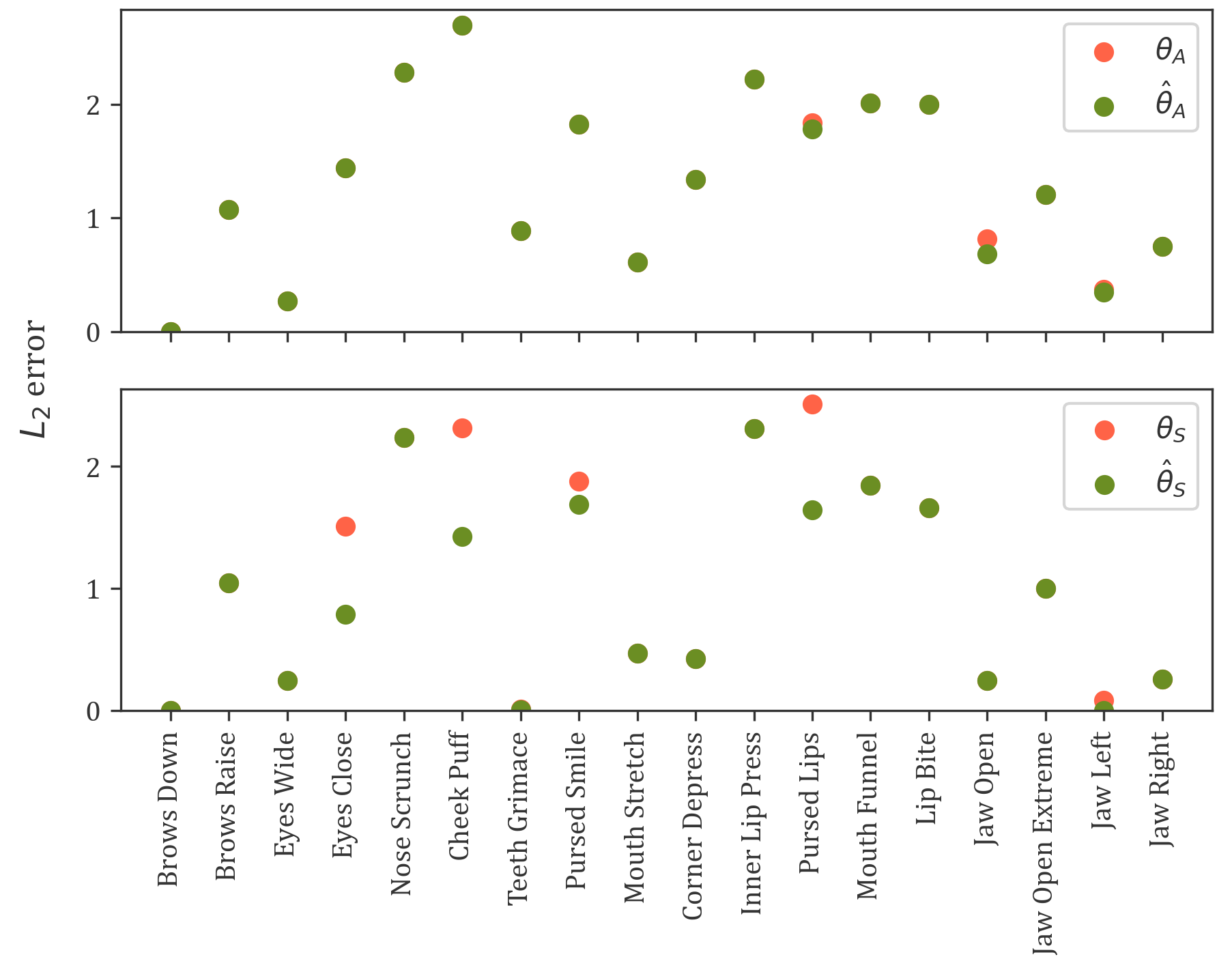}
\vspace*{-5ex}
\caption{$L_2$ errors (according to all $\gamma_1^{\mathcal{H}}$ terms) on the various expressions before (red) and after (green) optimizing $\theta_A \rightarrow \hat{\theta}_A$ (top) and $\theta_S \rightarrow \hat{\theta}_S$ (bottom) using the full rig while filtering the primary controls on each expression via $\gamma_1^{\mathcal{H}}$ and $\gamma_2^{\mathcal{H}}$ and additionally filtering various spurious controls via an additional $\gamma_1^{\mathcal{H}}$ term. Note that the minimization is still only considering the columns corresponding to the primary controls.}
\Description{DESCRIPTION}
\label{fig:closedsource_stage3}
\end{figure}

\begin{figure}[!htb]
\includegraphics[width=\linewidth]{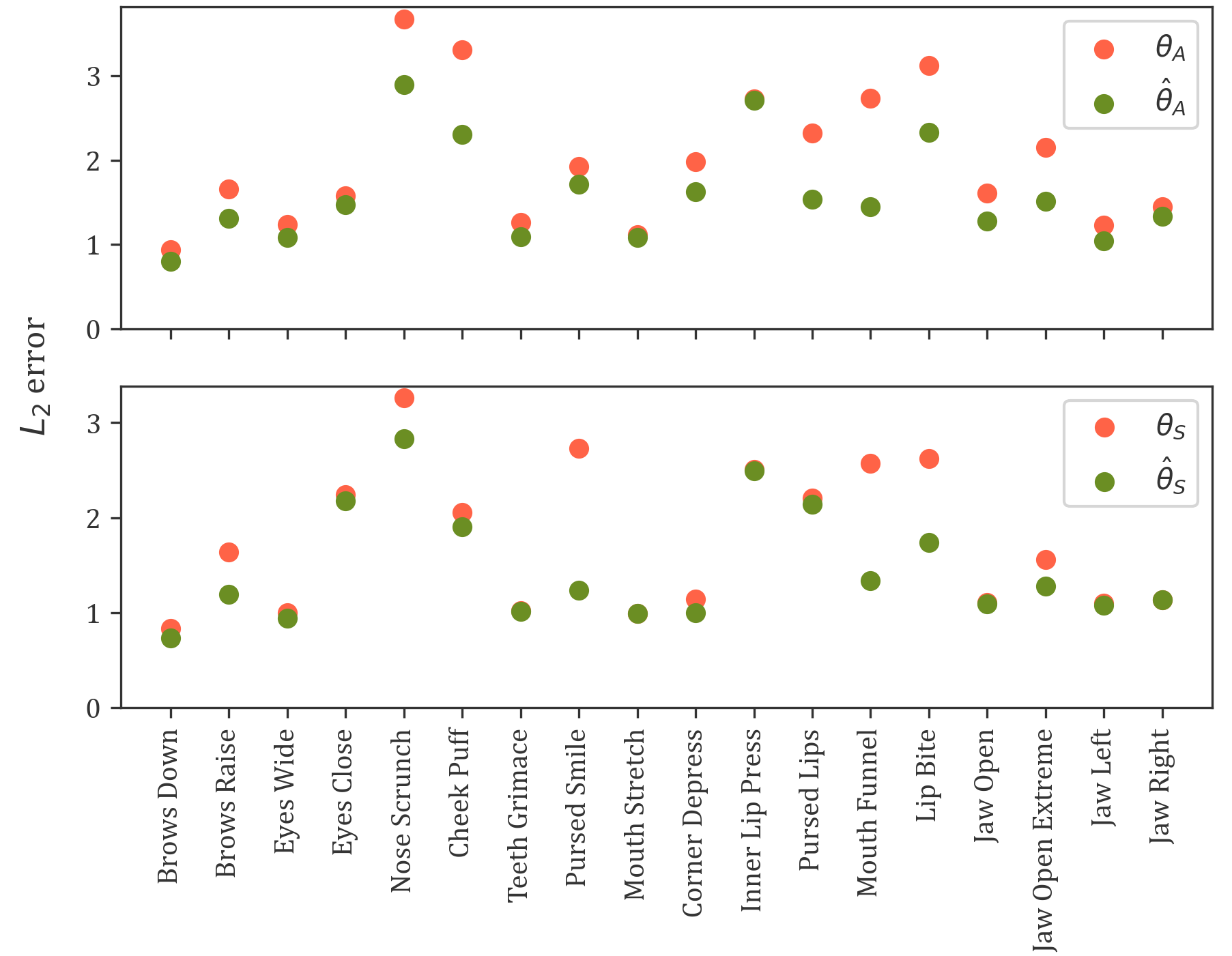}
\vspace*{-5ex}
\caption{$L_2$ errors (according to all $\gamma_1$ terms) on the various expressions before (red) and after (green) optimizing $\theta_A \rightarrow \hat{\theta}_A$ (top) and $\theta_S \rightarrow \hat{\theta}_S$ (bottom) using the full rig while filtering the primary controls on each expression via $\gamma_1^{\mathcal{H}}$ and $\gamma_2^{\mathcal{H}}$ and additionally filtering various spurious controls via an additional $\gamma_1^{\mathcal{H}}$ term. The difference between this figure and Figure \ref{fig:closedsource_stage3} is that the minimization is now considering the columns of the spurious controls while freezing the columns of the primary controls.}
\Description{DESCRIPTION}
\label{fig:closedsource_stage4}
\end{figure}

\begin{figure}[!htb]
\includegraphics[width=\linewidth]{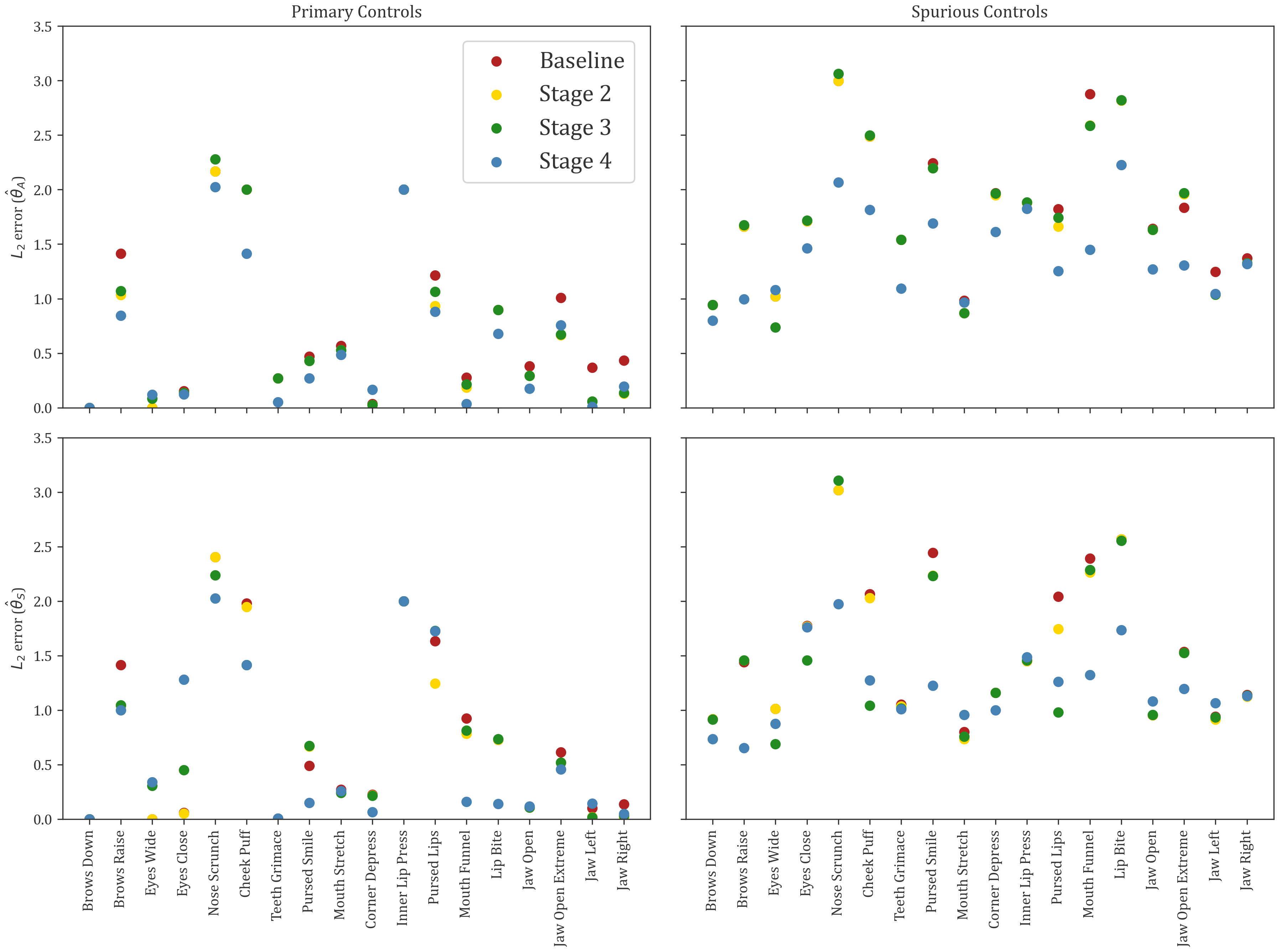}
\vspace*{-5ex}
\caption{Summary of the improvement in the $L_2$ errors (according to $\gamma_1$) throughout all three stages (there is no first stage on a closed-source tracker) of the process (see Figures \ref{fig:closedsource_stage2} to \ref{fig:closedsource_stage4}). The improvement in the primary controls is shown to the left, and the improvement in the spurious controls is shown to the right. Optimizing $\theta_A \rightarrow \hat{\theta}_A$ is shown on the top, and optimizing $\theta_S \rightarrow \hat{\theta}_S$ is shown on the bottom.}
\Description{DESCRIPTION}
\label{fig:closedsource_allstages}
\end{figure}

In this section, we consider a black-box tracker where one cannot make use of $\trkd$. Once again, we aim to improve the semantics of the tracker by modifying $\theta_A \rightarrow \hat{\theta}_A$ and $\theta_S \rightarrow \hat{\theta}_S$; in addition, all the trackers use the same reconstructed geometry from Titan. Figures \ref{fig:closedsource_geometryreconstruction} and \ref{fig:closedsource_geometryreconstruction_viz} show the effect of modifying $\theta_A \rightarrow \hat{\theta}_A$ and $\theta_S \rightarrow \hat{\theta}_S$ on the ability of the tracker to invert the rig and match the geometry. Although the first stage from Section \ref{sec:opensourcetracker} cannot be used to bootstrap the process, since one cannot make use of $\trk_D$, the other three stages are straightforward to implement. The results of the optimization are shown in Figures \ref{fig:closedsource_stage2}, \ref{fig:closedsource_stage3}, \ref{fig:closedsource_stage4}, and a summary of how the full tracker $\trk$ improves over all stages is shown in Figure \ref{fig:closedsource_allstages}. 


When considering a black-box tracker, it is important to note that a $\theta_S$ determined via Simon-Says may not work as well as the default $\theta_A$ because the internal parameters $\psi$ are often overfit to certain properties assumed to be intrinsic to $\theta_A$. Although the straightforward remedy would be to re-optimize the internal tracker parameters $\psi$ via Equation \ref{eq:trackerobj} aiming to increase the efficacy of $\theta_S$, this may not be possible with a black-box tracker. This means that our ability to optimize $\theta_S \rightarrow \hat{\theta}_S$ to obtain a viable $\hat{\theta}_S$ without re-optimizing the internal tracker parameters $\psi$ can be seen as an additional contribution from our approach. That is, our approach facilitates the use of custom rigs and black-box trackers.

Perhaps one of the most important things to keep in mind when perturbing $\theta_A \rightarrow \hat{\theta}_A$ or $\theta_S \rightarrow \hat{\theta}_S$ without the ability to re-optimize $\psi$ via Equation \ref{eq:trackerobj} or even to understand how $\psi$ was optimized is that sensitive threshold conditions may have been used. For example, a lip sealing constraint may activate when the lips are close enough together but otherwise do nothing. This means that a small perturbation of the rig could cause lips that were previously sealed via this constraint to instead be noticeably open even though their actual position prior to the constraint activation only changes by a small amount. 

\section{Retargeting}
\label{sec:retargeting}

Having shown that the optimization framework presented in Section \ref{sec:optimizing_the_tracking_rig} can be used to optimize the default MetaHuman rig parameters $\theta_A$ or Simon-Says (see Section \ref{sec:simonsays}) rig parameters $\theta_S$ used by an open-source (see Section \ref{sec:opensourcetracker}) or closed-source (see Section \ref{sec:blackboxtracker}) tracker and that the optimized results still typically reconstruct reasonable geometry when inverting the rig (see Figures \ref{fig:opensource_geometryreconstruction}, \ref{fig:opensource_geometryreconstruction_viz}, \ref{fig:closedsource_geometryreconstruction}, \ref{fig:closedsource_geometryreconstruction_viz}), this section demonstrates the comparable efficacy of using $\theta_A$, $\hat{\theta}_A$, $\theta_S$, and $\hat{\theta}_S$ in the tracker in the context of retargeting where differing rig controls can produce noticeably different results. In order to do this, we choose a target subject, reconstruct their neutral geometry, fit it with an animation rig, and calibrate that rig via Simon-Says as discussed in Section \ref{sec:simonsays}. For the sake of evaluation, we show images and geometric reconstructions of both the performer and the target making semantically equivalent but geometrically different expressions. As a baseline (i.e. before any of our improvements), we show the results obtained using the default MetaHuman rig (i.e. $\theta_A$) in the tracker mapped to the default MetaHuman rig ($\theta_A$ instead of $\theta_S$) for the target. The results shown in this section were all chosen from the pangram used in Figures \ref{fig:opensource_geometryreconstruction}, \ref{fig:opensource_geometryreconstruction_viz}, \ref{fig:closedsource_geometryreconstruction}, and  \ref{fig:closedsource_geometryreconstruction_viz}  with data that remained unseen in the Simon-Says calibration (see Section \ref{sec:simonsays}) and tracker optimization (see Sections \ref{sec:opensourcetracker} and \ref{sec:blackboxtracker}) making it quite useful for predicting generalization. 

\begin{figure*}[p]
\includegraphics[width=\linewidth]{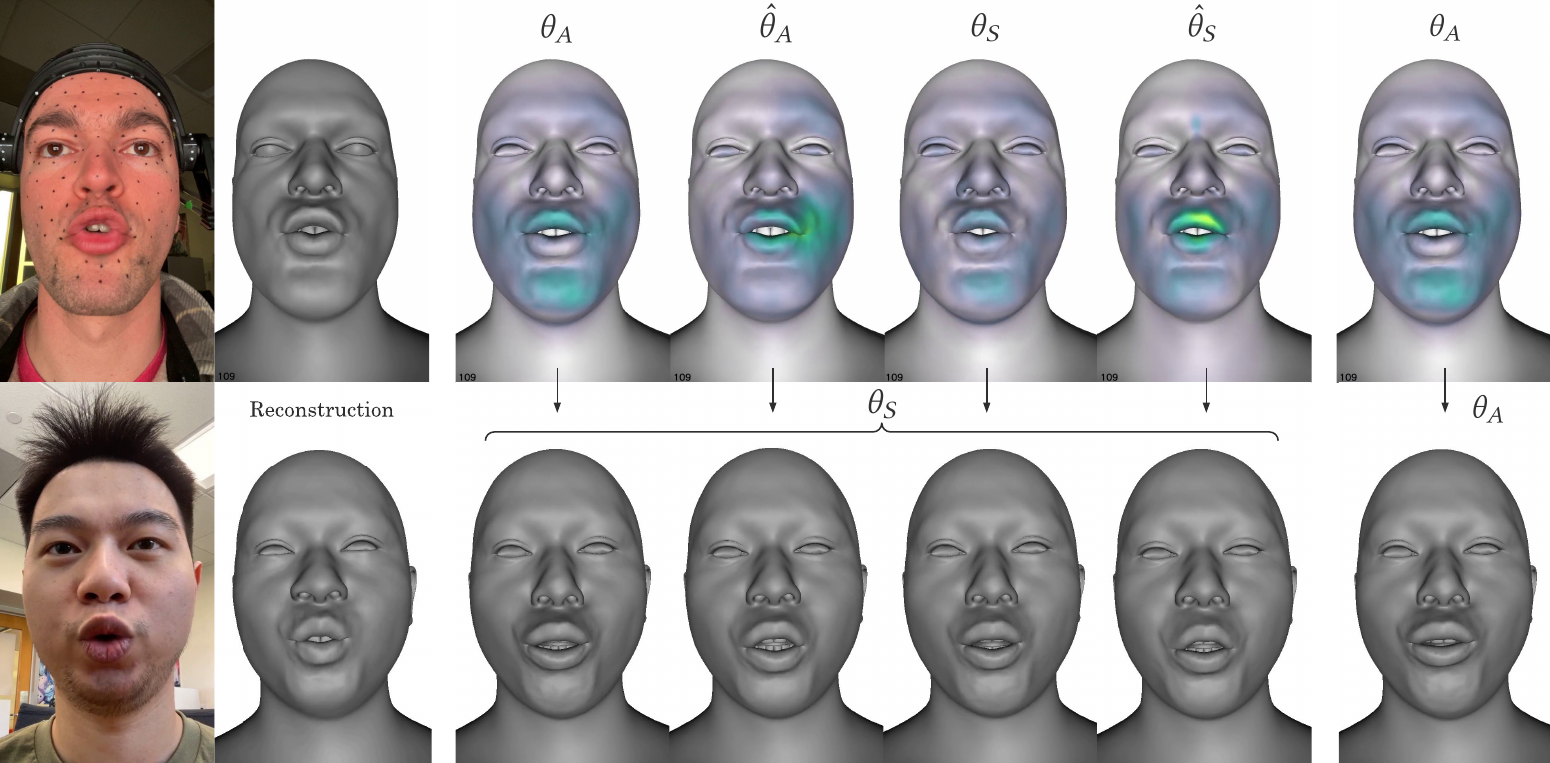}
\vspace*{-3ex}
\caption{Open-source tracker: retargeting to another identity. The top row shows geometry errors for the different trackers, and the bottom row shows the retargeted results. Both subjects are speaking the same word but have a different motion signature. Modifying $\theta_A \rightarrow \hat{\theta}_A$ and $\theta_S \rightarrow \hat{\theta}_S$ both improve the retargeted results. The last column shows a state-of-the-art off-the-shelf baseline, not using any of our approaches, for the sake of comparison.}
\label{fig:retargeting_haodi1}
\Description{DESCRIPTION}
\end{figure*}

\begin{figure*}[p]
\includegraphics[width=\linewidth]{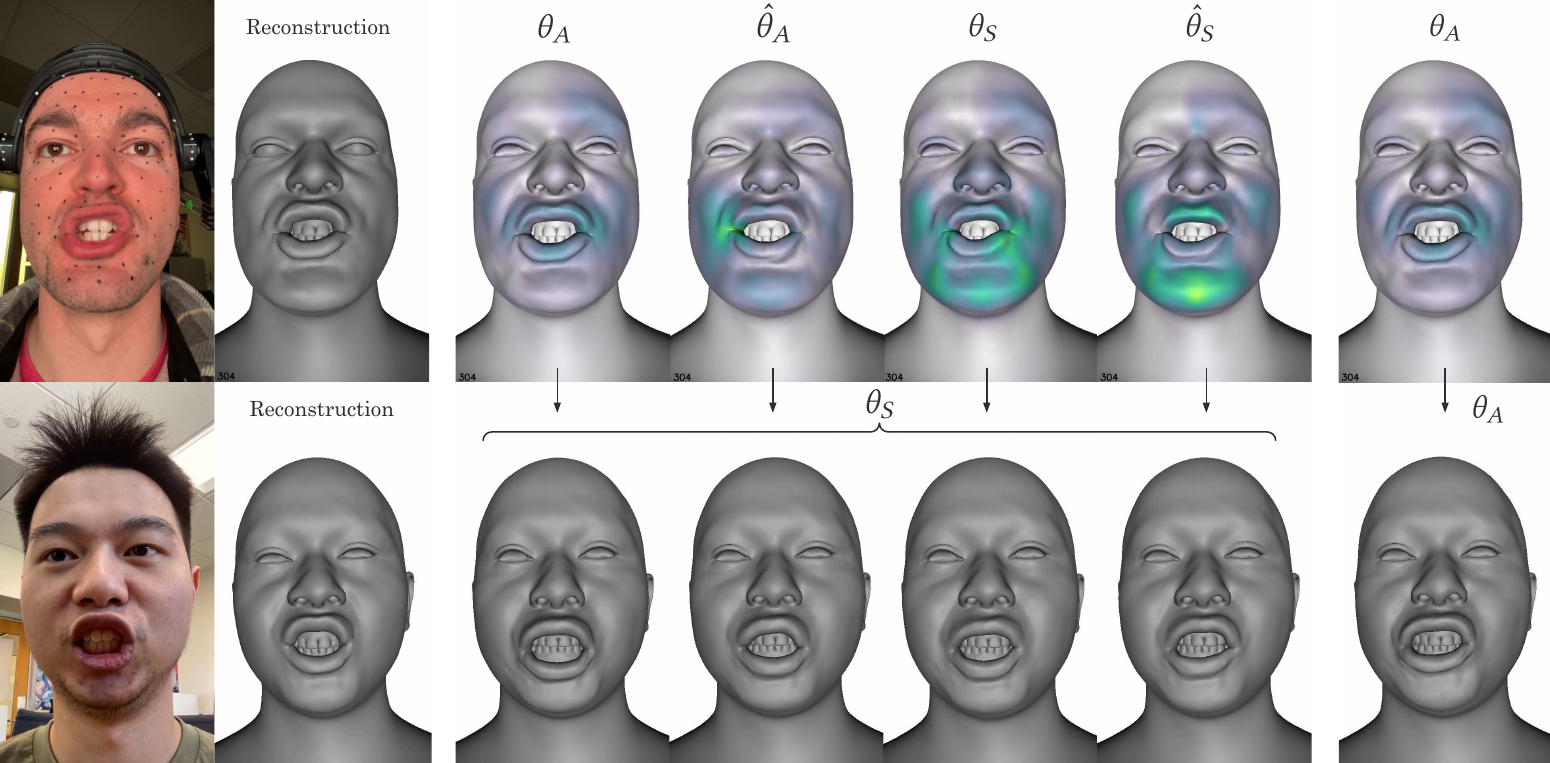}
\vspace*{-3ex}
\caption{Open-source tracker: retargeting to another identity. The top row shows geometry errors for the different trackers, and the bottom row shows the retargeted results. On this funnel-type expression, modifying $\theta_A \rightarrow \theta_S$ improves the severity of the expression due to the difference in motion signatures, while modifying $\theta_A \rightarrow \hat{\theta}_A$ and $\theta_S \rightarrow \hat{\theta}_S$ improves lip shape as compared to the reference. The last column shows a state-of-the-art off-the-shelf baseline, not using any of our approaches, for the sake of comparison.}
\label{fig:retargeting_haodi2}
\Description{DESCRIPTION}
\end{figure*}

\begin{figure*}[p]
\includegraphics[width=\linewidth]{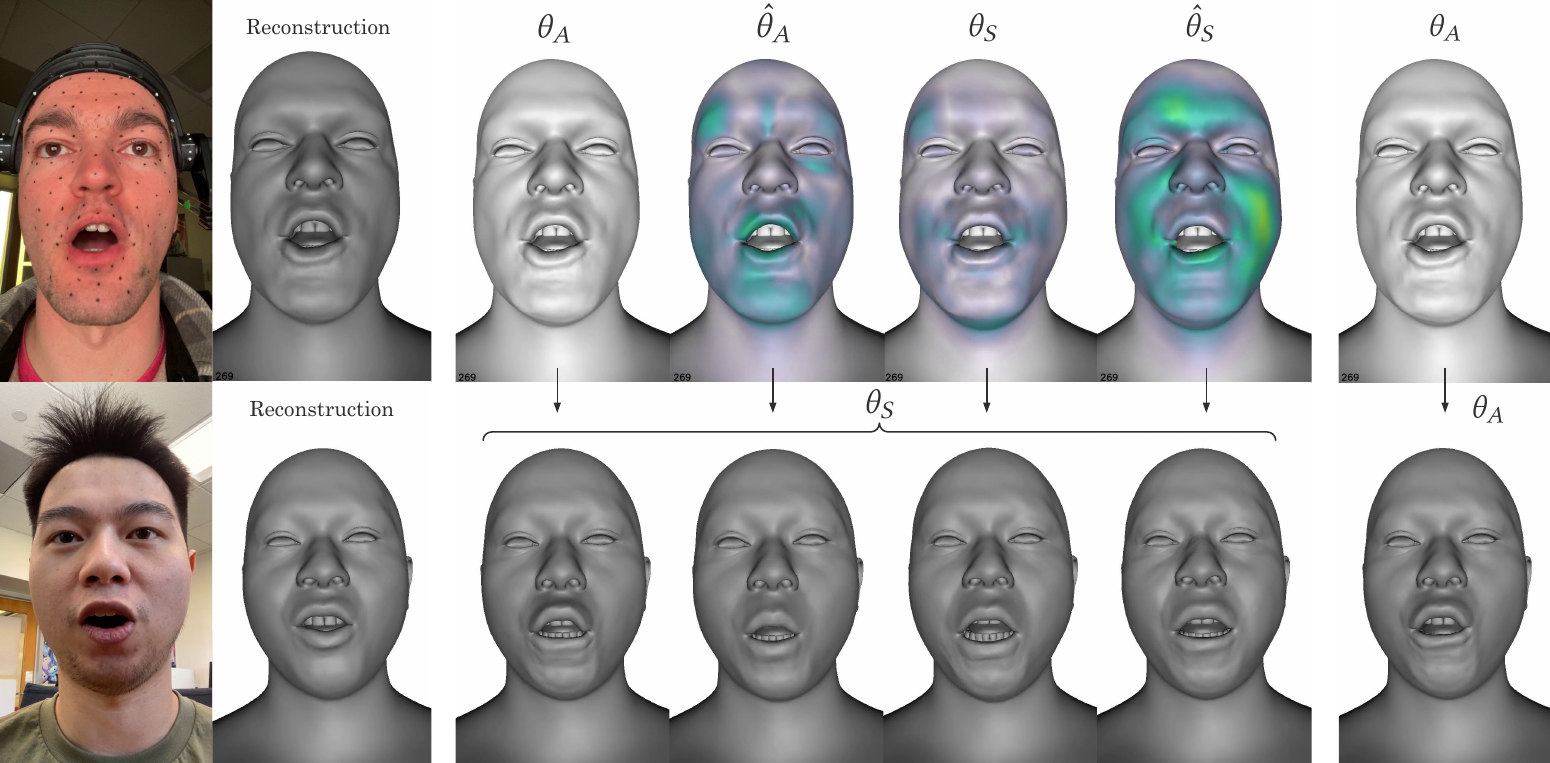}
\vspace*{-3ex}
\caption{Closed-source tracker: retargeting to another identity. The top row shows geometry errors for the different trackers, and the bottom row shows the retargeted results. This example was chosen to highlight the plausible behavior of our approach under asymmetries, even though both the Simon-Says calibration and the tracker optimization only considered expressions that were aspirationally symmetric. The last column shows a state-of-the-art off-the-shelf baseline, not using any of our approaches, for the sake of comparison.}
\label{fig:retargeting_blackbox_haodi1}
\Description{DESCRIPTION}
\end{figure*}

Figures \ref{fig:retargeting_haodi1}, \ref{fig:retargeting_haodi2}, \ref{fig:retargeting_blackbox_haodi1} show some representative results. As expected, the performer geometry (top rows) remains similar regardless of whether $\theta_A$, $\hat{\theta}_A$,  $\theta_S$, or $\hat{\theta}_S$ was used. In fact, as compared to the so-called reconstruction (as can be seen by the vertex color shading), the geometry can sometimes get worse when modifying $\theta_A \rightarrow \hat{\theta}_A$, $\theta_S \rightarrow \hat{\theta}_S$, or even $\theta_A \rightarrow \theta_S$; however, these differences are unimportant, since the focus is on obtaining good results on the target. As far as the target is concerned (bottom rows), modifying $\theta_A \rightarrow \hat{\theta}_A$, $\theta_S \rightarrow \hat{\theta}_S$, and $\theta_A \rightarrow \theta_S$ all improve the results, illustrating the efficacy of our approach. In general, key poses in performance or speech are better explained by relevant primary controls and less explained by undesired tweaker controls by using our approach, leading to cleaner and more accurate expressions.

\section{Puppeteering}
\label{sec:puppeteering}

\begin{figure}[!htb]
\includegraphics[width=\linewidth]{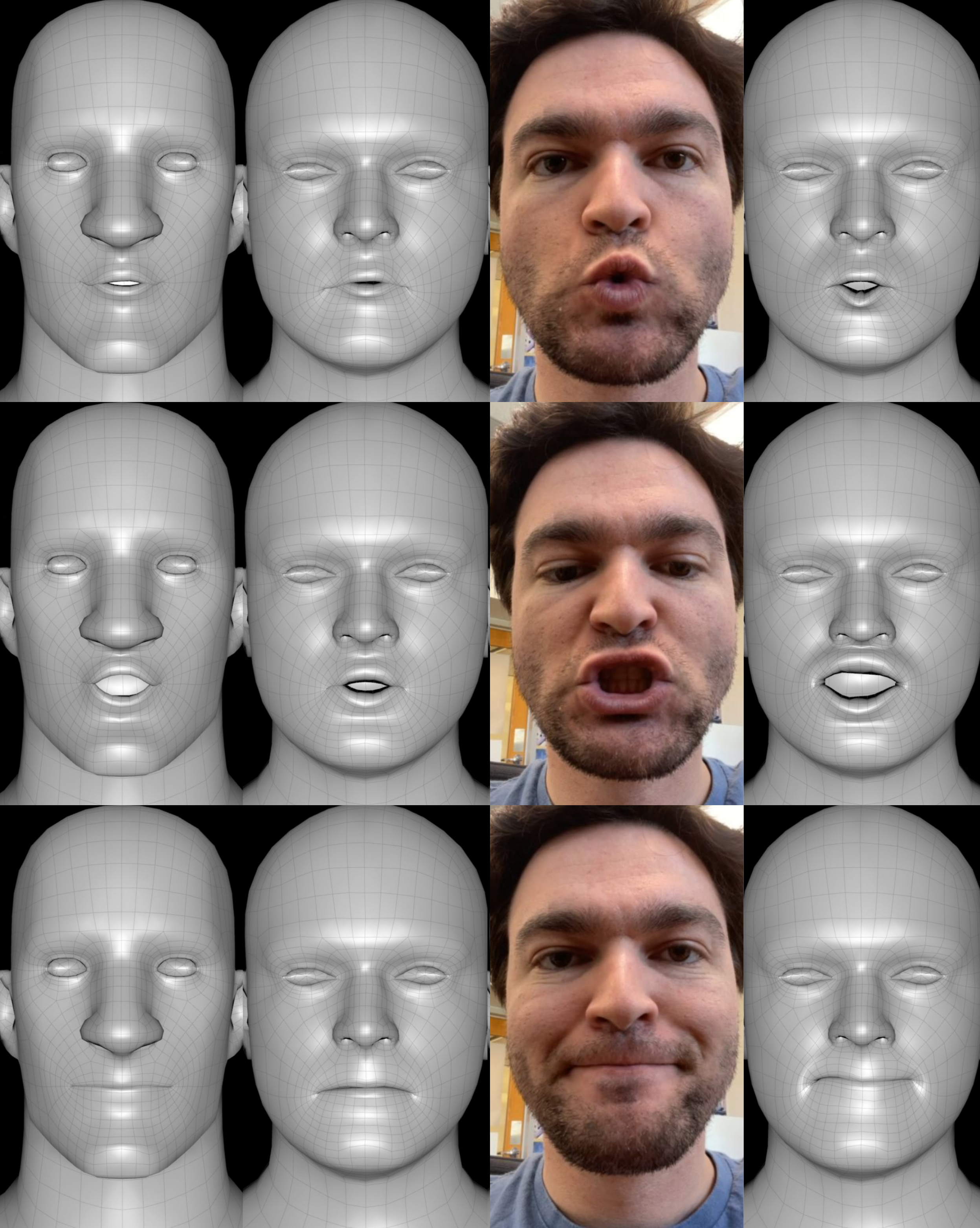}
\vspace*{-1ex}
\caption{Simon-Says process on specific phoneme-level controls (``OO'', ``CH'', ``M/B/P'') unique to the game/VR character rig (one on each row). Left to right: Game/VR character rig, morphed rig, performer's expression, Simon-Says results. Note that the first two columns provide little usable input for the performer, so they need to rely more heavily on the description of each desired expression.}
\label{fig:puppeteering_simonsays}
\Description{DESCRIPTION}
\end{figure}

Whereas Section \ref{sec:retargeting} retargeted to another human, here we consider retargeting to a synthetically generated humanoid character. Various rig frameworks, especially those for game/VR characters, will differ significantly in their semantic control space. Given the difficulties associated with remapping controls between two disparate rigs, it seems prudent to utilize the game/VR character rig on the performer as well. Thus, we create a game/VR character representation for the performer. This is accomplished by first creating a correspondence between the game/VR character's triangle mesh and the performer's triangle mesh in order to assign a game/VR character triangle mesh to the performer's neutral geometry. Note that the teeth were ignored, since they were not used in the tracking of the performer. Then, the game/VR character rig was volumetrically morphed to the performer's new game/VR character consistent neutral geometry. Next, the morphed rig was calibrated via Simon-Says. It is inappropriate to use the same Simon-Says expression set as was used for person-to-person retargeting. Instead, we designed a set of basis expressions that better covered the rig controls of the game/VR character. In total, fifteen expressions
\begin{itemize}
   \item neutral, eyes close, eyes wide, brows raise, brows down, teeth grimace, pursed smile, corner depress, jaw open, jaw left, jaw right
   \item ``OO'' phoneme, ``CH'' phoneme, ``M/B/P'' phoneme, ``F/V'' phoneme
 \end{itemize}
were used (including the four phoneme-level expressions). As shown in Figure \ref{fig:puppeteering_simonsays}, the low expressivity of the game/VR character rig meant that the Simon-Says results did not well match the characteristics of the performer; nevertheless, the vast improvement of the basis shapes led to more accurate semantic interpretation in our retargeting experiments.

\begin{figure*}[p]
\includegraphics[width=0.9\linewidth]{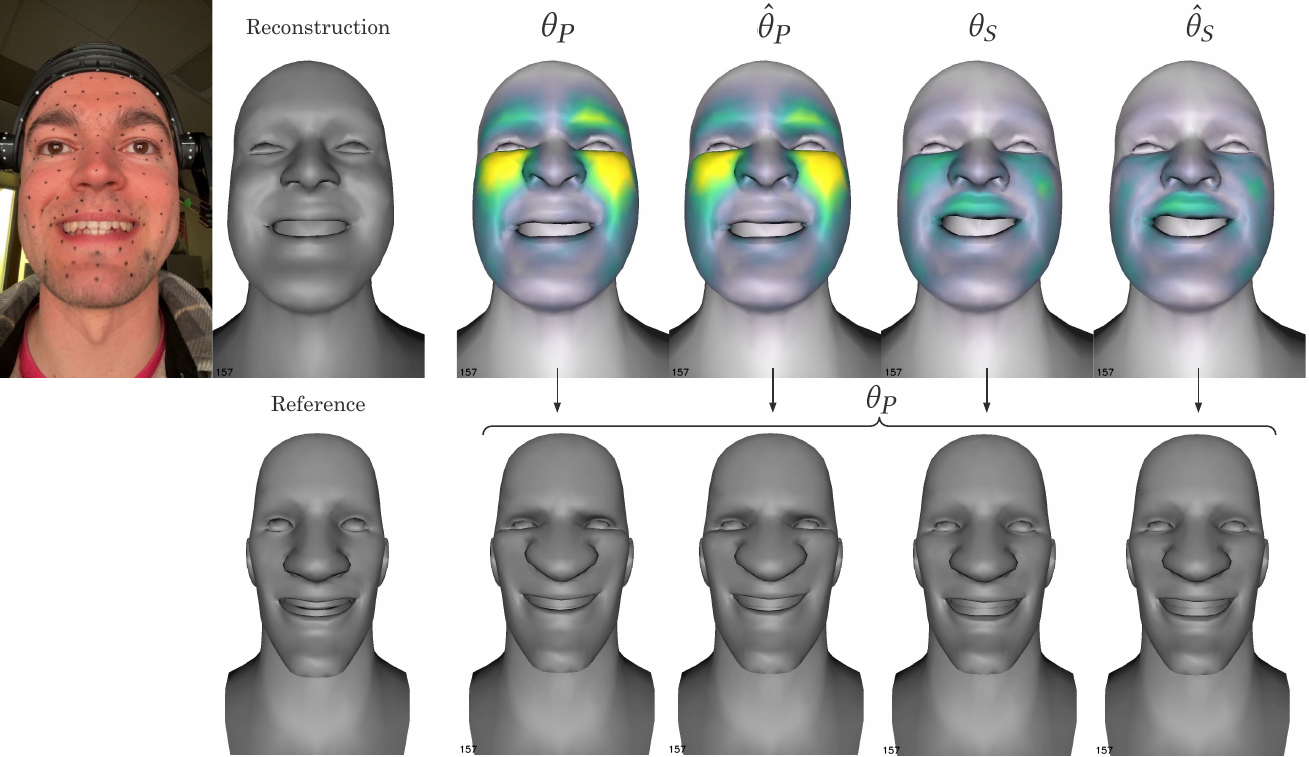}
\vspace*{-2ex}
\caption{Open-source tracker: puppeteering. The top row shows geometry errors for the different trackers, and the bottom row shows the retargeted results. In this expression, extraneous controls in the eye region activated by the uncalibrated rig parameters $\theta_P$ are remedied by $\theta_S$. In addition, $\hat{\theta}_S$ retargets to a mouth shape that better matches the reference for this phoneme, despite additional reconstruction errors being introduced around the mouth.}
\label{fig:puppeteering_jonesy1}
\Description{DESCRIPTION}
\end{figure*}

\begin{figure*}[p]
\includegraphics[width=0.9\linewidth]{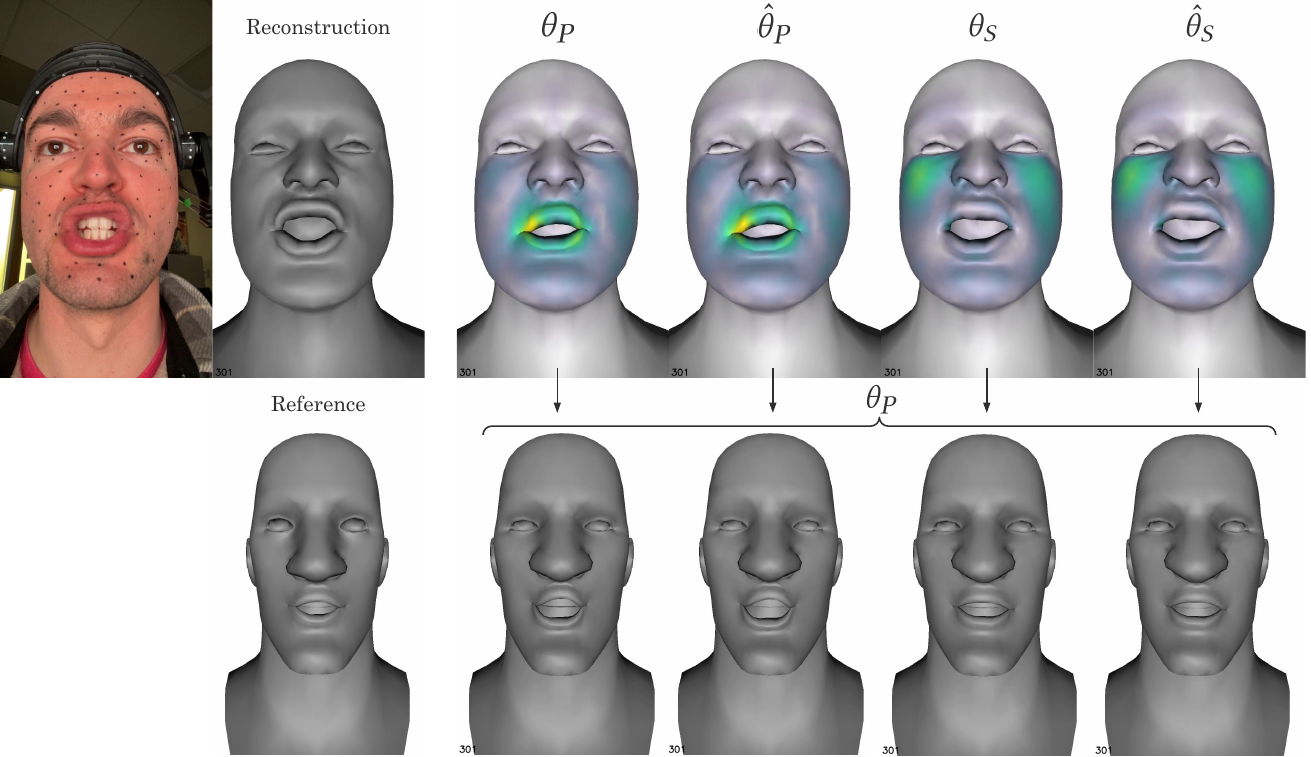}
\vspace*{-2ex}
\caption{Open-source tracker: puppeteering. The top row shows geometry errors for the different trackers, and the bottom row shows the retargeted results. The ``CH'' phoneme control is activated correctly using $\theta_S$ and $\hat{\theta}_S$, resulting in the retargeted expression more closely matching the reference. The $\theta_P$ and $\hat{\theta}_P$ results over-emphasize the lip movement in the retarget since the their motion signatures are significantly different.}
\label{fig:puppeteering_jonesy2}
\Description{DESCRIPTION}
\end{figure*}

\begin{figure*}[p]
\includegraphics[width=0.9\linewidth]{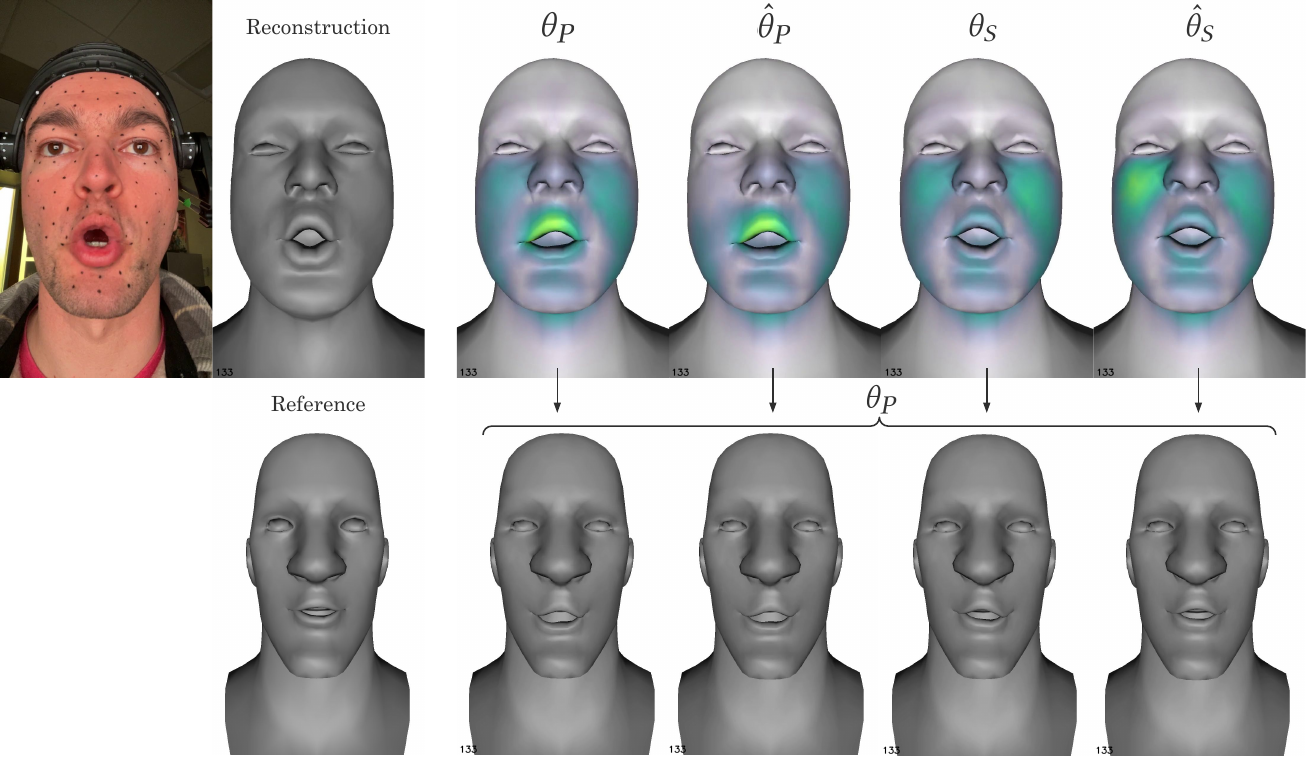}
\vspace*{-2ex}
\caption{Open-source tracker: puppeteering. The top row shows geometry errors for the different trackers, and the bottom row shows the retargeted results. The ``OO'' phoneme control is activated more correctly when using $\theta_S$, and even more correctly when using $\hat{\theta}_S$.}
\label{fig:puppeteering_jonesy3}
\Description{DESCRIPTION}
\end{figure*}

\begin{figure*}[p]
\includegraphics[width=0.9\linewidth]{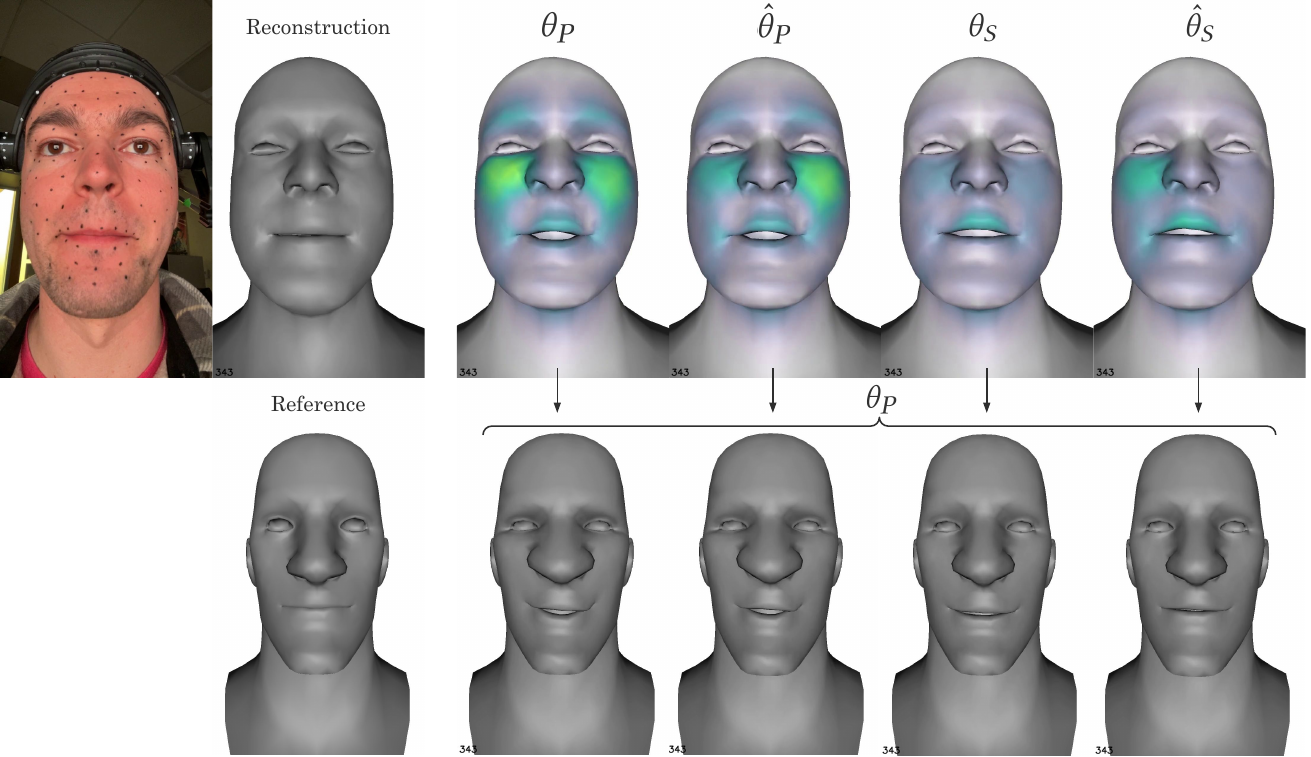}
\vspace*{-2ex}
\caption{Open-source tracker: puppeteering. The top row shows geometry errors for the different trackers, and the bottom row shows the retargeted results. The ``M/B/P'' phoneme control improves when using $\theta_S$ and improves even more when using $\hat{\theta}_S$, resulting in the better lip sealing. This is especially important for avoiding the uncanny valley.}
\label{fig:puppeteering_jonesy5}
\Description{DESCRIPTION}
\end{figure*}

\begin{figure*}[p]
\includegraphics[width=0.9\linewidth]{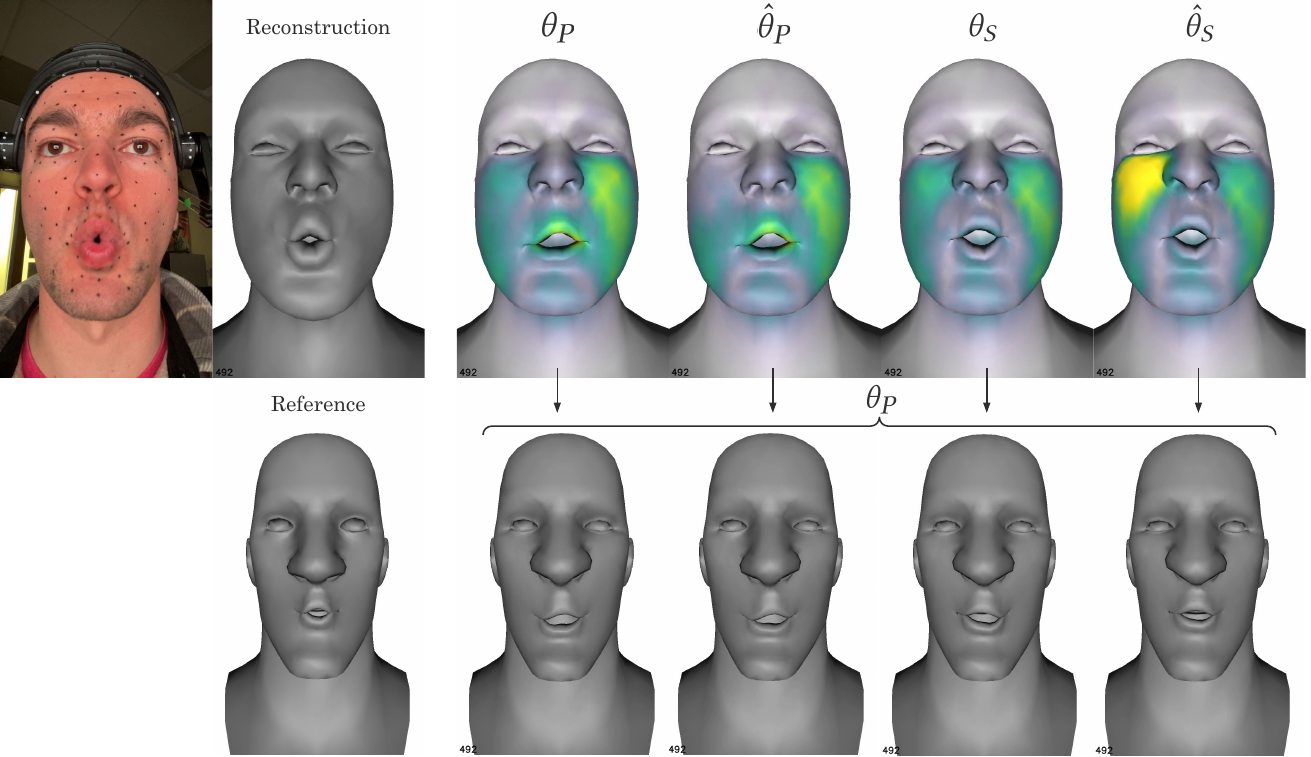}
\vspace*{-2ex}
\caption{Open-source tracker: puppeteering. The top row shows geometry errors for the different trackers, and the bottom row shows the retargeted results. Lip tightness and shape are improved during the word ``quacking'' when using $\theta_S$ and even much more so when using $\hat{\theta}_S$.}
\label{fig:puppeteering_jonesy6}
\Description{DESCRIPTION}
\end{figure*}

Since the closed-source MetaHuman Animator tracker is heavily designed to work specifically with the MetaHuman rig framework, our experiments only use our custom open-source tracker (see Section \ref{sec:opensourcetracker}). In addition, unlike the examples shown in Section \ref{sec:retargeting}, there is no convenient off-the-shelf solver that can be used as a baseline for comparison. Although some of the prior works discussed in Section \ref{sec:relatedwork} could potentially be utilized, it would not be straightforward to implement them on this example. The fact that it is straightforward to apply our methodology to the puppeteering of a highly crafted game/VR character rig emphasizes the efficacy of our approach. For results on a pangram, see Figures \ref{fig:puppeteering_jonesy1} through \ref{fig:puppeteering_jonesy6}. Although we did not expect the puppet rig to be useful without Simon-Says calibration, we show $\theta_P$ and $\hat{\theta}_P$ results as a baseline for comparison. 

\section{Conclusion}
\label{sec:conclusion}

Unfortunately, too many would-be trackers produce animation curves that require subsequent significant cleanup effort by artists in order to obtain results that can be used effectively for retargeting. Our mathematical analysis illustrated that a tracker can be thought of as geometric reconstruction followed by rig inversion highlighting the fact that those who merely stress the efficacy of their geometric reconstructions are missing a key component of the problem. Moreover, we showed that it can even be beneficial to relax the accuracy of the geometric reconstruction as a tradeoff for obtaining better and more usable animation curves via the rig inversion. 

Although we did build an open source tracker in order to test our hypotheses, our real goal is to improve off-the-shelf closed source trackers specialized for use in industry. In the latter case, one does not necessarily have access to the internal tracker parameters or have the ability to optimize them or to rewrite the tracker to be differentiable (even if one could modify such a tracker, it may no longer work as desired). However, all of these trackers do accommodate variations in rig parameters, since rig parameters (and geometry) vary across individuals and characters. Thus, we set out to improve the animation curve output of the tracker by modifying the accessible rig parameters. In order to do this without access to the internal workings of the (closed source) tracker itself, we leveraged the implicit function theorem and a reformulation of Broyden's method in order to implicitly differentiate the closed source tracker with respect to the parameters of the animation rig. Our results showed that the derivative estimates were valid enough to minimize the various objective functions one might consider when aiming to improve the animation curve output of the tracker. 

Compared to the baseline results obtained without using our approach, we showed that the rig parameters could be modified (either $\theta_A \rightarrow \hat{\theta}_A$ in Section \ref{sec:retargeting} or $\theta_P \rightarrow \hat{\theta}_P$ in Section \ref{sec:puppeteering}) in order to obtain improved results when retargeting from one individual to another (in Section \ref{sec:retargeting}) or when utilizing performance capture for puppeteering (in Section \ref{sec:puppeteering}). It is important to stress that both of these results are also just baselines for the sake of comparison, since we strongly advocate using personalized animation rigs in order to capture the varying motion signatures particular to each individual. In order to do this, we utilize a Simon-Says calibration process in order to personalize the performer's animation rig (obtaining $\theta_S$) before modifying it (i.e. $\theta_S \rightarrow \hat{\theta}_S$) via our implicit differentiation approach. The results obtained using $\hat{\theta}_S$ for the rig parameters seen by the tracker represent our proposed method. Note that a personalized animation rig is also preferable for the target in person-to-person retargeting and that the animation rig for a character (i.e. $\theta_P$) is already highly personalized by design. 

We obviously did not exhaust all possible approaches for improving rig parameters, nor did we incorporate an exhaustive list of all possible hand-tuned constraints on rig parameters and rig motion. This can be seen as either limitations or future work. That is, although we show the efficacy of our approach, we did not build the ultimate rig or the ultimate tracker. Given the high level of expertise of those who work in this field, we felt that providing an additional tool was more important than providing some sort of claim about an ultimate solution. One interesting avenue for future work consists of generalizing the concepts presented in this paper to a wider set of approaches. For example, consider decomposing the latent code obtained via GAN inversion of a specific individual into two components, where a parallel can be drawn between one component and a triangulated surface representation of the neutral pose and a second parallel can be drawn between the other component and animation rig parameters.

\begin{acks}
Research supported in part by ONR N00014-24-1-2644, ONR N00014-21-1-2771, and ONR N00014-19-1-2285. We would like to acknowledge both SONY and Epic Games for additional support.
\balance
\end{acks}


\bibliographystyle{ACM-Reference-Format}
\bibliography{reference}

\end{document}